\magnification = \magstep0
\hsize=11.3truecm  \hoffset=3.0truecm  \vsize=18.6truecm  \voffset=2.5truecm
\font\sc=cmr8
\def\dblbaselines{\baselineskip=12pt    \lineskip=0pt   \lineskiplimit=0pt}
\def\vsl{\vskip\baselineskip}   \def\vs{\vskip 5pt} \def\vsss{\vskip 3pt}
\parindent=10pt \nopagenumbers

  
\def\col#1{\empty} 
\def\bw#1{#1}       

\def\col#1{#1}      
\def\bw#1{\empty}
     

\def\t#1{#1} 
\def\t#1{\empty}
\def\omit#1{\empty}
  \def\s{\null}  
  \def\ba{\kern -1pt}
\parskip = 0pt 
\def\ts{\thinspace} \def\cl{\centerline}
\def\ni{\noindent}  
\def\nnhi{\noindent \hangindent=0.42truecm}
\def\hhhi{\parindent 0.42truecm \indent \hangindent=0.42truecm}

\def\iihi{\noindent \kern 0.45truecm \hangindent=0.45truecm}
  
\def\makeheadline{\vbox to 0pt{\vskip-30pt\line{\vbox to8.5pt{}\the
                               \headline}\vss}\nointerlineskip}
\def\toppageno{\headline={\hss\tenrm\folio\hss}}
\def\footnoterule{\kern-3pt \hrule width \hsize \kern 2.6pt \vskip 3pt}
\output={\plainoutput}    \pretolerance=10000   \tolerance=10000


\def\sup1{$^{\rm 1}$} \def\sup2{$^{\rm 2}$}
\def\r0{$\rho_0$}   
 \def\0{\phantom{0}} \def\bb{\kern -2pt}
\def\1{\phantom{1}}         
\def\etal{{\it et~al.~}}
\def\gapprox{$_>\atop{^\sim}$} \def\lapprox{$_<\atop{^\sim}$}
          
\newdimen\sa  \def\sd{\sa=.1em \ifmmode $\rlap{.}$''$\kern -\sa$
                               \else \rlap{.}$''$\kern -\sa\fi}
\newdimen\sb  \def\md{\sb=.02em\ifmmode $\rlap{.}$'$\kern -\sb$
                               \else \rlap{.}$'$\kern -\sb\fi}

\def\ss{\ifmmode ^{\prime\prime}$\kern-\sa$ \else $^{\prime\prime}$\kern-\sa\fi}
\def\mm{\ifmmode ^{\prime}$\kern-\sa$ \else $^{\prime}$\kern-\sa \fi}
\def\msun {M$_{\odot}$~}  
  
\def\m31{M{\ts}31} \def\mm32{M{\ts}32} \def\mmm33{M{\ts}33} \def\M87{M{\ts}87} 

\def\nhi2{\noindent \hangindent=2.79cm}
\font\bigbig=cmbx12 scaled 1200
\font\big=cmr12 scaled 1200

 


\cl{\null} \vskip 21pt    
 
\ni {\bigbig SECULAR EVOLUTION AND THE} \vsss

\ni {\bigbig FORMATION OF PSEUDOBULGES} \vsss

\ni {\bigbig IN DISK GALAXIES}

\vsl
 
\ni{\big John Kormendy} 
\vsl
\ni {Department of Astronomy, University of Texas,}

\ni {Austin, Texas 78712, USA; email: kormendy@astro.as.utexas.edu}

\vsl

\ni{\big Robert C.~Kennicutt, Jr.}
\vsl
\ni{Department of Atronomy, Steward Observatory, University of Arizona,}

\ni{Tucson, Arizona 85721, USA; email: rkennicutt@as.arizona.edu}
 
\vsl

\nhi2 KEY WORDS: \quad galaxy dynamics, galaxy structure, galaxy evolution

\vsl\vsl

\dblbaselines

\ni ABSTRACT

      The Universe is in transition.  At early times, galactic evolution was
dominated by hierarchical clustering and merging, processes that are violent 
and rapid.\ts~In the far future, evolution will mostly~be~secular:~the slow
rearrangement of energy and mass that results from interactions involving
collective phenomena such as bars, oval disks, spiral structure, and triaxial
dark halos.  Both processes are important now.  This~paper reviews internal
secular evolution, concentrating on one important consequence, the buildup of
dense central components in disk galaxies that look like classical, merger-built
bulges but that were made slowly out of disk gas.  We call these pseudobulges.

      We begin with an ``existence proof'' -- a review of how bars rearrange
disk gas into outer rings, inner rings, and stuff dumped onto~the~center.
Simulation results correspond closely to the morphology of barred
galaxies.  In the simulations, gas transported to small radii reaches
high densities that plausibly feed star formation. In the observations,
many SB and oval galaxies have dense central concentrations of gas and star
formation.  Optical colors and spectra often imply young stellar populations.
So the formation of pseudobulges is well supported by theory and observations.
It is embedded in a broader evolution picture that accounts for much of the
richness observed in galaxy structure.  

      If secular processes built dense central components that masquerade
as bulges, how can we distinguish them from merger-built bulges?  Observations
show that pseudobulges retain a memory of their~disky~origin.  That is, they
have one or more characteristics of disks:~(1) flatter shapes than those of
classical bulges, (2) correspondingly large ratios of ordered to random
velocities indicative of disk dynamics, (3) small velocity dispersions
$\sigma$ with respect to the Faber-Jackson correlation between $\sigma$
and bulge luminosity, (4) spiral structure or nuclear bars in the ``bulge''
part of the light profile, (5) nearly exponential brightness profiles, and (6)
starbursts.  These structures occur preferentially in barred and oval galaxies
in which secular evolution should be most rapid.  So the cleanest examples of
pseudobulges are recognizable. 

      Are their formation timescales plausible?  We use measurements of
central gas densities and star formation rates to show that pseudobulges of
the observed densities form on reasonable timescales of a few billion years.


      Thus a large variety of observational and theoretical results contribute
to a new picture of galaxy evolution that complements hierarchical clustering
and merging.  Secular evolution consists of more than the aging of stellar
populations.  Every galaxy is dynamically evolving.

\pageno=2\toppageno

\vsl\vsss
\ni {\bf 1.~INTRODUCTION}
\vsl

\ni This paper reviews internal processes of secular evolution in disk galaxies.
We concentrate on one important consequence:~the buildup of dense central
components that look like classical, i.{\ts}e., merger-built bulges but that
were made slowly by disks out of disk material.  These are called pseudobulges.
Our discussion updates a review by Kormendy (1993).

      The relative importance of the different physical processes of galaxy 
evolution (Figure 1) changes as the Universe expands.  Rapid processes that
happen in discrete events are giving way to slow, ongoing processes.  

      At early times, galactic evolution was dominated by a combination of
dissipative collapse (Eggen, Lynden-Bell, \& Sandage 1962; Sandage 1990) and
mergers (Toomre 1977a) of galaxies that virialized out of the density
fluctuations of cold dark matter.  These are the top processes in Figure~1.
The evolution timescale was short, on the order of the dynamical
time of an individual halo, $t_{\rm dyn} \sim (1 / G \rho)^{1/2}$, where $\rho$
is the mean density and $G$ is the gravitational constant (Binney \& Tremaine
1987, \hbox{Equation 2-30)}.  The processes were violent.  Many present-day
galaxies owe their properties to this violence.  Because mergers scramble
disks and induce dissipation and starbursts, they are thought to make classical
bulges and elliptical galaxies.  Here, we do not review classical bulges other
than to contrast them to pseudobulges.  Most work on galaxy evolution in the
past 25 years has concentrated on hierarchical clustering and mergers.  As the
Universe expands, and as galaxy clusters virialize and acquire large internal
velocities, mergers get less common (Toomre 1977a; Conselice et al.~2003).

     In the distant future, internal secular processes will become dominant. By
these, we mean slow processes, ones that have timescales that are much longer
than the dynamical time $t_{\rm dyn}$.  To be interesting, they must operate
over long times.  Some secular processes, such as disk heating via stellar
enounters with molecular clouds, are well known  
(Spitzer \& Schwarzschild 1951, 1953).  But star-star relaxation is
exceedingly slow almost everywhere in almost every galaxy.  Therefore, relevant
secular processes generally involve the interactions of individual stars or gas
clouds with collective phenomena such as bars, oval distortions,
spiral structure, and triaxial dark matter halos.  Also important are the interactions of these collective phenomena~with~each~other.  Given that
hierarchical clustering continues today, has there been time for secular
processes to have any significant effect?  A clue that the answer is ``yes''
is provided by galaxies with superthin -- and fragile -- disks but apparently no
bulges (e.{\ts}g., de Vaucouleurs 1974a; Goad \& Roberts 1981;  Matthews, $\phantom{000000000000000000000000}$

\vfill

\bw{\includegraphics{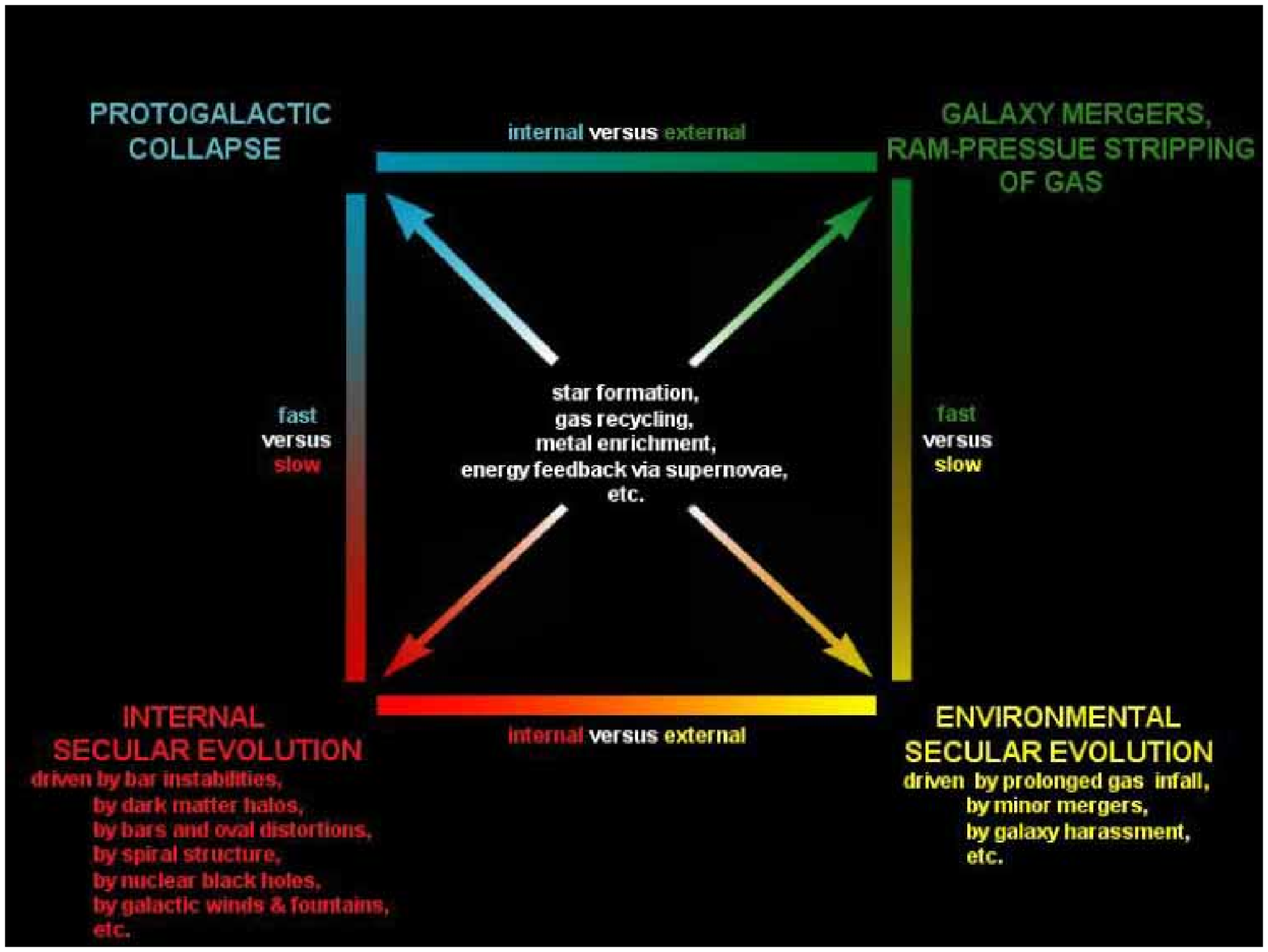}}

\col{\includegraphics{secularnew.ps}}

      {\it Figure 1}\quad Morphological box (Zwicky 1957) of processes of
galactic evolution updated from Kormendy (1982a).  Processes are divided 
vertically into fast (top) and slow (bottom).  Fast evolution happens on a
free-fall (``dynamical'') timescale, $t_{\rm dyn} \sim (G\s\rho)^{-1/2}$, where
$\rho$ is the density of the object produced and $G$ is the gravitational
constant.  Slow means many galaxy rotation periods.  Processes are divided
horizontally into ones that happen purely internally in one galaxy (left) and
ones that are driven by environmental effects such as galaxy interactions
(right).  The processes at center are aspects of all types of galaxy evolution.
This paper is about the internal and slow processes at lower-left.

\eject

\ni Gallagher, \& van Driel 1999b; Freeman 2000; van der Kruit et al.~2001).
It may seem counterintuitive to use bulgeless galaxies to argue that there
has been time for secular evolution to make pseudobulges, but it is quite 
remarkable that such galaxies can exist within the paradigm of hierarchical
clustering.  They again provide us with an ``existence proof''.  They show that
some galaxies have suffered no major merger violence since the onset of star
formation in the disk (Toth \& Ostriker 1992).~So there has been time for
secular evolution to be important in at least some galaxies. Given that mergers
make bulges and ellipticals, these tend to be late-type galaxies.  Actually,
since the latest-type galaxies are (pseudo)bulgeless, secular evolution is
likely to be most important in intermediate-late type galaxies, i.{\ts}e., Sbcs.
But even some S0 and Sa galaxies contain pseudobulges. Secular processes
have received less attention than galaxy mergers.

      Still, this subject has made rapid progress.  Many reviews discuss
secular evolution in barred and oval galaxies
(Kormendy 1979a,{\ts}b; 1981, 1982a; 
Norman 1984;
Combes 1991, 1998, 2000, 2001;
Bosma 1992;
Martinet 1995;
Friedli \& Benz 1995;
Pfenniger 1996a, b, 2000;
Sellwood \& Debattista 1996;
Buta 1995, 1999, 2000;
Athanassoula 2002;
Maciejewski 2003;
Wada 2003;
Sellwood \& Shen 2004;
and especially Sellwood \& Wilkinson 1993, hereafter SW93, and
Buta \& Combes 1996).
Discussions of pseudobulge formation include, besides the above,
Pfenniger \& Norman (1990);
Courteau (1996b);
Wyse, Gilmore, \& Franx (1997);
Carollo, Ferguson, \& Wyse (1999);
Kormendy \& Gebhardt (2001);
Balcells (2002);
and especially
Kormendy (1993)
and Carollo (2003).

      Theory and observations point to a variety of processes that redistribute
energy in disks.  Section 2 reviews evidence that bars and ovals rearrange disk 
gas into outer rings, inner rings, and central mass concentrations.  Crunching
gas makes stars (Schmidt 1959); this star formation produces a central stellar 
subsystem that has the high density and steep density gradient of a bulge but
that was not formed by galaxy mergers. 

    Secular evolution is not confined to barred and oval galaxies.  Bars can
self-destruct by building up the central mass concentration, so secular
evolution may have happened even if no bar is seen today.  Global spiral
structure also makes galaxies evolve, albeit more slowly than~do~bars.

    These processes are manifestations of very general dynamical principles.
Disks spread in radius -- the inner parts shrink and the outer parts expand --
because this lowers the total energy for fixed total angular momentum 
(Lynden-Bell \& Kalnajs 1972; Tremaine 1989).~Two-dimensional spreading by
angular momentum transport is as fundamental for rotation-dominated disks as is
three-dimensional spreading by energy transport in the core collapse of
ellipsoidal systems dominated by random motions.  The reason is the
same, too.  Self-gravitating systems have negative specific heats, so increasing
the central density by flinging away the periphery lowers the total energy
(Lynden-Bell \& Wood 1968; Binney \& Tremaine 1987).  What
makes evolution important in some systems and not in others?~The determining
factor is whether any evolution process is fast enough.  Core collapse
requires short relaxation times.~Galaxy disks have long relaxation times, so
their evolution is interesting only if they have an alternative to relaxation.
Non-axisymmetries provide the engine for rapid evolution.

      We will argue that pseudobulges are one result of this evolution.  Of key
importance is the observation that they retain some memory of their disky
origin.  We review this subject in detail (Section 4), because it is central to
any conclusion that evolution has happened, and because it is the only way that
we can recognize pseudobulges.

      Next, we review observations of gas content and star formation rates.
From these, we estimate pseudobulge growth times. These prove to be in good
agreement with stellar population ages.  So the picture of pseudobulge growth
from rearranged disk gas is internally consistent.

      The purpose of this paper is to connect up the large number of disparate
results in this subject into a well developed and (as we hope to show) a well
supported paradigm.  Still, many questions remain unanswered.  We
especially need a better understanding of the relative importance of mergers 
and secular evolution as a function of galaxy type and luminosity.  We hope 
that this paper will provide a concrete context that will allow efficient
progress in this subject.

\vsl\vsss
\ni {\bf 1.1.~What is a Bulge?  Classical and Physical Morphology}
\vsl

      Before we proceed, we need to be clear on what we mean by a ``bulge''.
This will indicate why, for some galaxies, we use the term ``pseudobulge''.


      Renzini (1999) clearly states the canonical interpretation of
Hubble-Sandage-de Vaucouleurs classifications: ``It appears legitimate to look
at bulges as ellipticals that happen to have a prominent disk around them [and]
ellipticals as bulges that for some reason have missed the opportunity to 
acquire or maintain a prominent disk.''  We adopt this point~of~view.  However,
as observations improve, we discover more and more features that make it
difficult to interpret every example of what we used to call a ``bulge'' as an
elliptical living in the middle of a disk.  This leads authors to agonize, ``Are
bulges of early-type and late-type spirals different?  Are their formation
scenarios different?  Can we talk about bulges in the same way for different
types of galaxies?'' (Fathi \& Peletier 2003).

      We will conclude that early- and late-type galaxies generally do make
their dense central components in different ways.   This is not recognized in 
classical morphology, because it defines classification bins -- deliberately
and with good reason -- without physical interpretation.  Sandage and Bedke
(1994) describe how, in the early stages of investigating a subject, a
classifier should look for ``natural groups'' (Morgan 1951) of objects with
similar features.  Sandage emphasizes that it is important not to be led astray
by preconceptions: ``The extreme empiricist claims that no whiff of theory may
be allowed into the {\it initial} classification procedures, decisions, and
actions.''  Nevertheless, some choice of which features to consider as important
and which to view as secondary must be made.  After all, the goal is 
to understand the physics, and the exercise is useful only if classification 
bins at least partly isolate unique physics or order galaxies by physically
relevant parameters.  The Hubble-Sandage-de Vaucouleurs classification scheme
has done these things remarkably well.  

      However, it is reasonable to expect that improved understanding of 
galaxies will show that the classification missed some of the physics. Also,
some features of galaxies could not be observed well enough in the photographic
era to be included.  These include high-surface-brightness disky substructures
in galaxy centers.  Consistent with physical morphology as discussed in
Kormendy (1982a), we wish to distinguish components in galaxies that have
different origins. 

      At the level of detail that we nowadays try to understand, the time
has passed when we can make effective progress by defining morphological bins 
with no guidance from a theory.  Disks, bulges, and bars were different enough
that we could do this.  Afterward, robust conclusions could be reached, 
e.{\ts}g., about the relative timescales of collapse and star formation
(Eggen, Lynden-Bell, \& Sandage 1962).  But even inner rings and
spiral arms -- which are not subtle -- do not scream the appropriate message,
which is that spiral arms are details that would disappear quickly and without
a trace if the driving mechanism switched off, whereas we will see that rings
are a permanent rearranging of disk material.  Inner rings are, in this 
sense, more fundamental than spiral arms.  Years ago, people commonly reacted
badly to a classification as complicated as (R)SB(r)b.  The reason, we believe,
was that the phenomenology alone did not sell itself.  People did not see why 
this level of detail was important.  Now, we will show that every letter in the
above classification has a clearcut meaning in terms of formation physics.  
This is the goal of physical morphology.

      We adopt the view that bulges are ellipticals living in the middle of a 
disk.  Ellipticals formed via mergers (Toomre 1977a, Schweizer 1990).  Therefore
we do not use the name ``bulge'' for every central component that is in excess 
of the inward extrapolation of an exponential fitted to the disk brightness
profile.  If the evidence suggests that such a component formed by secular
processes, we call it a ``pseudobulge''.  In practice, we cannot be certain
about formation mechanisms. If the component in question is very E-like, we call
it a bulge, and if it is disk-like, we call it a pseudobulge. We comment on the
difficulty in classifying intermediate cases in \S\S\ts4, 7, 9.1.

      Finally, we comment on one of the biggest problems in this subject.
It is exceedingly easy to get lost in the details.  Many of the papers that
we review interpret observations or simulations in much more detail than we
will do here.  For example, it is common for observers to distinguish 
nuclei, nuclear bars, nuclear disks, nuclear spiral structure, boxy bulges, exponential bulges (sometimes more than one per galaxy, if the brightness
profile is piecewise exponential), and star formation rings.  We will discuss
all of these features, because they are central to the developing picture of
what secular evolution can accomplish.  But we will consider them all to be 
features of pseudobulges, because the evidence is that all of them are built
by secular evolution out of disk material.  In the same way, global spiral
structure, flocculent spiral structure, and no spiral structure in S0 
galaxies are all features of disks.  

\vsl\vsss
\ni {\bf 2.~SECULAR EVOLUTION OF BARRED GALAXIES}
\vsl

      We see only snapshots of galaxy evolution, so it is difficult to study
slow processes.\footnote{$^1$}{\kern -5pt Mergers are an easier problem --
transient phenomena such as tidal tails are readily recognizable (e.{\ts}g.,
Toomre \& Toomre 1972; Schweizer 1990).}  Why do we think that secular evolution
is happening?  We begin with an ``existence proof'' -- a review of
how $n$-body simulations account for the morphological features seen in barred
galaxies.  Our suggestion that pseudobulges are constructed out of rearranged
disk gas is embedded in this larger picture of SB secular evolution.

      Barred galaxies are reviewed in detail by SW93.
They are a rich subject; whole conferences have been devoted to them (e.{\ts}g.,
Buta, Crocker, \& Elmegreen 1996; Sandqvist \& Lindblad 1996).   We cover only
those subjects that relate to our theme.  

\vsl\vsss
\ni {\bf 2.1.~Morphology of Barred Galaxies}
\vsl

Barred galaxy morphology is discussed by de Vaucouleurs (1959, 1963), Sandage
(1961, 1975), Kormendy (1979b), Sandage \& Bedke (1994), Buta \& Crocker (1991),
Buta \& Combes (1996),
and Buta (1995, 1999).  We use these diagnostic features: 

\vs
\nnhi 1.~Barred spiral galaxies are divided into subclasses SB(s), in
      which the spiral arms begin at the ends of the bar, and SB(r), in which
      a complete ``inner ring'' of stars connects the ends of the bar.  In the
      latter case, the spiral arms start somewhere on the ring, ``often 
      downstream from the ends of the bar'' (Sandage \& Bedke 1994).
      SB(r) and SB(s) galaxies are contrasted in Figure 6; additional SB(r)
      galaxies are shown in Figures 3 and 5, and additional SB(s) galaxies are
      shown in Figure~7.

\vs
\nnhi 2.~Some barred and oval galaxies have ``outer rings'' (R) that are 
      $2.2 \pm 0.1$ times the diameter of the bar or inner disk. Outer rings
      in barred and unbarred galaxies are similar (Fig.~2, 5). Inner and
      outer rings are different; there is no size overlap.  Some galaxies
      contain~both~(Fig.~5).

\vs
\nnhi 3.~At intermediate Hubble types, when the bar is made mostly of old
      stars and the disk contains many young stars, the stellar population 
      of inner and outer rings is like that of the disk, not like that of the
      bar (Figures 2 and 3).  Inner and outer rings generally contain gas.

\vs
\nnhi 4.~In SB(s) galaxies, an almost-straight dust lane parallels the ridge
      line of the bar but is displaced slightly forward in the direction of
      galactic rotation.  Such dust lanes are analogous to and connect up with
      the prominent dust lanes seen on the trailing side of the arms in
      global-pattern spirals.  Examples are shown in Figures 6, 7, and 8.
      These dust lanes are almost never present in SB(r) galaxies 
      (Sandage 1961). 
      NGC 1512 in Figure 3 is a rare exception.

\vs
\nnhi 5.~Many barred and oval galaxies have very active star formation near
      their centers, in what is conventionally identified as the bulge.
      Often the star formation is concentrated in a ring.
      Figures 3, 7, and 8 show examples.

\vs
\nnhi 6.~Many barred galaxies have ``bulges'' that are themselves elongated
      into a structure resembling a bar.  Examples are shown in Figure 14.

\vs
\nnhi 7.~Many early-type SB galaxies contain a ``lens'' in the disk -- a shelf
      of slowly decreasing surface brightness with a sharp outer
      edge.  Lenses have intrinsic axial ratios of $\sim 0.85$; the bar usually
      fills the longest dimension.  These properties are discussed in Kormendy
      (1979a,{\ts}b, 1981, 1982a) and in Athanassoula et al.~(1982). Lenses are
      sometimes seen in unbarred galaxies; NGC 1553 is the best example (Freeman
      1975; Kormendy 1984).  Lenses in early-type galaxies look similar to oval
      disks in late-type galaxies (Section 3.2); it is not clear whether or not
      they are physically similar.  Lenses are illustrated in Figures 2 and 5.

\vs

      These features can be understood at least qualitatively as results of
secular evolution driven by nonaxisymmetric gravitational potentials.  An
exact correspondence between $n$-body simulations and observations cannot be
expected, because real galaxies have a complicated interplay between gas, star
formation, and energy feedback from massive young stars back into the
interstellar medium.  Such effects, along with the self-gravity of the gas, are
often omitted from simulations and at best are included only approximately.
Nevertheless, $n$-body simulations have been conspicuously successful in
reproducing the structure of barred galaxies.


\vfill\eject

\cl{\null}

\vskip 2.04truein \vfill

\col{\includegraphics{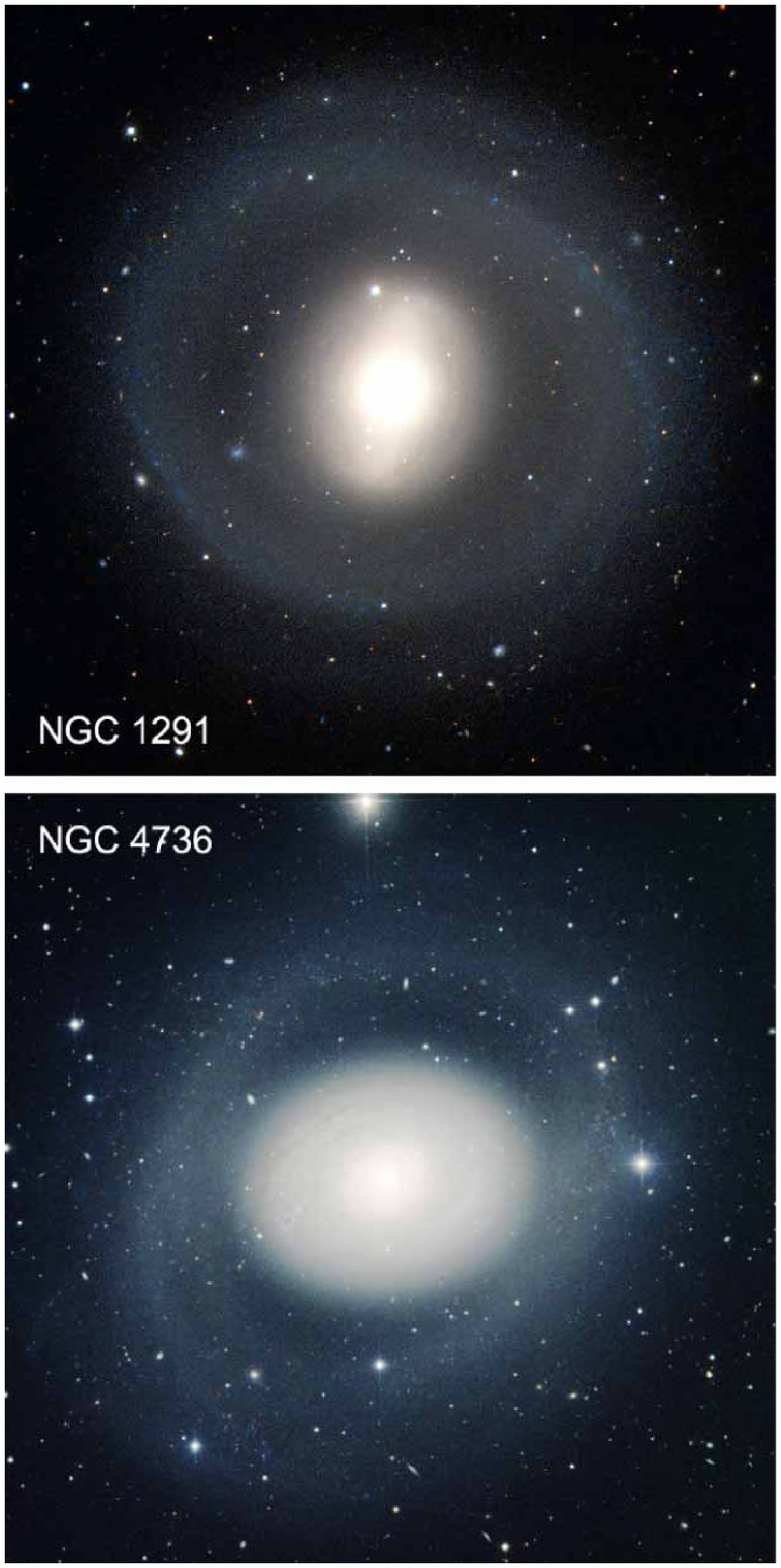}}

     {\it Figure 2}\quad Prototypical outer rings in barred and unbarred
galaxies.  NGC 1291 is an (R)SB(lens)0/a galaxy -- it has a bar embedded
in a lens of the same major-axis diameter (see also Kormendy 1979b).  NGC 4736
is classified (R)SA(r)ab.  The purpose of this figure is to show how blue the
outer rings are: they are dominated by young stars.  Both rings also contain
H{\ts}I gas (van Driel et al.~1988; Bosma et al.~1977b).
Sources: NGC 1291 -- Buta, Corwin, 
\& Odewahn (2003); NGC 4736 -- NOAO.

\eject

\cl{\null}

\vskip 4.2truein 

\col{\includegraphics{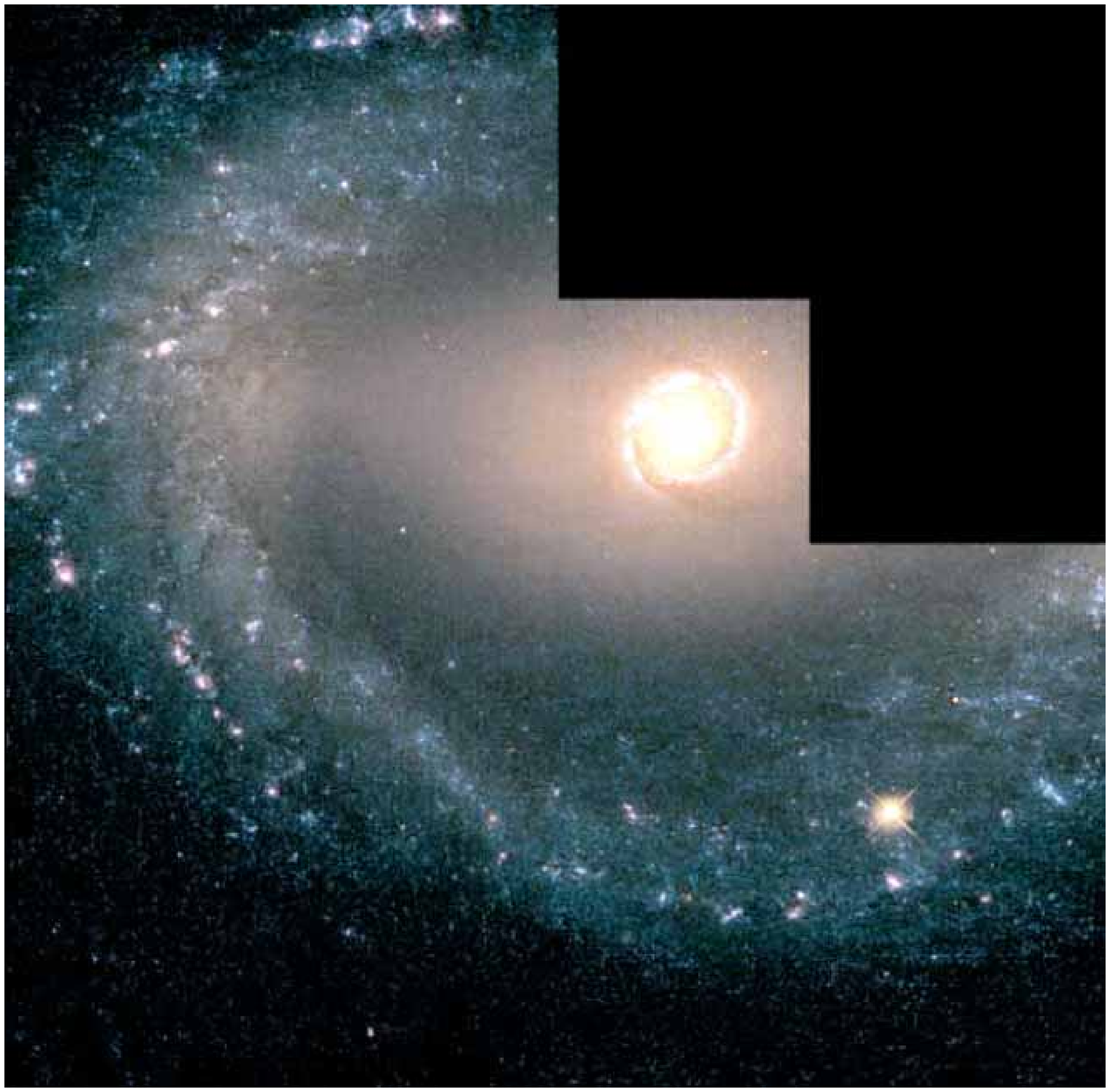}}

     {\it Figure 3}\quad NGC 1512, an SB(r)ab galaxy imaged with HST by
Maoz et al.~(2001).  This figure (courtesy NASA and ESA) illustrates
the stellar population of inner rings.  As is common in intermediate-Hubble-type
galaxies, the bar in NGC 1512 is made of old, red stars and the disk is made of
young, blue stars.  The point of this figure is that the inner ring has the same stellar population as the disk, not the bar.  Also seen at center is a nuclear 
star formation ring that is shown at higher magnification in Figure 8 and the
start of a
well developed, curved dust lane \hbox{(cf.{\ts}Figures 6 -- 8)} that extends
out of the field of view to the right.  The corresponding dust lane on the other
side is visible near the central ring but not at larger radii.  The outer parts
of NGC 1512 are illustrated by Sandage \& Bedke (1994), who note that NGC 1512
is morphologically normal except for some distortion of its outer spiral 
structure (not shown here) by a tidal encounter with neighboring NGC 1510.


\vfill\eject

\vsl\vsss\vsss\vsss
\ni {\bf 2.2.{\ts}Dynamics of Barred Galaxies:~The Importance of Resonances}
\vsl

      To understand bar-driven evolution, we need to dip briefly into the
dynamics of bars.  An in-depth review is provided by SW93.  Here we need
a primer on the nature and importance of orbital resonances.  

      Seen from an inertial frame, an orbit in a galactic disk is an unclosed
rosette.~That is, there are a nonintegral number of radial oscillations for
every revolution around the center.  However, in a frame of reference that
rotates at the average angular velocity of the star, the star's mean
position is fixed and its radial oscillation makes it move in a small
ellipse around that mean position\footnote{$^2$}{It is an ellipse and not just
radial motion because the star revolves faster than average near pericenter and
slower than average near apocenter.}.  Any global
density pattern such as a bar that rotates at the above angular velocity will
pull gravitationally on the star in essentially the same way at all times and
will therefore make large perturbations in its orbit.   ``Corotation'' is the
strongest of a series of resonances in which the pattern repeatedly sees the
star in the same way.  

      For example, there is another rotating frame in which the star executes
two radial oscillations for each circuit around the center.  If a bar rotates
at this angular velocity, it sees the stellar orbit as closed, roughly
elliptical, and centered on the galactic center (Figure 4).  This is called
inner Lindblad resonance (ILR).  It occurs when the pattern speed of the bar is
$\Omega_p = \Omega - \kappa/2$, where $\Omega$ is the average angular velocity
of revolution of the star and $\kappa$ is its frequency of radial oscillation.
The limit of small radial oscillations is called the epicyclic approximation;
then \hbox{$\kappa^2 = (2V/r)(V/r + dV/dr)$,} where $V(r)$ is the circular-orbit
rotation curve (Mihalas \& Routly 1968 provide a particularly transparent
discussion).

      Outer Lindblad resonance (OLR) is like inner Lindblad resonance, except
that the star drifts backward with respect to the rotating frame while it
executes two radial oscillations for each revolution: $\Omega_p = \Omega +
\kappa/2$.

      Resonances are important for several reasons.  Figure 4 shows generic
frequency curves and the most important periodic orbit families in a barred
galaxy.  We can begin to understand how a self-consistent bar is constructed by
exploiting the fact that $\Omega - \kappa/2$ varies only slowly with radius
(except near the center, if there is an ILR).  Calculations of orbits in a
barred potential show that the main ``$x_1$'' family of orbits is elongated
parallel to the bar between ILR and corotation.  Bars are largely made of $x_1$
orbits and similar, \hbox{non-periodic} orbits that are trapped around them by 
the bar's self-gravity.  Typical $x_1$ orbits are shown in the bottom panel
of Figure 4.  They are not nearly circular, but the essence of their behavior 
is captured if we retain the language of the epicyclic approximation and say
that orbits of different radii look closed in frames that rotate at
different angular velocities $\Omega - \kappa/2$.  But if $\Omega - \kappa/2$
varies only a little with radius, 

\vfill\eject

\cl{\null}

\vfill

\col{\includegraphics{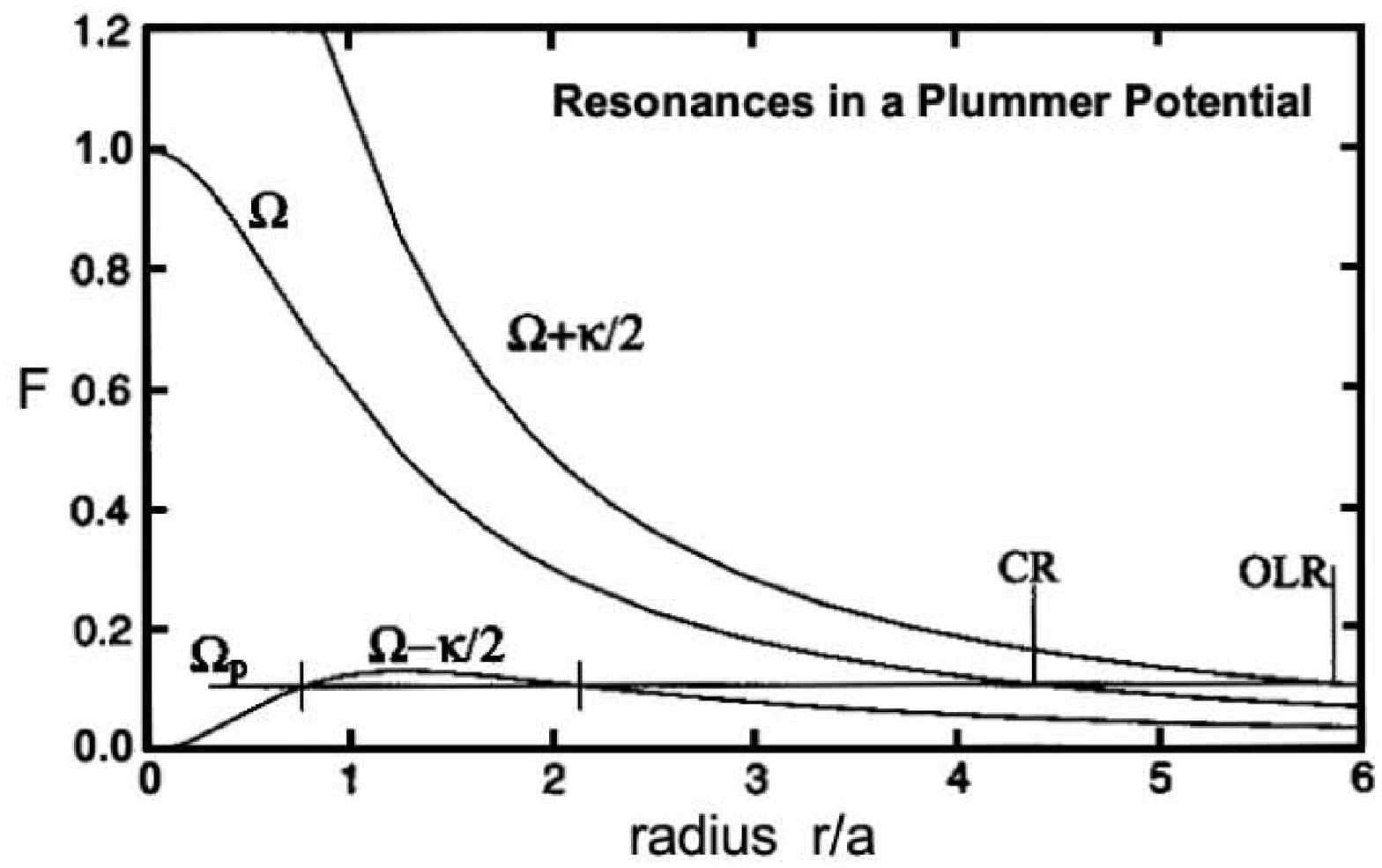}}

\col{\includegraphics{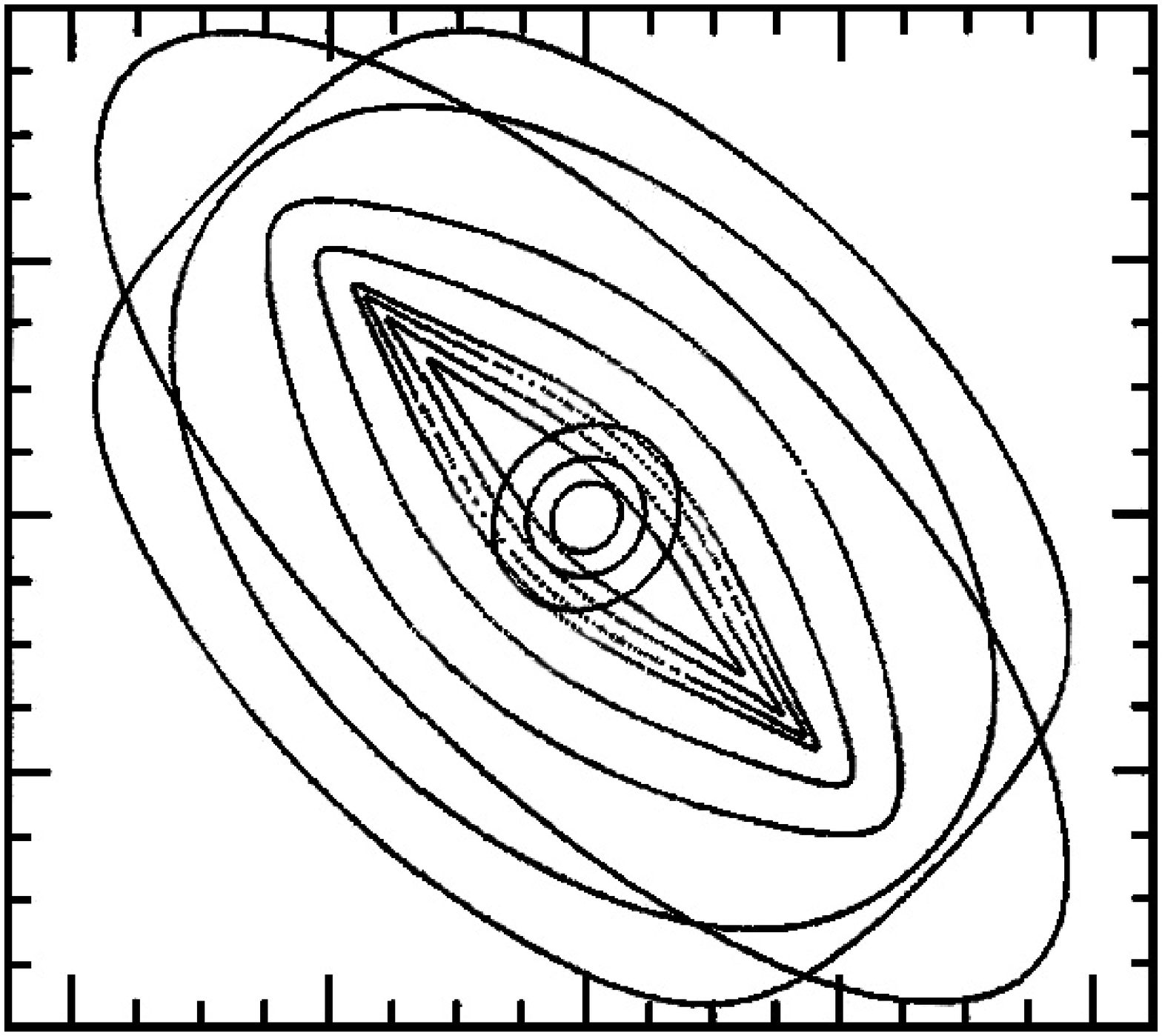}}

     {\it Figure 4}\quad (Top) Frequencies $\Omega(r) = V(r)/r$ and 
$\Omega \pm \kappa/2$, where \hbox{$\kappa^2 = (2V/r)(V/r + dV/dr)$} is the 
epicyclic frequency of radial oscillations for almost circular orbits.
This figure (Sparke \& Gallagher 2000) is for a Plummer potential, but the
behavior is generic.  \hbox{For a pattern speed $\Omega_p$,} the most important
resonances occur where $\Omega_p = \Omega$ (corotation), where 
$\Omega_p = \Omega + \kappa/2$ (outer Lindblad resonance OLR) and where
$\Omega_p = \Omega - \kappa/2$ (two inner Lindblad resonances ILR, marked with
vertical dashes).  (Bottom) From Englmaier \& Gerhard (1997), examples of the
principal orbit families for a bar oriented at 45$^{\circ}$ as in Figure 7.  
The elongated orbits parallel to the bar are the $x_1$ family out of
which the bar is constructed.  Interior to ILR (or outer ILR, if there are two
LRs), the $x_2$ family is perpendicular to the bar.  Near corotation 
is the 4:1 ultraharmonic resonance; the almost-square orbit makes 4 radial
oscillations during each circuit around the center.  Since the principal orbits 
change orientation by 90$^{\circ}$ at each resonance shown, they must cross near
the resonances.  
  
\eject

\ni then it is possible to pick a single pattern speed $\Omega_p$ in which the
orbits precess almost together.  If they precessed exactly together, then one
could make a bar by aligning elongated orbits as in the bottom panel of the
figure.  Since $\Omega - \kappa/2$ is not quite constant, it is the job of
self-gravity to make the orbits precess not approximately but exactly together.
This idea was used to understand self-consistent bars by Lynden-Bell \& Kalnajs
(1972) and by Lynden-Bell (1979) and to demystify spiral structure by Kalnajs
(1973) and by Toomre (1977b).  They were following in the pioneering footsteps
of Bertil Lindblad (1958, 1959:~see \S\ts20).

      Calculations of orbits in barred potentials reveal other orbit families (e.{\ts}g., Contopolous \& Mertzanides 1977; Athanassoula 1992a,{\ts}b; SW93),
only a few of which are relevant here.  Next in importance is the $x_2$ family, 
which lives interior to ILR and which is oriented perpendicular to the bar
(Figure 4, bottom).  Between corotation and OLR, the principal orbits are
elongated perpendicular to the bar, and outside OLR, they are again oriented
parallel to the bar.  Near corotation is the 4:1 ultraharmonic resonance in
which a star executes 4 radial oscillations for every revolution: $\Omega_p =
\Omega - \kappa/4$.  We will need these results in the following sections.

      The important consequence is emphasized by SW93: ``Not only do the
eccentricities of the orbits increase as exact resonance is approached, but the
major axes switch orientation across all three principal resonances, making the
crossing of orbits from opposite sides of a resonance inevitable'' (bottom
panel of Figure 4).  This is important mainly when the orbits are very
noncircular, as in strongly barred galaxies.
Now, orbits that cross are no problem for stars.  But gas clouds that move
on such orbits must collide near resonances.  Dissipation is inevitable, and
the consequence is an increase in the gas density and hence star formation.
This heuristic discussion helps to explain the numerical results reviewed in
the following sections, in which gas tends to build up in rings and to form 
stars there.

\vsl\vsss
\ni {\bf 2.3.~Bar-Driven Radial Transport of Gas:~The Formation of Rings}
\vsl

Theory (Binney \& Tremaine 1987; SW93; Lynden-Bell 1979; 1996) and
\hbox{$n$-body} simulations (Sellwood 1981; Sparke
\& Sellwood 1987; Pfenniger \& Friedli 1991; Athanassoula 2003) show that bars
grow by transferring angular momentum to the outer disk, thereby driving spiral
structure.  As a result, stellar orbits in the bar get more elongated, and the
bar grows in amplitude.  Its pattern speed slows down. 

      The essence of the response of gas to a bar is captured in Figure 3 of
Simkin, Su, \& Schwarz (1980), reproduced here as Figure 5.  Outside corotation,
gas is driven outward by the angular momentum transfer from bar to disk that 
makes the bar grow.  This gas collects into an ``outer ring''near OLR.  As
discussed earlier, outer rings are oriented perpendicular
to the bar when they are interior to OLR; this is the usual situation
(Kormendy 1979b; Buta 1995).  At radii well inside corotation, gas falls
toward the center.  This is the gas that is believed to make pseudobulges.
Within an annular region around corotation, gas is collected into an ``inner
ring'' near corotation or near the 4:1 ultraharmonic resonance. 
 
      This behavior is seen in a variety of simulations starting as early 
as Prendergast (1964), Duus \& Freeman (1975), and S\o rensen et al.~(1976). 
By the early 1980s, there was already an extensive literature on the subject
(see Kormendy 19982a\ts~and Prendergast 1983 for reviews). 

\vfill





\col{\includegraphics{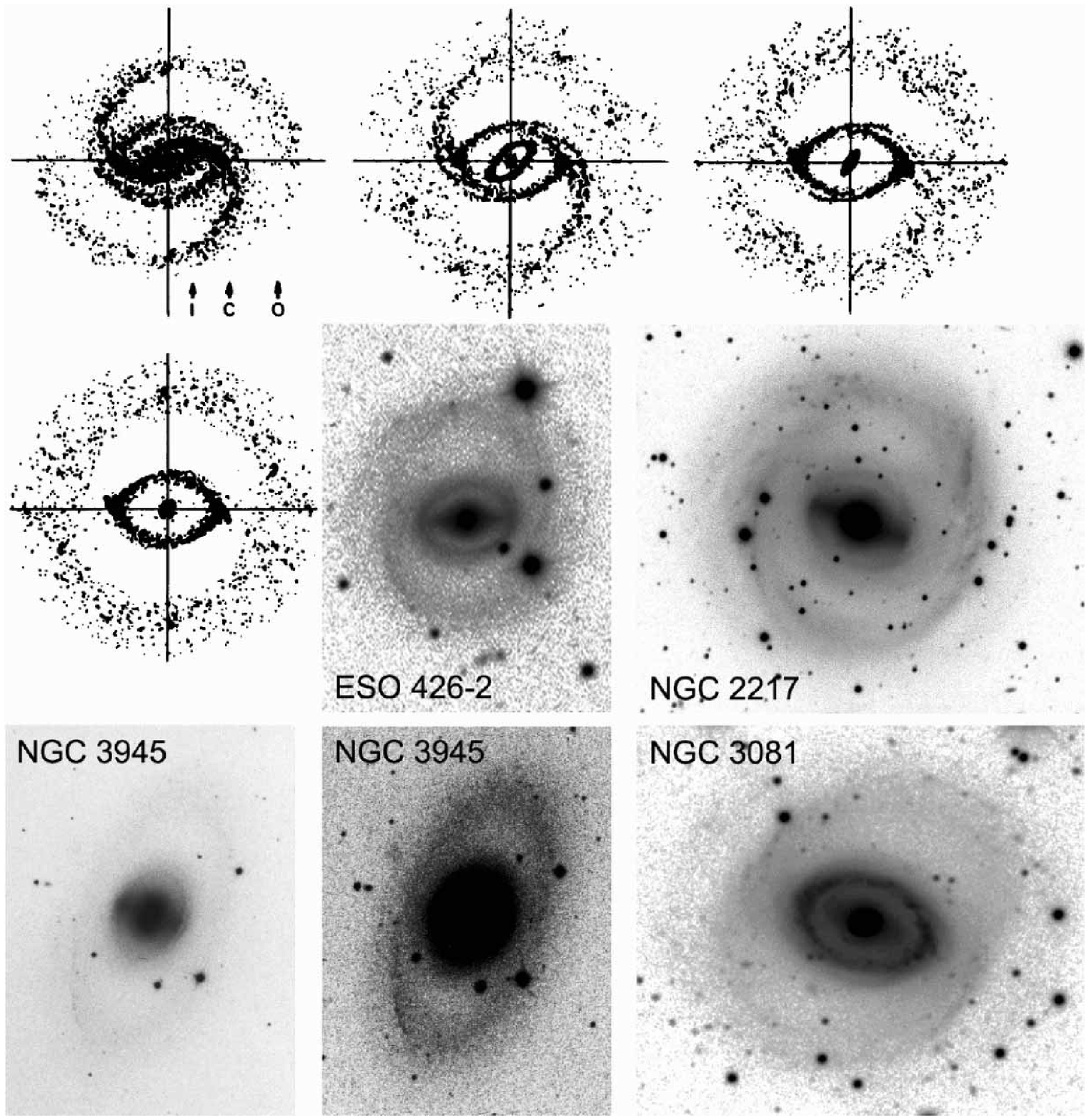}}

     {\it Figure 5}\quad Evolution of gas in a rotating oval potential 
(Simkin, Su, \& Schwarz 1980; see also Schwarz 1981, 1984).  The gas
particles in this sticky-particle \hbox{$n$-body} model are shown after 2, 3, 
5, and 7 bar rotations (top-left through center-left).  Arrows show the radii 
of ILR, corotation, and OLR.  Four SB0 or SB0/a galaxies are shown that have
outer rings and a lens (NGC 3945) or an inner ring (obvious in ESO 426-2 and
in NGC 3081 but poorly developed in NGC~2217).\ts~Sources: NGC 3945{\ts}--{\ts}Kormendy (1979b); NGC 2217, 
NGC 3081{\ts}--{\ts}Buta et al.~(2003); ESO 426-2{\ts}--{\ts}Buta \& Crocker
(1991).

\eject

      The reason why SB(r) and SB(s) galaxies are different was investigated
by Sanders \& Tubbs (1980).  They simulated the response of
gas to an imposed, rigid bar potential that they grew inside a disk
galaxy. Examples of the steady-state gas response are shown in Figure 6. In the
top two rows of panels, the strength of the bar increases from left to right,
either because the ratio of bar mass to disk mass increases (top row), or
because the bar gets more elongated (second row). In both cases, weak bars tend
to produce an SB(s) response while strong bars produce ring-like structures that
resemble SB(r) galaxies.  If the bar gets too strong (top-right panel), the
result does not look like a real galaxy. The bottom row of simulations explores
the effect of varying the bar's pattern speed.  Rapid pattern speeds produce
dramatically SB(s) structure. Slower pattern speeds in which corotation is near
the end of the bar produce inner rings. Very slow pattern rotation (right panel,
in which corotation is at 3 bar radii) produce responses that do not look like
real galaxies.  This is because $\Omega_p$ is now so small that the radius of
ILR is large.  Inside ILR, closed gas orbits align perpendicular to the bar.
These can never have substantially the same radius as the bar, as they do in 
the bottom-right simulation in Figure 6.  If the response to the bar were
perpendicular to the bar over most of the radius of the bar, it would be
impossible to make that response add up to a self-consistent bar.  Pattern
speeds are never so slow that corotation radii are so far out in the disk that
the entire bar is inside ILR. This was possible in Sanders \& Tubbs (1980) only
because the bar was inserted by hand and given a chosen (not a self-consistent)
pattern speed.  Theoretical arguments tell us that bars end inside or near
corotation (Contopoulos 1980; SW93).  Observations agree 
(Kent 1987b;
Sempere et al.~1995;
Merrifield \& Kuijken 1995;
Gerssen, Kuijken, \& Merrifield 1999, 2003;
Debattista \& Williams 2001;
Gerssen 2002;
Debattista, Corsini, \& Aguerri 2002;
Aguerri, Debattista, \& Corsini~2003;
Corsini, Debattista, \& Aguerri~2003;
Corsini,  Aguerri, \& Debattista~2003;  
see Elmegreen 1996 for a review)
except in late-type galaxies in which $V\propto r$ rotation curves imply that
the bar is safely clear of ILR anyway
(Elmegreen 1996; Elmegreen, Wilcots, \& Pisano 1998).

      On the other hand, Sanders \& Tubbs (1980) share a number of technical
problems with other early simulations of gas response to bars.  Their beam
scheme (Sanders \& Prendergast 1974) code has coarse spatial resolution and
unphysical numerical viscosity (see Athanassoula 1992b and SW93).  In fact, 
there are conflicting views on whether viscosity is important at~all; Combes
(1998) suggests that it is negligible compared to gravitational torques, while
SW93 at least consider the possibility that it is important.  
Gas infall timescales are very uncertain in early simulations.

      Still, the main conclusion reached by Sanders \& 
Tubbs\ts(1980)\ts--{\ts}that weak, fast bars favor SB(s) structure and that
strong, slow bars favor SB(r) structure -- has largely been confirmed by
higher-quality simulations \phantom{0000000000000000000000}
     
\cl{\null}

\vfill



\col{\includegraphics{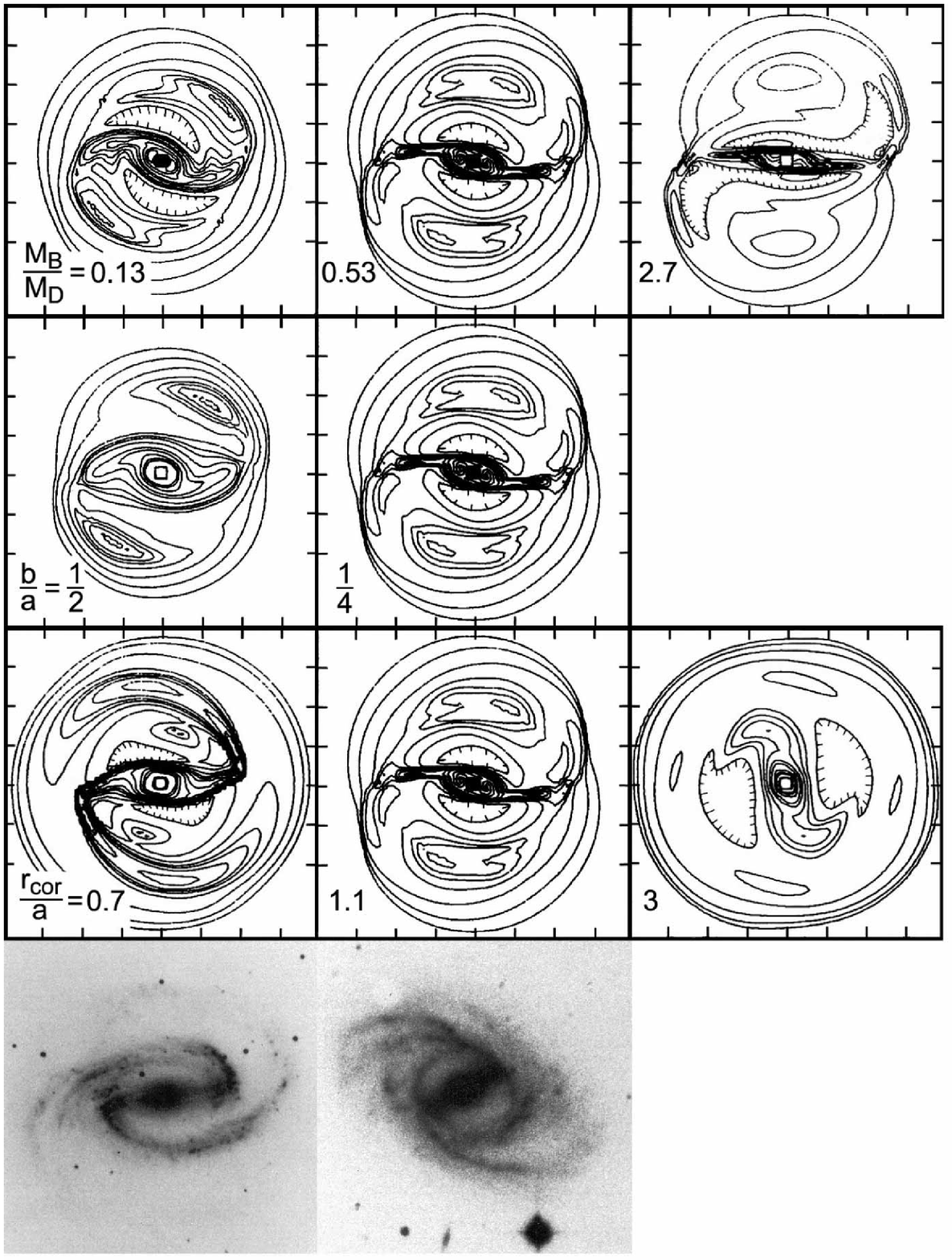}}

     {\it Figure 6}\quad Contours of steady-state gas density in response to
a bar (adapted from Sanders \& Tubbs 1980, who also show intermediate cases).
The bar is horizontal and has a length equal to four axis tick marks.  The top
row explores the effect of varying the ratio $M_B/M_D$ of bar mass to disk mass.
The second row varies the bar's axial ratio $b/a$.  The third row varies the
bar pattern speed, parametrized by the ratio $r_{\rm cor}/a$ of the corotation radius to the
disk scale length.  The middle column is the same standard model in each row; 
it approximates an SB(r) galaxy such as NGC 2523 (bottom center).  The left
panels resemble SB(s) galaxies such as NGC 1300 (bottom left). The right panels
carry the parameter sequences to unrealistic extremes; they do not resemble
real galaxies.

\eject

\ni (e.{\ts}g., Schwarz 1984; Combes \& Gerin 1985; Byrd et al.~1994; Englmaier
\& Gerhard 1997; Salo et al.~1999; Weiner, Sellwood, \& Williams 2001a).  Well
motivated hints of these results came much earlier (Freeman 1970b).


      Therefore a widespread feeling has developed that we understand the
essentials of ring formation.  We share this feeling.  However, we also share a
concern expressed by K.~C.~Freeman (private communication):  Why do sticky
particle simulations make rings so much more clearly than do other (e.{\ts}g.,
hydrodynamic) simulations?  There may be physics in this.  For example, the gas
really may be in discrete clouds that collide inelastically.  
Still, we cannot help but notice that, as simulations have improved, features
such as the ones we discuss next -- radial dust lanes and nuclear star 
formation rings -- have improved dramatically, but ring formation has made
less progress.  The subject deserves to be revisited.  

      Nearly radial dust lanes in bars (see \S\ts2.1 and Figures 3, 6, 7, 8) 
are a particularly important diagnostic of SB evolution.  They are widely
believed to be the observational signatures of shocks that drive gas infall.
The idea was proposed by Prendergast (1964); other early studies include 
S\o rensen, Matsuda, \& Fujimoto (1976); Roberts, Huntley, \& van Albada (1979),
and, as discussed above, Sanders \& Tubbs (1980).

      In an important paper, Athanassoula (1992b) explored the response of
inviscid gas to a bar using a high-resolution code.  Her main focus was gas
shocks and their relation to dust lanes.
Typical results are shown in Figure 7.  If and only if the mass distribution
is centrally concentrated enough to result in an inner Lindblad resonance,
the dust lanes are offset in the forward (rotation) direction from the ridge
line of the bar.  Because of the presence of the $x_2$ orbits -- the ones 
that align perpendicular to the bar inside ILR -- the offset is largest near
the center, as it is in many galaxies, including the two shown in Figure 7.
The models also reproduce the observation that the dust lanes in many bars 
curve around the center of the galaxy at small radii and become nearly
azimuthal.  Athanassoula found that the dust lanes are more curved into an open
S-shaped structure when the bar is weak; this is confirmed observationally  by
Knapen, P\'erez-Ram\'\i rez, \& Laine (2002).  One shortcoming in the models is
that, for weak central concentrations (galaxies with no $x_2$ orbits), the
shocks are essentially on the ridge line of the bar.  Such dust lanes are not
observed.  But the main conclusion, as Athanassoula notes, is that ``the 
resemblance between [the models and the observations] is striking''.

     The important consequence for our discussion is this:~Shocks inevitably 
imply that gas flows toward the center.  Because the shocks are
nearly radial, gas impacts them at a steep angle.  Therefore much of the
velocity that is lost in the shock is azimuthal.  This robs the gas of energy
and forces it to fall toward the center.

\vfill\eject
      
\cl{\null}

\vfill

\col{\includegraphics{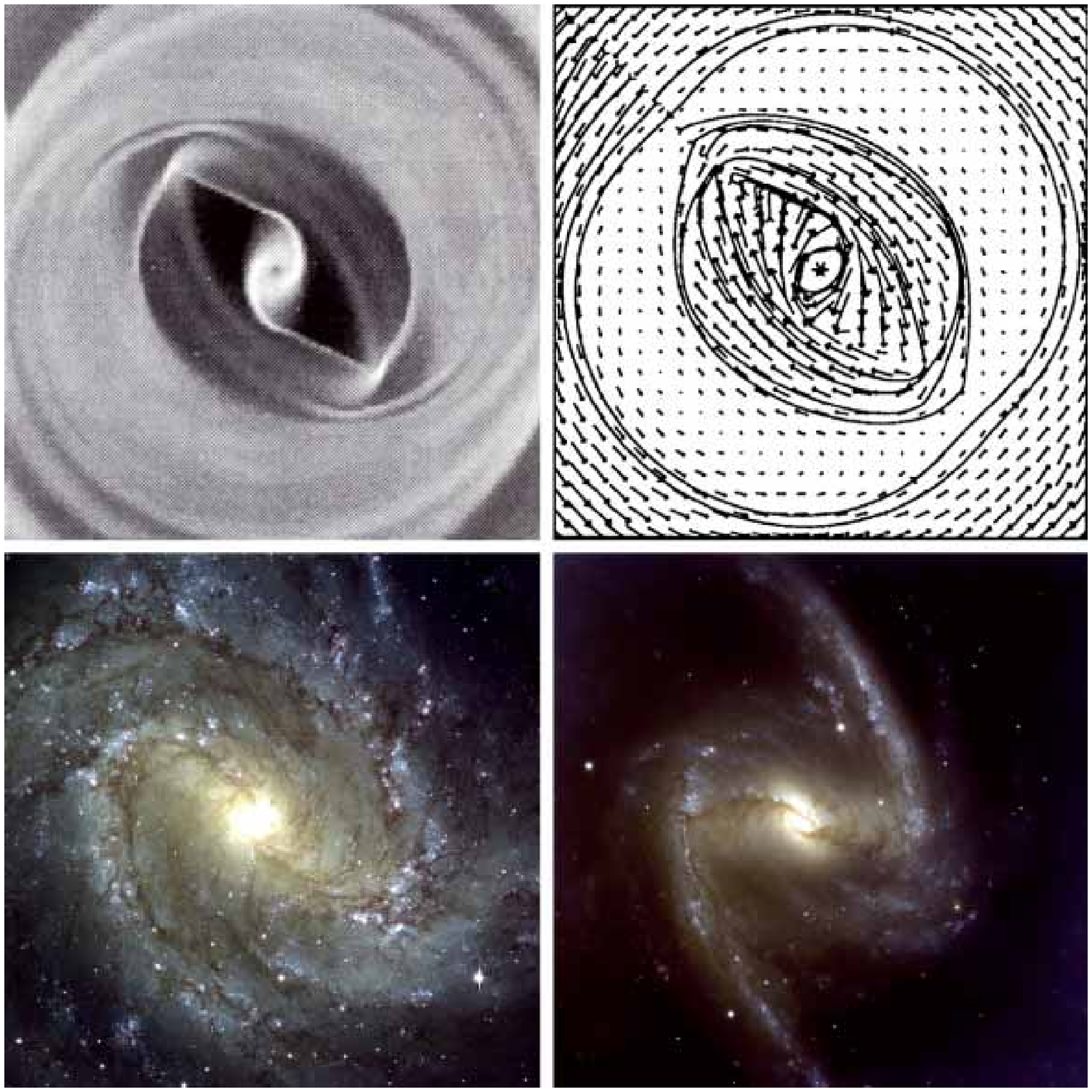}}

     {\it Figure 7}\quad Comparison of the gas response to a bar (Athanassoula
1992b model 001) with NGC 5236 (left) and NGC 1365 (right).~The galaxy images
were taked with the VLT and are reproduced courtesy of ESO.  In the models, the
bar potential is oriented at 45$^{\circ}$ to the horizontal, parallel to the bar
in NGC 5236.  The bar axial ratio is 0.4 and its length is approximately half of
the box diagonal.  The top-right panel shows the velocity field;
arrow lengths are proportional to flow velocities.  Discontinuities in gas
velocity indicate the presence of shocks; these are where the gas density is high in the density map at top-left. High gas densities are identified with dust
lanes in the galaxies.  The model correctly reproduces the observations (1) that
dust lanes are offset in the forward (rotation) direction from the ridge line of
the bar; (2) that they are offset by larger amounts nearer the center, and
(3) that very near the center, they curve and become nearly azimuthal.  As
emphasized by the velocity field, the shocks in the model and the dust lanes in
the galaxy are signs that the gas loses energy.  Therefore it must
fall toward the center.  In fact, both galaxies have high gas densities and
active star formation in their bright centers (e.{\ts}g., Crosthwaite et
al.~2002; Curran et al.~2001a, b).

\eject

      Athanassoula estimated that azimuthally averaged gas sinking rates are
typically 1 km s$^{-1}$ and in extreme cases up to $\sim 6$ km s$^{-1}$.
Since viscosity is not an issue in her models, these estimates are more
realistic than earlier ones.  And since 1 km s$^{-1}$ = 1 kpc ($10^9$
yr)$^{-1}$, the implication is that most gas in the inner part of the disk --
depleted by star formation but augmented by mass loss during stellar evolution
-- finds its way to the vicinity of the center over the course of several
billion years, if the bar lives that long.

      In recent years, simulations have continued to concentrate on these inner
regions of barred galaxies where dust lanes and star formation are most
important
(Friedli \& Benz 1993, 1995;
Piner, Stone, \& Teuben 1995;
Lindblad, Lindblad, \& Athanassoula 1996;
Englmaier \& Gerhard 1997;
Salo et al.~1999;
Weiner, Sellwood, \& Williams 2001a;
Maciejewski et al. 2002;
Regan \& Teuben 2003).
Details differ, but these conclusions are robust:
(1) Everbody~agrees that gas flows toward the center. (2) Star formation fed
by the inflow is often concentrated in a narrow nuclear ring.  (3)
The inflow is a result of gravitational torques produced by the bar, but its
immediate cause is the shocks.  In essence, these are produced because gas 
accelerates as it approaches and decelerates as it leaves the potential minimum
of the bar.  So it tends to pile up near the ridge line of the bar.  Incoming 
gas overshoots a little before it plows into the departing gas, so the shocks
are nearly radial but offset from the ridge line of the bar in the forward
(rotation) direction.  More recent simulations confirm Athanassoula's conclusion
that offsets happen when the central mass concentration is large enough to allow
a ``sufficient'' range of $x_2$ orbits.  The agreement in morphology between the
simulated shocks and the observed dust lanes has continued to improve.  But
there is an even better reason to think that they are connected.  Compelling
support is provided by the observation of large velocity jumps across the dust
lanes 
(Pence \& Blackman 1984;
Lindblad, Lindblad, \& Athanassoula 1996;
Regan, Sheth, \& Vogel 1999;
Weiner et al.~2001b;
and especially Regan, Vogel, \& Teuben 1997).  

      What happens to the infalling gas? Star formation is almost inevitable.
The simulations, expectations from the Schmidt (1959) law, observations of young
stars in SB nuclei, and star formation indicators (\S\ts5) all point to
enhanced star formation, often in substantial starbursts near the center.
Examples are shown in Figure 8.  NGC 4314 is a 
barred galaxy whose central star formation is also illustrated in the {\it
Hubble Atlas} (Sandage 1961).  NGC 1512 is an SB(rs) galaxy whose outer parts
are shown in Figure 2.  The dust lane in the bar is best seen in the {\it 
Carnegie Atlas of Galaxies} (Sandage \& Bedke 1994).  NGC 6782 contains an oval
disk with an embedded bar; Athanassoula (1992b) predicts very curved dust lanes
like those in NGC 6782 when the potential is not very barred.  Finally, NGC 4736
is a prototypical unbarred oval galaxy.  It is included to illustrate the theme
of the next section -- barred and oval galaxies evolve similarly.

\vfill\eject
      
\cl{\null}

\vfill




\col{\includegraphics{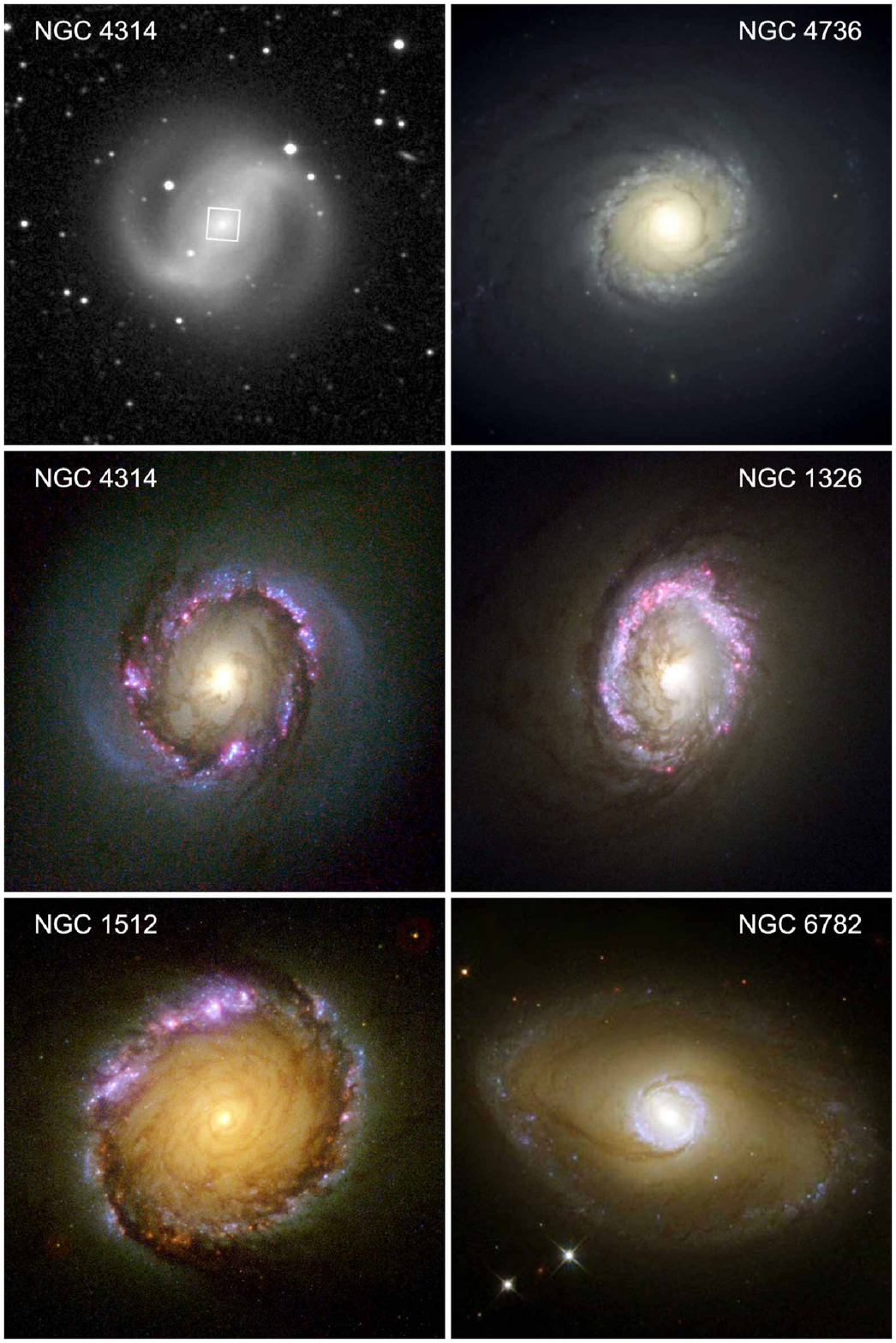}}

     {\it Figure 8}\quad Nuclear star formation rings in barred and oval
galaxies.  For NGC 4314, a wide-field view is at top-left; for NGC 4736,
the wide-field view is in Figure 2. Sources: NGC 4314 -- Benedict et al.~(2002);
NGC 4736 -- NOAO; NGC 1326 -- Buta et al.~(2000) and Zolt Levay (STScI);
NGC 1512 -- Maoz et al.~(2001); NGC 6782 --
Windhorst et al.~(2002) and the Hubble Heritage Program.

\eject

\omit{
\cl{\null}
\vskip 2.04truein \vfill
\col{\includegraphics{N4321-M51-v3-hicontrast.ps}}
     {\it Figure 9}\quad Nuclear star formation in the unbarred galaxies
M{\ts}51 and NGC 4321 (M{\ts}100).  Dust lanes on the trailing side of the
global spiral arms reach in to small radii.  As in barred spirals, they are
are indicative of gas inflow.  Both galaxies have concentrations of star
formation near their centers that resemble those in Figure 8.  These images 
are from the {\it Hubble Space Telescope\/} and are courtesy STScI.  
\eject
}

      The examples shown in Figure 8 all have star formation concentrated in
tiny rings with mean radii $\sim 0.5$ kpc (Buta \& Crocker 1993).  The physics
that determines their radii is complicated and not well understood. Many
authors suggest that they are caused by gas stalling at ILR or between ILRs
(if there are two of them), but this is disputed by Regan \& Teuben (2003), and
in reality, it is likely that star formation physics and not just inflow physics
is involved. In any case, nuclear star formation rings are a reasonably common 
phenomenon; the most prominent examples have been known for a long time (Morgan
1958; Burbidge \& Burbidge 1960, 1962; Sandage 1961; S\'ersic \& Pastoriza 1965,
1967).  Kennicutt (1994, 1998a) provides reviews, contrasting the global star
formation in barred galaxies, which is indistinduishable from that in unbarred
galaxies, with the nuclear star formation, which is enhanced over that in
unbarred galaxies.  As predicted by simulations, there is plenty of fuel -- the
central concentration of molecular gas is higher in barred than in unbarred
galaxies (Sakamoto et al.~1999).  Additional examples of nuclear star formation
rings -- and multiple discussions of the best cases -- can be found in 
van der Kruit (1974, 1976);
Rubin, Ford, \& Peterson (1975);
Sandage \& Brucato (1979);  
Hummel, van der Hulst, \& Keel (1987);
Gerin, Nakai, \& Combes (1988);
Benedict et al.~(1992, 1993, 1996, 2002);
Buta (1986a,{\ts}b, 1988, 1995);
Pogge (1989);
Garc\'\i a-Barreto et al.~(1991);
Devereux, Kenney, \& Young (1992);
Buta \& Crocker (1993);
Pogge \& Eskridge (1993: NGC 1819);
Forbes et al.~(1994a, b);
Quillen et al.~(1995);
Phillips et al.~(1996);
Buta \& Combes (1996);
Maoz et al.~(1996);
Regan et al.~(1996);
Elmegreen et al.~(1997);
Colina et al.~(1997);
Contini et al.~(1997);
Vega Beltr\'an et al.~(1998);           
Buta \& Purcell (1998);
Buta, Crocker, \& Byrd (1999);
Martini \& Pogge (1999);
Buta et al.~(1999, 2000, 2001);         
P\'erez-Ram\'\i rez et al.~(2000, 2001);
Wong \& Blitz (2000, 2001);
Waller et al.~(2001);
Lourenso et al.~(2001);                  
Alonso-Herrero, Ryder, \& Knapen (2001); 
D\'\i az et al.~(2002);
Knapen, P\'erez-Ram\'\i rez, \& Laine (2002);
Windhorst et al.~(2002);
Erwin \& Sparke (2003);
Eskridge et al.~(2003);
Martini et al.~(2003a);
Combes et al.~(2003);
Kohno et al.~(2003); and
Fathi et al.~(2003).  

      Many galaxies discussed in the above papers are barred. The ones that are
classified as transition objects (SAB) or as unbarred (SA), have created some
uncertainty about how much the star formation depends on bars. However, many SAB
and some SA objects are prototypical oval galaxies such as NGC 2903, NGC
3504, NGC 4736 (Figures 2, 8), NGC 5248, and NGC 6951 (see Sandage 1961).  
We will see in the next section that barred and oval galaxies are
essentially equivalent as regards gas inflow, star formation, and pseudobulge
building.  Section 3.4 suggests that similar evolution happens in unbarred
spirals that do not have an ILR.

\omit{  This is the long version. 
      We argue in later sections, as did some of the above authors, that the
nuclear star formation is building pseudobulges.  Here we address one issue
raised by the observation that the star formation is frequently in a
ring.  Is this likely to form a stellar ring and not a pseudobulge?  There are
several reasons to think that the answer is ``no''.  (1) If the rings are
associated with ILR, then their radii should change as the galaxy evolves.  The
radius of outer ILR should increase with time, because the central concentration
increases and because $\Omega_p$ decreases. Simulations predict the opposite -- 
that rings shrink as they age (e.{\ts}g., Regan \& Teuben 2003).  However, this
happens when the potential is not allowed to evolve.  For the moment, we do not
know whether rings expand or shrink, but it is unlikely that they stay fixed
in radius.  Therefore we expect that the ring of star formation ``burns'' its
way through the pseudobulge as it grows.  (2) The spiral dust lanes interior to
the star formation rings (Figure 8) suggest that gas continues to sink inside
ILR (Elmegreen et al.~1998).  (3) We chose to illustrate star-forming rings,
because they most clearly make the connection between star formation and
bar-driven secular evolution.  However, in many galaxies, the star formation is
spread throughout the central region.  An example is NGC 1365 (Figure 7; Knapen
et al.~1995a, b; Sakamoto et al.~1995; Lindblad 1999).
}

      We argue in later sections, as did some of the above authors, that the
nuclear star formation is building pseudobulges.  Note that, although the star
formation is frequently in a ring, it is not likely to form a ring of stars.
If the star-forming ring is associated with ILR, then its radius should change
as the central concentration of the galaxy evolves.  We expect that the ring of
star formation ``burns'' its way through the pseudobulge as it grows.  Also,
the spiral dust lanes interior to the star formation rings (Figure 8) suggest 
that gas continues to sink inside ILR (Elmegreen et al.~1998).  Finally, we
chose to illustrate star-forming rings, because they most clearly make the
connection between star formation and bar-driven secular evolution.  However,
in many galaxies, the star formation is spread throughout the central region.
An example is NGC 1365 (Figure 7; Knapen et al.~1995a, b; Sakamoto et al.~1995;
Lindblad 1999).

      In summary, a comprehensive picture of the secular evolution
of barred galaxies has emerged as simulations of gas response to bars have
succeeded with increasing sophistication in matching observations of galaxies.
Bars rearrange disk gas to make outer rings, inner rings, and central mass concentrations.  SB(s) structure is favored if the bar is weak or rotating
rapidly; SB(r) structure is favored if the bar is strong or rotating slowly.
Since bars grow stronger and slow down as a result of angular momentum transport
to the disk, we conclude that SB(r) galaxies are more mature than SB(s) 
galaxies.  Consistent with this, dust lanes diagnostic of gas inflow are seen
in SB(s) galaxies but only rarely in SB(r) galaxies.  By the time an inner ring
is well developed, the gas inside it has been depleted. Embedded in this larger
picture is the most robust conclusion of both the modeling and the observations
-- that a substantial fraction of the disk gas falls down to small galactocentric radii in not more than a few billion years.  Star formation is
the expected result, and star formation plausibly associated with bars
(concentrated near resonance rings) is seen.  These results provide part of the
motivation for our conclusion that secular evolution builds pseudobulges, that 
is, dense but disk-like central components in spiral and S0 galaxies that are
not made by galaxy mergers.  

\vsl\vsss\vsss\vsss
\ni {\bf 3.~THE SECULAR EVOLUTION OF UNBARRED GALAXIES}
\vsl\vsss

      How general are the results of the previous section?~We reviewed the
effects of bars on disks as the most clearcut example of internal secular
evolution.\ts~But we do not mean to create the impression that such evolution
is important only in the $\sim 1/3$ of all disk galaxies that look barred at
optical wavelengths.  In this section, we first review evidence that many
apparently unbarred galaxies clearly show bars in the infrared.  With the
previous section as a guide, we then argue that similar evolution happens in
unbarred but oval galaxies and at slower rates in global-pattern spirals. 
In fact, any nonaxisymmetry in the gravitational potential plausibly 
rearranges disk gas.

\vfill\eject

\vsl\vsss\vsss\vsss\vsss
\ni {\bf 3.1.~Many ``Unbarred'' Galaxies Show Bars In the Infrared}
\vsl

      Near-infrared images penetrate dust absorption and are insensitive to the
low-$M/L$ frosting of young stars in Sb{\ts}--{\ts}Sm disks.  We then see the
underlying old stars that trace the mass distribution.  The most important
revelation is that bars are hidden in many galaxies that look unbarred in the
optical
(Block \& Wainscoat 1991;
Spillar et al.~1992;
Mulchaey \& Regan 1997;
Mulchaey et al.~1997; 
Seigar \& James 1998; 
Knapen et al.~2000;
Eskridge et al.~2000, 2002;
Block et al.~2001;
Laurikainen \& Salo 2002;
Whyte et al.~2002).~About 
two-thirds of all spiral galaxies look barred in the infrared.
Quantitative measures of bar strengths based on infrared images (Buta \& Block
2001; Block et al.~2001; Laurikainen \& Salo 2002) should prove useful in
gauging the consequences for secular evolution.

      Some bars are weak in amplitude.  But secular evolution can be more
important than this suggests, because many of bars are embedded in oval disks
(\S\ts3.2) that contribute at least as much to the nonaxisymmetric potential as
do the bars.  NGC 1068 is one example (Scoville et al.~1988; Thronson et
al.~1989; Pompea \& Rieke 1990); for others, see Hackwell \& Schweizer (1983);
Block et al.~(2002); Jarrett et al.~(2003).

      Hidden bars are the first reason why the results of \S\ts2 are relevant
to more than just the galaxies that look barred at optical wavelengths.

\vsl\vsss
\ni {\bf 3.2.~Oval Galaxies}
\vsl

      A strong bar has an axial ratio of $\sim$\ts1/5 and a mass of $\sim$\ts1/3
of the disk mass.  In this section, we discuss unbarred but globally oval
galaxies in which the whole inner disk has an axial ratio of $\sim$ 0.85. Ovals
are less elongated than bars, but more of the disk mass participates in the
nonaxisymmetry.  As a result, barred and oval galaxies evolve similarly.

      Strongly oval galaxies can be recognized independently by photometric 
criteria (Kormendy \& Norman 1979; Kormendy 1982a) and by kinematic criteria
(Bosma 1981a, b).  The diagnostics are illustrated in Figure 9.  

      \underbar{Brightness distributions:} In prototypical ovals, the disk
consists of two nested ovals, each with a shallow surface brightness gradient 
interior to a sharp outer edge.  The inner oval is much brighter than the outer
one.  The two ``shelves'' in the brightness distribution have different axial
ratios and position angles, so they must be oval if they are coplanar.  But the
flatness of edge-on galaxies shows that such disks really are oval.  Warped
disks are common, but they occur at lower surface brightnesses. 

      Nested ovals in unbarred galaxies are analogous, in barred galaxies, to
lenses with embedded bars interior to outer rings.  For the purposes of this
paper, lenses in early-type galaxies and oval disks in late-type galaxies are
functionally equivalent.  Both are elliptical shelves in the disk density,
and both are nonaxisymmetric enough to drive secular evolution. 


\eject

\cl{\null}

\vskip 2.75truein

\col{\includegraphics{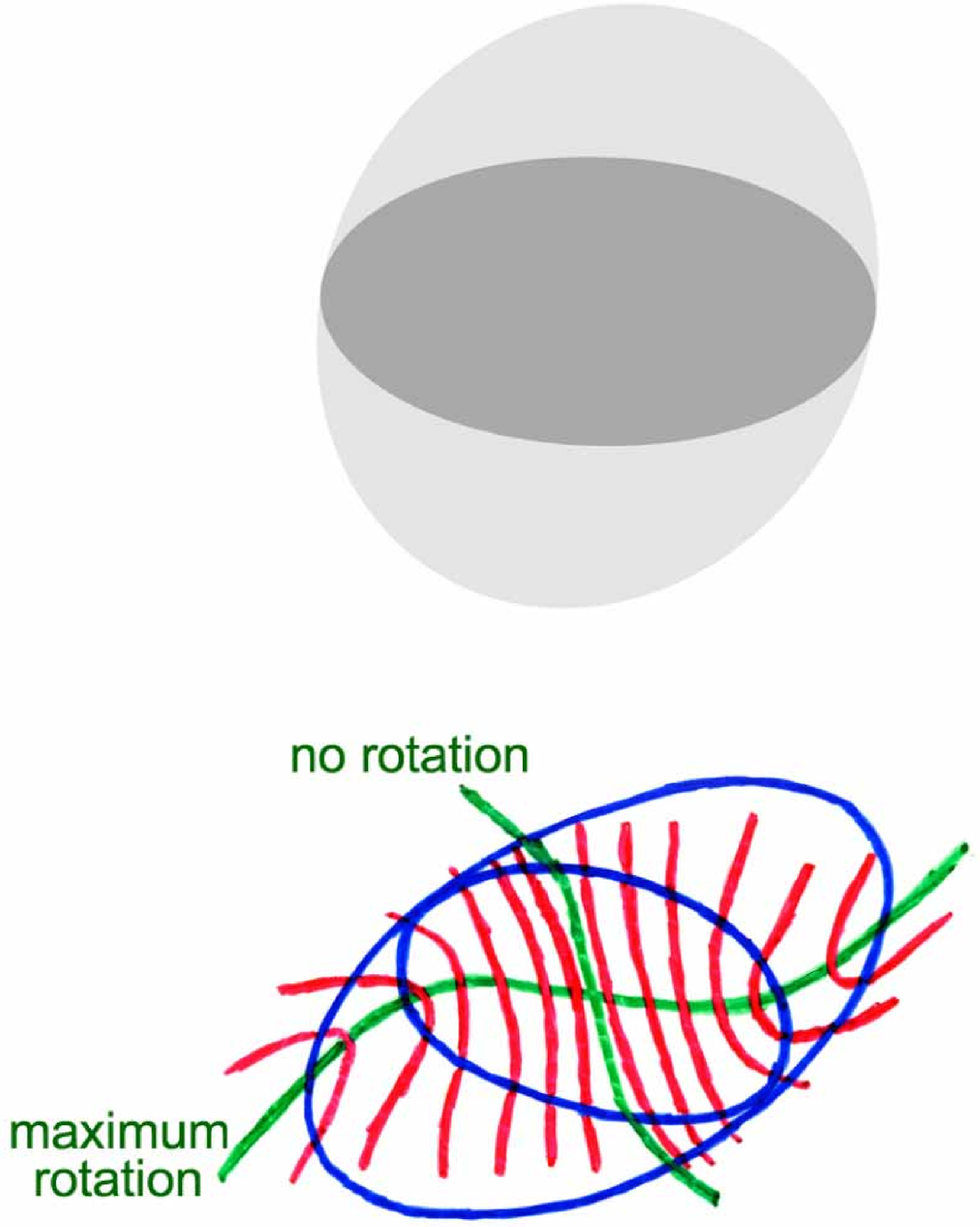}}

\col{\includegraphics{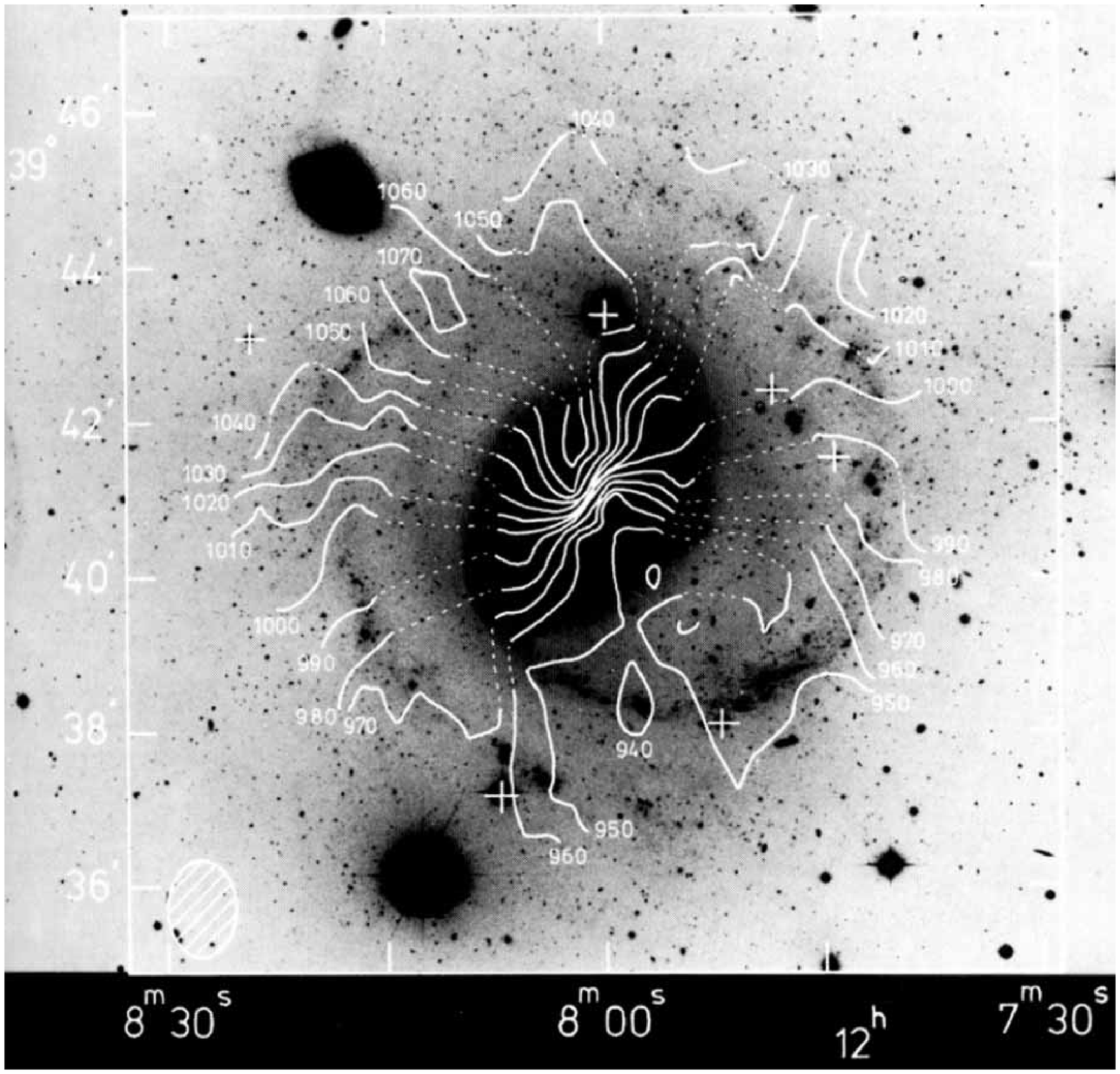}}

     {\it Figure 9}\quad Criteria for recognizing strongly oval but unbarred 
galaxies shown schematically at left and with observations of NGC 4151 at right.
This figure is adapted from Kormendy (1982a).  The NGC 4151 H{\ts}I velocity 
field is from Bosma, Ekers, \& Lequeux (1977a).

\vs\vs

      Besides NGC 4736 (Fig.~2) and NGC 4151 (Fig.~9), oval disks illustrated
in the {\it Hubble Atlas} (Sandage 1961) include NGC 4457 (Sa), NGC 3368
(Sa), NGC 4941 (Sa/Sb), NGC 1068 (Sb), NGC 210 (Sb), NGC 4258 (Sb), NGC 5248
(Sc), and NGC 2903 (Sc). Their similarity to barred galaxies can be seen
by comparing the two shelves in their brightness distributions 
with similar ones in NGC 1291 (Fig.~2), NGC 3945 and NGC 3081 (Fig.~5), 
and, in the {\it Hubble Atlas}, NGC 2859 (SB0), NGC 5101 (SB0), NGC 5566 (SBa),
NGC 3504 (SBb), and NGC 1097 (SBb).

      \underbar{Kinematics:} Velocity fields in oval disks are symmetric and
regular, but (1) the kinematic major axis twists with radius, (2) the optical
and kinematic major axes are different, and (3) the kinematic major and
minor axes are not perpendicular.  Twists in the kinematic principal axes are
also seen when disks warp.  Warps in H{\ts}I disks are common, but Bosma
(1981a, b) points out that they happen at larger radii and lower surface
brightnesses than oval structures, which are obvious in Figures 2 and 9 even 
at small radii.  Also, observations (2) and (3) imply ovals, not warps.

      As pointed out in Kormendy (1982a), the photometric and kinematic 
criteria for recognizing ovals are in excellent agreement.  These strong ovals
are expected to evolve similarly to barred galaxies, because the nonaxisymmetry 
in the potential is similar to that in barred galaxies.  In fact, many
simulations of the response of gas to ``bars'' actually assumed (presumably for
computational convenience) that all of the potential is somewhat oval rather
than that part of the potential is strongly barred and the rest is not.  NGC
4736 is representative of the many unbarred but oval galaxies with strong 
evidence for secular evolution (see Figure 8 for star formation and Figure 17 
for dynamical evidence).  

      So strongly oval galaxies are readily recognizable.  Many are classified
SAB; some are SA.  However, statistical analyses of large samples of galaxies
show that even unbarred disks are slightly oval.~The scatter in the Tully-Fisher
relation implies that the ellipticity in the potential that controls the disk
lies in the range \hbox{0\ts--\ts0.06.} (Franx \& de Zeeuw~1992).  The
corresponding axial ratio of the density distribution is 0.84\ts--\ts1.0.
Analyses of the velocity fields of individual galaxies give similar
results (e.{\ts}g., Andersen et al.~2001).  And, in a study of 18 face-on
spiral galaxies using K$^\prime$-band photometry, Rix \& Zaritsky (1995) showed
that the typical disk has axial ratio 0.91.  Not surprisingly, typical disks are
more circular than easily recognized ovals.  But they are not round.  This is 
plausible, since disks live inside cold dark matter halos that are predicted to
be very triaxial (Frenk et al.~1988; Warren et al.~1992; Cole \& Lacey 1996).
We now need an investigation of how much secular evolution is driven by the
above, small nonaxisymmetries. 


\omit{The origin of oval disks has no immediate bearing on their subsequent
secular evolution, so it is not directly a challenge to the themes of
this paper.  But it is a major puzzle.  There is a big problem in
making the obvious and canonical assumption that oval disks are a
response to triaxial halos.  In the absence of statistics on barred
galaxies, this would be a plausible assumption, and we cannot exclude
its relevance to unbarred ovals.  But the outer, oval disks of barred
galaxies -- including especially outer rings and pseudorings -- are not
aligned randomly with respect to the bar.  As shown by Buta (1995; see
also Kormendy 1979), outer rings are aligned either perpendicular to
the bar (in 2/3 of the cases) or parallel to the bar (in about 1/3 of
the cases).  The pattern speeds of bars are almost certainly much
faster than the pattern speeds of dark matter halos -- see the
heuristic discussion of pattern speeds in the section ``A Short Primer
on Resonances''.  So bars and halos should generally be misaligned.
How can the outer disk of an SB galaxy have its orientation controlled
by the bar unless the bar potential overwhelms the halo potential at
the large radii of outer rings.  This circumstance is exceedingly
surprising.  And if bars form outer rings as we have argued, while
unbarred ovals are oval because of triaxial dark matter halos, why are
their properties so similar to the ones of barred galaxies?  In fact,
outer disks and rings of unbarred ovals such as NGC 4736 also are
aligned parallel or (more commonly) perpendicularly to the inner oval,
suggesting that even in these cases the inner oval controls the
evolution.  We regard the lack of observational evidence that outer rings
``see'' triaxial halos as a major puzzle.}

\vsl\vsss
\ni {\bf 3.3.~The Demise of Bars}
\vsl

      Bars commit suicide if they drive gas inward and build up too
large a central mass concentration 
(Hasan \& Norman 1990; 
Freidli \& Pfenniger 1991;
Friedli \& Benz 1993;
Hasan, Pfenniger, \& Norman 1993;
Norman, Sellwood, \& Hasan 1996;
Heller \& Shlosman 1996;
Berentzen et al.~1998;
Sellwood \& Moore 1999;
Shen \& Sellwood 2004).
This is another example of an internal secular evolution process.
For example, Norman et al.~(1996) grew a point mass at the center of an $n$-body
disk that previously had formed a bar.  Before they switched on the point mass,
they checked that the bar was stable and long-lasting.  As they gradually turned
on the point mass, the bar amplitude weakened.  It weakened more for larger
point masses; central masses of 5\ts--\ts7\ts\% of the disk mass were enough to
dissolve the bar completely.  The result was a nearly axisymmetric galaxy.

      Why?  A heuristic understanding is provided by \S\ts2.2.  Inward gas
transport increases the circular-orbit rotation curve and the associated
epicyclic frequency $\kappa(r)$ of radial oscillations near the center.  
As~a~result, $\Omega - \kappa/2$ increases more rapidly toward small
radii.  That is, it is less nearly constant.  So it is more difficult for
self-gravity to persuade $x_1$ orbits with different radii to precess together
at $\Omega_p$ and not almost~together~at~$\Omega(r) - \kappa(r)/2$.  
Furthermore, while $\Omega - \kappa/2$ increases
because the central mass concentration increases, the bar slows down because it
transfers angular momentum to the outer disk.  That is, $\Omega_p$ and $\Omega 
- \kappa/2$ evolve in opposite directions.  This makes it still harder 
for $\Omega_p$ to be approximately equal to $\Omega - \kappa/2$.  And as the
radius of ILR grows, the radius range of the $x_2$ orbits that are perpendicular
to the bar and that cannot support it also grows.  Real bars get
nonlinear as their amplitude grows, so the epicyclic approximation on which this
discussion is based eventually breaks down\footnote{$^3$}{\kern -2pt For
example, in Norman et al.~(1996), $\Omega_p$ increases slightly late in the
simulation as the internal structure of the dissolving bar changes (Sellwood \&
Debattista 1993a).}.  Nevertheless, it provides a plausibility argument for the
result found in the simulations, which is that more and more orbits become
chaotic and cease to support the bar. 

      How much central mass is required to destroy the bar differs in different
papers.  In some simulations, a central mass of 2\ts\% of the disk already
weakens the bar (Berentzen et al.~1998).  Shen \& Sellwood (2004)
investigate this problem and find that great care is needed to make the time
step short enough near the central mass; otherwise, the bar
erodes erroneously quickly.  They also find that ``hard'' central
masses -- ones with small radii, like supermassive black holes -- destroy bars
more easily than ``soft'' masses -- ones with radii of several hundred parsec, 
like molecular clouds and pseudobulges.  A bar can tolerate a soft central mass
of 10\ts\% of the disk mass, although its amplitude is reduced by a factor of
$\sim$\ts2.


     What does a defunct bar look like?  Kormendy (1979b, 1981, 1982a) suggested
that some bars evolve into lens components.  The suggestion was based partly on
the observation (point 7 in Section 2.1) that, when they occur together, the bar
almost always fills the
lens in its longest dimension.  At the time, no reason for such evolution was
known.  However, the large velocity dispersion observed in the lens of NGC 1553
(Kormendy 1984) is consistent with the above idea, as follows.  Elmegreen \& 
Elmegreen (1985) find that early-type galaxies tend to have bars with flat
brightness profiles, while late-type galaxies tend to have bars with exponential
profiles.  Therefore, azimuthal phase-mixing of an early-type bar would
produce a hot disk with a brightness distribution like that of a lens, while
azimuthal phase-mixing of a late-type bar would produce a brightness
distribution that is indistinguishable from that of a late-type unbarred galaxy.
Lenses do occur preferentially in early-type galaxies (Kormendy 1979b).  To
test whether bars evolve into lenses, we need an $n$-body simulation
in which a bar with a flat profile and a sharp outer edge is destroyed by
growing a central mass. The bars in published simulations have steep
density profiles.


     Therefore, secular evolution tends to kill the bar that drives it.
The important implication is this: Even if a disk galaxy does not currently 
have a bar, bar-driven secular evolution may have happened in the past.

\vfill\eject

\vsl\vfill
\ni {\bf 3.4.~Global Pattern Spirals}
\vsl

      Our picture of global spiral structure in galaxies is by now well
developed (Toomre 1977b).  Global spirals are density waves that
propagate through the disk.  Like water waves in an ocean but unlike
bars, they are not made of the same material at all times.  In general,
stars and gas revolve around the center faster than the spiral arms, so
they catch up to the arms from behind and pass through them.  Central
to our understanding of why young and bright but short-lived stars are
concentrated in the arms is the concept that star formation is
triggered when gas passes through the arms.  As in the bar case, the gas
accelerates as it approaches the arms and decelerates as it leaves them.  
Again, shocks form where the gas piles up.  This time the shocks have a spiral
shape.  Their observational manifestations are dust lanes located on the 
concave side of the spiral arms (e.{\ts}g., NGC 5236 in Figure 7).
The strength of the shocks can be predicted
from the rotation curve:~the mass determines the rotation velocity,
and the central concentration determines the arm pitch angle and hence
the angle at which the gas enters the arms.  The results (Roberts,
Roberts, \& Shu 1975) provide the basis of our understanding of van
den Bergh (1960a,{\ts}b) luminosity classes of galaxies.  More massive
galaxies tend to have more differential rotation and stronger shocks,
so star formation is enhanced and the arms seen in young stars are
thinner and more regular.

      Gas loses energy at the shocks and sinks toward the center.
The effect is weaker than in barred galaxies, because the pitch angles of spiral
arms are much less than $90^\circ$.  The gas meets the shocks obliquely rather
than head-on.  Nevertheless, it must sink.  Where it stalls depends on the mass
distribution.  In early-type spirals with big classical bulges, the spiral
structure has an ILR at a large radius.  The spiral arms become azimuthal at 
ILR and stop there.  As the arm pitch angle approaches $0^\circ$ and as
the arm amplitude gets small, the energy loss drops to zero.  The 
gas stalls near ILR.  It may form some stars, but the bulge is already large,
so the relative contribution of secular evolution is minor.

      In contrast, in late-type galaxies, there is no ILR, or the ILR
radius is small.  The gas reaches small radii and high densities; the result 
is expected to be star formation.  If the process is fast enough, it can build
a pseudobulge.  Moreover, galaxies with no ILR are late in type.  They have
little or no classical bulge.  Therefore, secular processes can contribute a
central mass concentration that we would notice in just those galaxies in which
the evolution is most important.

      Is the evolution rapid enough to matter?  Theoretical timescales are
uncertain but look interestingly short.  Gnedin, Goodman, \& Frei (1995) measure
spiral arm torques from surface photometry of NGC 4321.  They estimate
that the timescale for the outward transport of angular momentum is
5{\ts}--{\ts}10 Gyr.  Thus, even the stellar distribution should have evolved
significantly if the spiral structure has consistently been as strong as it is
now. NGC 4321 has unusually regular and high-amplitude spiral arms; weak
spiral structure can easily imply angular momentum transport timescales that
are an order of magnitude longer (Bertin 1983; Carlberg 1987).  However, shocks
speed up the sinking of gas; Carlberg (1987) estimates that it takes
place on a Hubble timescale even for the weak spiral structure in his simulation.  Zhang (1996, 1998, 1999, 2003) derives even shorter timescales.
Apart from such disagreements, we do not know how long the spiral structure has
been as we observe it, a problem that Gnedin et al.~(1995) understood.  Despite
the uncertainties, more studies like Gnedin et al.~(1995) would be valuable.

      Whatever the theoretical uncertainties, observations show that star
formation takes place.  Timescales are discussed in Section 5.  Excellent
examples of nuclear star formation in unbarred galaxies can be seen in M{\ts}51
and NGC 4321 (Kormendy \& Cornell 2004 show illustrations).  Both galaxies have
exceedingly
regular global spiral structure.  The spiral arms and their dust lanes wind 
down very close to the center, where both galaxies have bright regions of star
formation.  NGC 4321 is studied by Arsenault et al.~(1988); Knapen et al.~(1995a, b), Sakamoto et al.~(1995), and  Garc\'\i a-Burillo et al.~(1998).
It is classified as Sc in Sandage (1961).  The RC3 (de Vaucouleurs et al.~1991)
classifies it SAB(s)bc; the spiral arms are distorted similar to pseudo-inner
and -outer rings.  There are signs of a weak bar in the infrared (see the above
references and Jarrett et al.~2003).  Nevertheless, NGC 4321 suggests that
secular evolution can be important even in galaxies that do not show prominent
bars.  


      Why doesn't every late-type galaxy have a pseudobulge?  Calculations
of spiral-arm shock strengths show that the shocks are weak if the rotation 
curve rises too linearly.  The lowest-luminosity galaxies have little shear;
it is not surprising that they do not make substantial pseudobulges. 

      In summary, late-type unbarred but global-pattern spirals are likely to
evolve in substantially the same way as barred galaxies, only more slowly. 

\vsl\vsss\vsss\vsss
\ni {\bf 3.4.~Conclusion}
\vsl

     Barred galaxies give us a rich picture of secular evolution at work.
One robust consequence is the buildup of the central mass concentration via 
the inward radial transport of gas.  Infrared imaging shows that the majority 
of spiral galaxies have bars.  Theoretical arguments and observational
evidence suggest that similar processes are at work in many unbarred
galaxies, especially in oval galaxies and in late-type, global-pattern spirals.
In late-type galaxies, it is relatively easy for the central mass concentration
that we see to be caused by secular processes, because the evolution happens
most readily if a galaxy does not already have a classical bulge.

\vsl\vsss\vsss\vsss
\ni {\bf 4.~THE OBSERVED PROPERTIES OF PSEUDOBULGES}
\vsl

      The suggestion that some ``bulges'' were built by the secular processes
discussed in \S\S\ts2 and 3 was first made by Kormendy (1982a, b).   A decade
later, both the evidence for evolution and the case that it can construct what
we now call pseudobulges had grown substantially (Kormendy 1993).  Now, after
another decade, it is a struggle to review the wealth of new evidence in a
single ARA\&A article.

      Other early papers that focused on the building of ``bulges'' by bars
include Combes \& Sanders (1981) and Pfenniger \& Norman (1990).  Two processes
were discussed. One is the inward transport of gas by bars and ovals. The other
involves dissipationless processes that can produce vertically thickened
central components when bars suffer buckling instabilities and when disk stars
scatter off of bars and are heated in the axial direction.  Both processes can
happen in the same galaxy and both make bulge-like components out of disk 
material.  Therefore we refer to the products of both processes as pseudobulges.
In this section, we discuss the observed properties of pseudobulges.  As
discussed in Section 1.1, we need the context of the above formation mechanisms
to make sense of the wealth (or plague) of detail in galactic centers. 

    How can we tell whether a ``bulge'' is like an elliptical or whether
it formed secularly?  The answer{\ts}--{\ts}and the theme of this
section{\ts}--{\ts}is that pseudobulges retain enough memory of their 
disky origin so that the best examples are easily recognizable. In the 
pre-{\it HST\/} era reviewed by Kormendy (1993), the cleanest evidence was
dynamical.  Pseudobulges are more dominated by rotation and less dominated by
random motions than are classical bulges and ellipticals.  This evidence 
remains compelling (Sections 4.6 and 4.7).  However, as a result of spectacular 
progress from {\it HST\/} imaging surveys, morphology and surface photometry 
now provide the best evidence for disk-like ``bulges''.   We begin with these
surveys. 

\vsl
\ni {\bf 4.1.~Embedded Disks, Spiral Structure, and Star Formation}
\vsss\vsss

   Central to our image of bulges as elliptical galaxies living in the middle of
a disk is their morphological resemblance to ellipticals.  Central to our
growing awareness that something different is going on in late-type galaxies is
the observation that their high-surface-brightness centers look nothing like
ellipticals.  Instead, they are dominated by young stars and by disk structure.
This is especially true in barred and oval galaxies, that is, in the objects in
which secular evolution should be most rapid. 

      What are we looking for?

  A clear statement that classical bulges are equivalent to ellipticals is
Sandage \& Bedke's~(1994) description of E/S0 galaxies: ``On short-exposure
plates showing only the central regions, no evidence of a disk~\dots~is seen;
the morphologies of the central regions are pure E.''  The section on elliptical
galaxies contains this caution:~``The presence or absence of dust is not
used as a classification criterion.  Some E galaxies~\dots~have dust patches and
remain classified as E types.''  The same is true for bulges; e.{\ts}g., S0
galaxies range from dustless (S0$_1$) to dusty (S0$_3$), but all have bulges. We
need to be careful that what we identify as pseudobulges are not just dust 
features or the outer disk extending inside a classical bulge all the way to
the center.  On the other hand, part of the definition of an elliptical, 
hence also of a bulge, is that ``There is no recent star formation, inferred 
from the absence of luminous blue and red supergiants''.  Even of Sab galaxies,
Sandage \& Bedke say that ``the central bulge is~\dots~nearly always devoid of
recently formed stars.''  Of course, old bulges must have contained young stars
in the past; these definitions -- and the Hubble sequence -- are understood to
apply to present-day galaxies and long after major mergers are completed.  But
ubiquitous ongoing star formation is a pseudobulge signature.

      Turning to pseudobulges, Kormendy (1993) noted that the prototypical oval
galaxy NGC 4736 has a disk-like ``bulge'':~``The central brightness
profile~\dots~is an $r^{1/4}$ law that reaches the high central surface
brightness characteristic of a bulge (Boroson 1981). However, the $r^{1/4}$-law 
component shows a nuclear bar and spiral structure to within a few arcsec of the
center. Bars are disk phenomena. More importantly, it is not possible to
make spiral structure in a bulge. Thus the morphology already shows that the
$r^{1/4}$-law profile belongs to the disk.''  This conclusion is consistent with
dynamical evidence shown in Figure 17 and and with the nuclear star formation
ring shown in Figure 8.

      Sandage (1961) comments similarly and presciently about flocculent
spirals, including NGC 4736, in his description of NGC 5055: ``The curious and
significant feature of [these galaxies] is the sharp discontinuity of surface
brightness of the spiral pattern between the inner and the outer regions [close 
to the center].  The spiral structure is as pronounced and well defined in the
bright region as in the outer parts.  The important point here is that, if the
inner arms were to coalesce and to lose their identity as spiral arms, the
region would look amorphous, would have a high surface brightness, and would
resemble the central regions of NGC 2841 [a classical bulge] \dots~and all
members of the E and S0 classes.''  The observation that the spiral structure 
is as pronounced in the bright region as in the outer parts has important 
implications.  If a high-surface-brightness classical bulge were projected in 
front of the (relatively faint) inward extrapolation of the outer disk, it would
dilute the spiral structure.  This is not seen.  Therefore it is the 
high-surface-brightness component that contains the spiral structure.
Again, this is a pseudobulge signature.

      {\it HST\/} spatial resolution reveals disk structure in the ``bulge''
regions of surprisingly many galaxies.  Carollo and collaborators have carried
out a snapshot survey of 75, S0 -- Sc galaxies with WFPC2 and the F606W filter
approximating $V$ band (Carollo et al.~1997, 1998; Carollo \& Stiavelli 1998;
Carollo 1999) and of 78 galaxies with NICMOS F160W approximating $H$ band
(Carollo et al.~2001, 2002; Seigar et al.~2002).  Figures 10 -- 13 show 
pseudobulges from these papers. 
What is remarkable about these generally Sb and Sbc galaxies is how often the
central structure looks like a smaller version of a normal, late-type disk.

     NGC 1353 (Figure 10) is one of the clearest examples.~The \hbox{top-right}
panel shows the central 18$^{\prime\prime}$\ts$\times$\ts18$^{\prime\prime}$ of
the PC image (Carollo et al.~1997, 1998).  The middle panel is the 
full WFPC2 field of view, and the bottom image is the 2MASS $JHK$-band 
composite.  The images show, as Carollo and collaborators concluded, that the
central structure in NGC 1353 is a disk with similar flattening and orientation
as the outer disk.  To make this quantitative, we measured the surface
brightness, ellipticity, and position angle profiles in the PC and 2MASS images
using the {\tt PROFILE} tool in the image processing system VISTA (Lauer 1985).
The left panels show that the apparent flattening at $2^{\prime\prime}$ \lapprox
\ts$r$ \lapprox \ts$4^{\prime\prime}$ is the same as that of the main disk at
large radii.  The position angle is the same, too.  So the part of the galaxy
shown in the top-right panel really is a disk.  The brightness profile shows
that this nuclear disk is responsible for much of the central rise in surface
brightness above the inward extrapolation of an exponential fitted to the outer
disk.  Presented only with the brightness profile or with the bottom two panels
of images, we would identify the central rise in surface brightness as a
bulge.  Given Figure 10, we identify it as a pseudobulge.

      We have decomposed the major-axis profile into an exponential outer disk
plus a S\'ersic (1968) function, $I(r) \propto e^{-K[(r/r_e)^{1/n} - 1]}$.
Here $n = 1$ for an exponential, $n = 4$ for a de Vaucouleurs (1948) $r^{1/4}$
law, and $K(n)$ is chosen so that radius $r_e$ contains half of the light
in the S\'ersic component.  In \S\ts4.2, we will discuss evidence that
``bulges'' in late-type galaxies are generally best described by S\'ersic
functions with $n \sim 1$.  That is, they are nearly exponential.  This behavior
is characteristic of many pseudobulges.  Here we note that NGC 1353 is an
example.  The best fit gives $n = 1.3 \pm 0.3$.

      The 2MASS image and the $\epsilon$ and PA profiles show that NGC 1353 
contains a weak bar with a projected radius of $\sim$\ts15$^{\prime\prime}$ and
an approximately NS orientation.  This is one example among many of the association between pseudobulges and nonaxisymmetric features 
that can drive secular evolution.  In visible light, the galaxy is classified
SBb by de Vaucouleurs et al.~(1991) 
and Sbc by Sandage \& Bedke (1994).

      Figure 11 shows another example.  NGC 5377 is classified SBa or Sa by
Sandage and (R)SBa by de Vaucouleurs, 
and it easily satisfies the photometric criteria for recognizing an oval outer
disk.  It is one of the earliest-type galaxies discussed in this paper.  An Sa should be dominated by a bulge.  Indeed, the brightness profile at $r$ \lapprox
\ts1$^{\prime\prime}$ and at about 6$^{\prime\prime}$ to 10$^{\prime\prime}$ is
bulge-like.  But the galaxy also contains a high-surface-brightness embedded
nuclear disk that is seen as the shelf in the brightness profile at $r \simeq
1^{\prime\prime}$ to $3^{\prime\prime}$.  Again, this has approximately the
same apparent flattening and position angle as the outer disk.  If a bulge is
defined to be the extra light at small radii above the inward extrapolation of
the outer disk profile, then that definition clearly includes the nuclear
disk.  We prefer not to adopt this definition but rather to identify NGC 5377 as
a galaxy with a substantial pseudobulge component.  Whether this is embedded in
a classical bulge or whether the whole of the central rise in surface brightness
is a pseudobulge, we cannot determine from the available data.

      Figure 12 shows a third case study, NGC 6384.  Its bar is subtle; the
galaxy is classified Sb by Sandage and SABbc by de Vaucouleurs. 
But it is clearly visible in the WFPC2 image (middle panel).  Sandage and Bedke
(1994) note that ``There is a smooth inner bulge \dots'' and the Carollo et
al.~(1998) image (top panel in Figure 12) confirms that the central brightness
distribution is smooth enough -- ignoring dust -- that one would ordinarily
identify this as a classical bulge.  However, photometry of the PC image shows
that the PA and apparent flattening are essentially the same at 
2$^{\prime\prime}$ \lapprox \ts$r$ \lapprox \ts12$^{\prime\prime}$  as in the
outer disk.  This ``bulge'' is quite flat.  Also, it is quite different from a
de Vaucouleurs $r^{1/4}$ law.  Carollo et al.~(1998) conclude that it is 
exponential.  We get $n = 2.2 \pm 0.2$, but this does not take into account the
light in the bar.  If bar stars that pass through the outer bulge were
subtracted from the profile, then $n$ would get smaller.  So the flatness of the
central component is enough to identify this as a pseudobulge, and its small
value of $n$ contributes to the identification of exponential profiles as a
pseudobulge characteristic (\S\ts4.2).  NGC 6384 demonstrates that pseudobulges
can be subtle enough so that photometry, and not just morphology, is needed to
recognize them.

     Further examples from Carollo et al.~(1997,{\ts}1998) of disky centers~in
Sa\ts--\ts{Sbc} galaxies are shown in Figure 13.~They look like miniature
\hbox{late-type} galaxies.  But they occur where the surface brightness rises
rapidly above the inward extrapolation of the outer disk profile.  This is not
obvious in Figure 13 because we use a logarithmic intensity stretch so that we
can show the structure over a large range in surface brightness.  Spiral
structure is a sure sign of a disk.  Carollo et al.~(1997) conclude that these
observations ``support scenarios in which a fraction of bulges forms relatively
late, in dissipative accretion events driven by the disk.''

      The statistics of the Carollo sample suggest that pseudobulges are
surprisingly common.  In the following summary, we distinguish classical bulges
that are well described by $r^{1/4}$ laws from pseudobulges that show at least
one of the following characteristics: they are flat or are dominated by disk
morphology such as spiral structure; they are vigorously forming stars; or they
have surface brightness profiles that are best described by S\'ersic functions
with $n$ \lapprox \ts2.  In a few cases, observing $n \simeq 1$ caused us to
reclassify a ``regular bulge'' in Carollo et al.~(1997, 1998) as a pseudobulge.
Also, we use the mean of the classifications given in the RC3 and in the
UGC/ESO-LV (Nilson 1973; Lauberts \& Valentijn 1989).  Then in the above sample
of 75 galaxies, classical bulges are seen in 69\ts\% of 13 \hbox{S0\ts--\ts{Sa}}
galaxies, 50\ts\% of 10 Sab galaxies, 22\ts\% of 23 Sb galaxies, 11\ts\%
of 19 Sbc galaxies, and 0\ts\% of 10 Sc and later-type galaxies.  Most of the
rest are pseudobulges or have a substantial pseudobulge component added to a
classical bulge.  In some cases, there is only a compact nuclear star cluster 
added to a late-type disk; it is not clear whether the same secular evolution
processes make these (see \S\ts4.9).  Distinguishing classical bulges from
pseudobulges is still an uncertain process.  Even the morphological types are
sometimes inconsistent between the RC3 and the UGC by several Hubble stages.
However, it is unlikely that the conclusions implied by the above statistics are
seriously wrong.  As noted by Carollo et al.~(1997, 1998), most early-type
galaxies appear to contain classical bulges; these become uncommon at types Sb
and later, and essentially no Sc or later-type galaxy has a classical bulge.
Kormendy (1993) reached similar conclusions.

      So an {\it HST} $V$-band survey shows that disky bulges are more common
than ground-based data suggested. Clearly it is desirable to check this result.
An $H$-band {\it HST} NICMOS survey by Carollo et al.~(2002) and by
Seigar et al.~(2002) complements the $V$-band survey in several ways.  The 
images are less affected by dust.  Classification of nuclear disks
is easier.  The infrared images are less sensitive to star formation, but
Carollo et al.~(2002) compensate by including $V - K$ color images.
The infrared survey confirms the $V$-band results.~Additional imaging~studies that reveal central disk structures, dust, 
or star formation in disk galaxies include Elmegreen et al.~(1998); van den
Bosch, Jaffe, \& van der Marel (1998); Peletier et al.~(1999); Erwin \& Sparke
(1999, 2002, 2003); Ravindranath et al.~(2001); Rest et al.~(2001); Hughes et
al.~(2003); Martini et al.~(2003a); Fathi \& Peletier (2003), Erwin et al.~2003,
and Erwin (2004).

      We do not mean to imply that a ``bulge'' is always either purely classical
or purely pseudo.  We cannot tell from available data how much of a classical 
bulge underlies the pseudobulge component in S0\ts--\ts{Sbc} galaxies.
Indications (e.{\ts}g., Figure 11) are that the classical bulge component in
many early-type galaxies is significant even when an embedded disky structure is
recognized.  If our formation picture is correct, then there is every reason to
expect that secular evolution often adds disky material to a classical bulge
that formed in a prior merger.  The relative importance of mergers and secular
evolution needs further investigation.

      At a more subtle level, some galaxies that are dominated by classical
bulges contain nuclear disks that contribute a negligible fraction of the galaxy
luminosity.  These may be cases in which secular evolution produced only a 
trace effect.~Alternatively, they may be later-type examples of the embedded
disks seen in elliptical galaxies.  If so, they cannot be a result of
disk-driven secular evolution.  They are discussed in Section 8.3.

\vfill\eject

\cl{\null}

\vfill

\col{\includegraphics{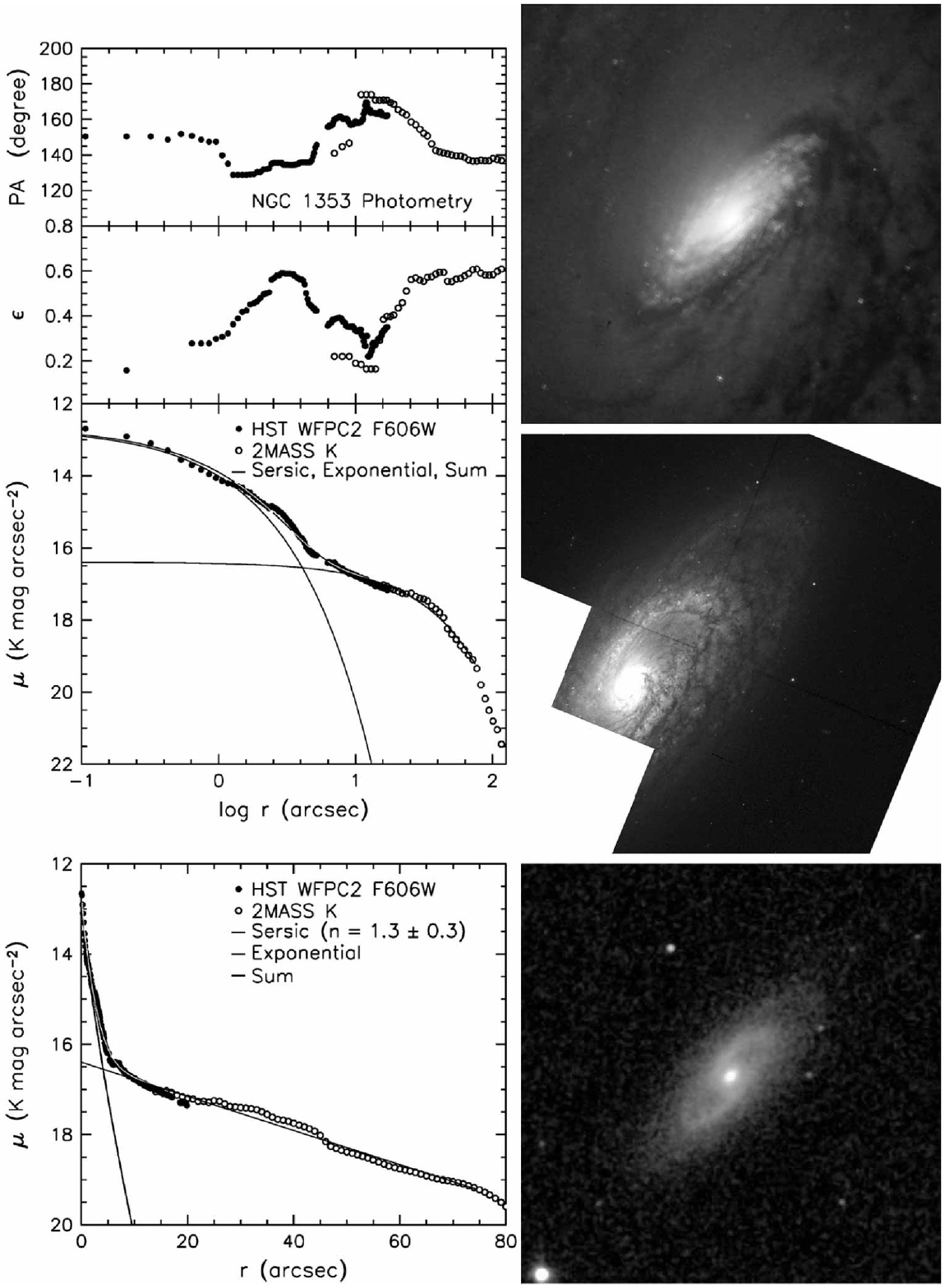}}

     {\it Figure 10}\quad NGC 1353 pseudobulge (top image: 18$^{\prime\prime}$ 
$\times$ 18$^{\prime\prime}$ zoom, and middle: full WFPC2 F606W image taken
with {\it HST\/} by Carollo et al.~1998).  The bottom panel is a 2MASS (Jarrett
et al.~2003) $JHK$
composite image with a field of view of 4\md4 $\times$ 4\md4.  The plots show
surface photometry with the {\it HST\/} profile shifted to the $K$-band
zeropoint.  The lines show a decomposition of the major-axis profile into a
S\'ersic (1968) function and an exponential disk.  The outer part of the
pseudobulge has the same apparent flattening as the disk.  This nuclear 
disk produces much of the rapid upturn in surface brightness toward the center. 

\eject

\cl{\null}

\vfill

\col{\includegraphics{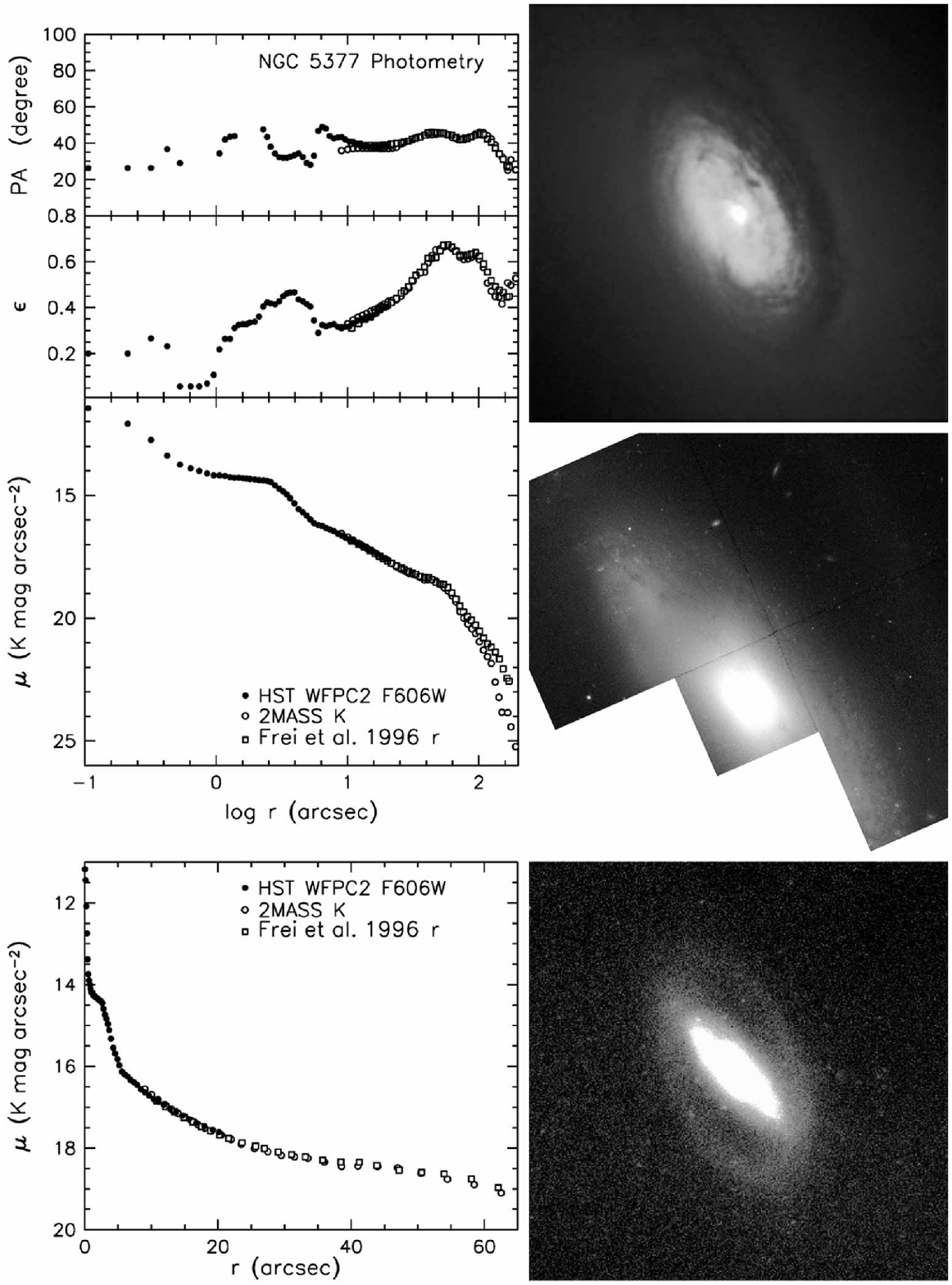}}

     {\it Figure 11}\quad NGC 5377 pseudobulge (top image: 18$^{\prime\prime}$ 
$\times$ 18$^{\prime\prime}$ zoom, and middle: full WFPC2 F606W image taken
with {\it HST\/} by Carollo et al.~1998).  At the bottom is a 
7$^{\prime}$\ts$\times$\ts7$^{\prime}$, $r$-band image of the outer ring (Frei
et al.~1996). The plots show surface photometry of the {\it HST\/}, $r$-band,
and 2MASS $JHK$ composite images, all shifted to the 2MASS, $K$-band zeropoint.
The two shelves in the brightness profile are the nuclear disk and inner oval.
The nuclear disk has the same apparent flattening and orientation as the outer
ring.  It may be embedded in a less obviously disky bulge, but it produces a
rapid upturn in surface brightness toward the center. 

\eject

\cl{\null}

\vfill

\col{\includegraphics{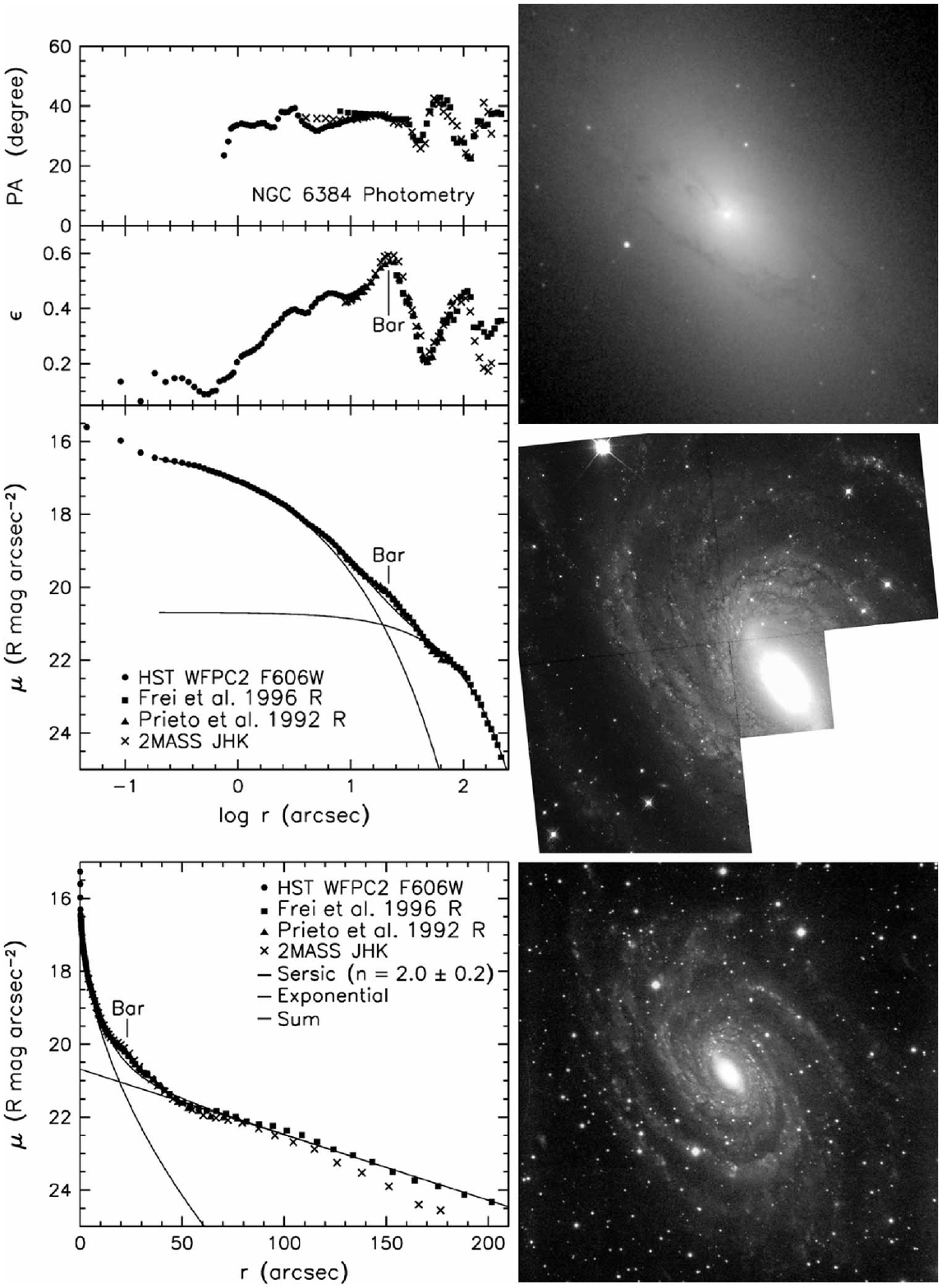}}

     {\it Figure 12}\quad NGC 6384 pseudobulge (top image: 18$^{\prime\prime}$ 
$\times$ 18$^{\prime\prime}$ zoom, and middle: full WFPC2 F606W image taken
with {\it HST\/} by Carollo et al.~1998).  At the bottom is the $B$-band image
from the {\it Carnegie Atlas of Galaxies\/} (Sandage \& Bedke 1994).  The top, 
middle, and bottom panels are shown with logarithmic, square root, and linear
stretches.  The plots show surface photometry of the {\it HST\/} and other
images identified in the key, all shifted to the $R$-band zeropoint.  The
decomposition into a S\'ersic function bulge and exponential disk is done over
a radius range that omits the region 
$12^{\prime\prime} < r < 40^{\prime\prime}$ affected by the bar.

\eject

\cl{\null}

\vfill


%

\col{\includegraphics{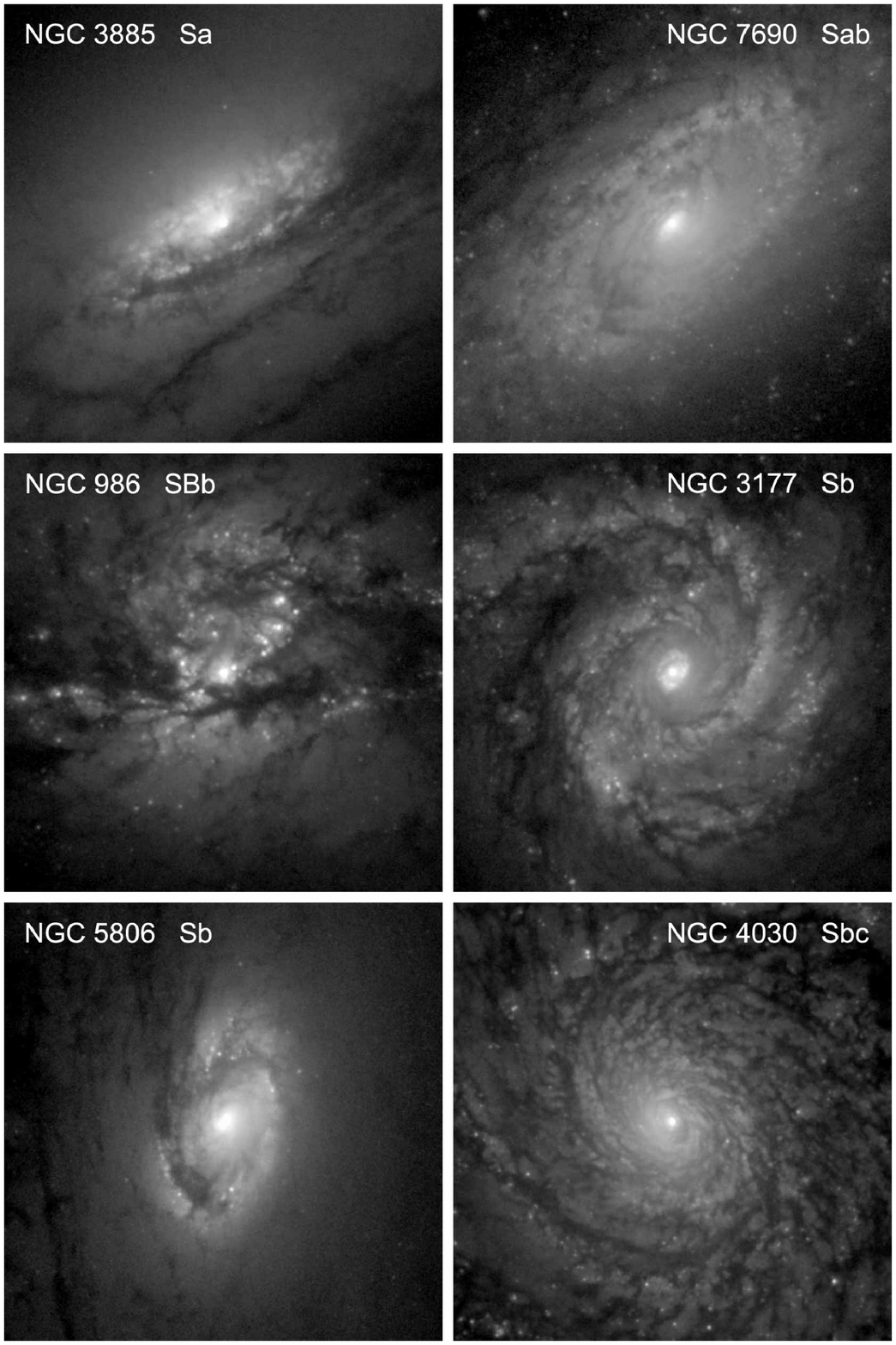}}

     {\it Figure 13}\quad Sa -- Sbc galaxies whose ``bulges'' have
disk-like properties.  Each panel shows an 18$^{\prime\prime}$ $\times$
18$^{\prime\prime}$ region centered on the galaxy nucleus and extracted from
{\it HST\/} WFPC2 F606W images taken and kindly provided by Carollo et
al.~(1998).  North is up and east is at left.  Displayed intensity is
proportional to the logarithm of the galaxy surface brightness.  Hubble
types are from Sandage \& Bedke (1994) except for NGC 4030; its type is
from the RC3 and was checked using high-quality images posted on NED.

\eject

\vsl\vsss
\ni {\bf 4.2.~Exponential Bulges}
\vsl

      The pseudobulge galaxies NGC 1553 and NGC 6384 (Figures 10, 12) have
nearly exponential bulge profiles.  Andredakis \& Sanders (1994) discovered
that this is a general phenomenon: The bulges of late-type galaxies are
better described by exponentials than by $r^{1/4}$-law surface brightness
profiles.  Andredakis, Peletier, \& Balcells (1995) generalized this
result and showed that the index $n$ of a S\'ersic (1968) function fitted to
the central profile varies from 
$n \simeq 3.7$ (standard deviation = 1.3) in S0 and S0/a bulges to 
$n \simeq 2.4$ (standard deviation = 0.66) in Sa -- Sb galaxies to 
$n \simeq 1.6$ (standard deviation = 0.52) in Sbc -- Sd galaxies.  
``For Sc and later, the profiles are very close to pure exponentials.''  
An example of an Sbc with an exponential bulge is our Galaxy (Kent,
Dame,~\&~Fazio~1991).  The above trend parallels the trend that pseudobulges
get more common in later-type galaxies.  Evidently small $n$ values are
pseudobulge signatures. 

      The time was right for S\'ersic functions to become the canonical fitting
function for bulges and ellipticals.  Caon, Capaccioli, \& D'Onofrio (1993) had
recently demonstrated that they fit ellipticals and bulges better than do
$r^{1/4}$ laws.  They note that this is not a surprise, because $r^{1/n}$
profiles have three parameters, while $r^{1/4}$ laws have only two.  The
argument that $n$ has physical meaning is the observation that it
correlates with the effective radius $r_e$ and total absolute magnitude
$M_{B,{\rm bulge}}$ of the elliptical or bulge.  This was confirmed by
D'Onofrio, Capaccioli, \& Caon (1994); Graham et al.~(1996); Binggeli \& Jerjen
(1998); Graham (2001); Trujillo et al.~(2002), and numerous subsequent papers.

      The idea that late-type ``bulges'' have $n \simeq 1$ to 2 immediately
gained acceptance and got simplified in many people's minds (and in our~title)
to the notion that they are exponential. One reason was that
confirmation followed quickly.  Courteau, de Jong, \& Broeils (1996) carried 
out bulge-disk decompositions for 243 galaxies from Courteau (1996a) and 86
galaxies from de Jong \& van der Kruit (1994) and from de Jong (1996a, b).  For
the Courteau sample, they conclude that ``about 85\ts\% of [the] Sb's and Sc's
are best fitted by the double exponential, while the remainder [are] better
fitted with an $r^{1/4}$ bulge profile.'' For the de Jong sample, they conclude 
that 60\ts\% of the galaxies are best modeled by a double exponential, $\sim 
25$\ts\% (mostly Sa's and Sb's) are best modeled with $n = 2$ and  only $\sim
15$\ts\% are best fitted by an $r^{1/4}$ law.  These results are broadly
consistent with the statistics in \S\ts4.1, which refer to a different galaxy
sample and which are partly based on morphology and partly on S\'ersic function
indices.  

      As a diagnostic of formation processes, Courteau et al.~(1996) go on to
examine the ratio $h_b/h_d$ of the scale lengths  of the inner and outer
exponentials. For the combined sample, they find that $h_b/h_d = 0.08 \pm 0.05$,
and for the de Jong sample, they find that $h_b/h_d = 0.09 \pm 0.04$. From this,
they conclude that: ``Our measurements of exponential stellar density profiles
[in bulges] as well as a restricted range of [bulge-to-disk] scale lengths
provide strong observational support for secular evolution models.
Self-consistent numerical simulations of disk galaxies evolve toward a double
exponential profile with a typical ratio between bulge and disk scale lengths
near 0.1 (D.~Friedli 1995, private communication) in excellent agreement with
our measured values'' (see Courteau et al.~1996 for details).
MacArthur, Courteau, \& Holtzman (2003) find that $h_b/h_d = 0.13
\pm 0.06$ for late-type spirals and again note the connection with secular
evolution.  We can add one more connection.  The above ratios of $h_b/h_d$,
together with the observation that bars are typically about 1 scale length
$h_d$ long (e.{\ts}g., Kormendy 1979b), imply that the scale length of the inner
exponential is similar to the radius of star-forming rings (Figure 8) discussed
in Section 2.3. We suggested there that these rings are building pseudobulges.

      There is a caveat: 
An examination of the above papers shows
that many bulges in late-type galaxies rise above the disk profile by only
small amounts.  
Leverage is limited.  Even the conclusion that some bulges are exponential
can be uncertain.

\omit{
      {\it HST\/} confirmation of the above results has therefore been very 
welcome.  Carollo et al.~(2002) provide the best statistics.  Their Table 1
classifies central components as ``$r^{1/4}$-law'', ''exponential'', or ``not
fitted'' based on the $V$-band images.  Galaxies were ``not fitted'' when the
brightness distribution was badly affected by dust, young stars, or patchiness.
One galaxy, NGC 2344, is classified as an Sc by the RC3 but has an $r^{1/4}$-law
bulge.  Images posted on the NASA/IPAC Extragalactic Database (NED) make it
clear that this galaxy is not an Sc.  We adopt the UGC classification, which is
Sb.  With this correction, the $V$-band statistics are as follows:
$r^{1/4}$-law bulges, exponential pseudobulges, and galaxies ``not fitted'' 
account for the following percentages of the Hubble types indicated: 
S0\ts$+$\ts{Sa} -- 50\ts\%, 10\ts\%, 40\ts\%; 
Sab -- 60\ts\%, 0\ts\%, 40\ts\%; 
Sb -- 17\ts\%, 11\ts\%, 72\ts\%; 
Sbc -- 0\ts\%, 28\ts\%, 72\ts\%; 
Sc -- 0\ts\%, 60\ts\%, 40\ts\%; 
Scd\ts{to}\ts{Sm} -- 0\ts\%, 50\ts\%, 50\ts\% of the galaxies.
When we classify the 45 galaxies that were not fitted in $V$-band using the
$H$-band images and $V - H$ images, we get 11 classical bulges and 34 
pseudobulges.  In most cases, the classification is clearcut; when it is not,
we try to err equally often in favor of classical bulges and pseudobulges.
The statistics on classical and pseudobulges~then~become:
S0\ts$+$\ts{Sa}{\ts}--{\ts}50\ts\%, 50\ts\%; 
Sab{\ts}--{\ts}60\ts\%, 40\ts\%;
Sb{\ts}--{\ts}44\ts\%, 56\ts\%;
Sbc{\ts}--{\ts}6\ts\%; 94\ts\%;
Sc\ts{to}\ts{Sm}{\ts}--{\ts}0\ts\%, 100\ts\%.  The differences between these
results and the ones quoted in \S\ts4.1 for the $V$-band survey alone give some
indication of the classification errors.  The $V$\null$+$\ts$H$-band results
are in satisfactory agreement with the optical results.  The majority of
early-type galaxies have classical bulges; there is a sharp transition at 
Hubble type Sb, and later-type galaxies mostly contain pseudobulges.
      Many other {\it HST} studies reach similar conclusions (e.{\ts}g.,
Phillips et al.~1996; Balcells et al.~2003; Fathi \& Peletier 2003).  Balcells
et al.~(2003) emphasize that 84\ts\% of their galaxies contain nuclei and that
the S\'ersic index $n$ is overestimated if these nuclei are mistakenly included
-- as they would be at ground-based spatial resolution -- in the profile
decompositions.  They find a mean index $<$\null$n$\null$>$ = $1.7 \pm 0.7$.
}

     {\it HST\/} confirmation of the above results has therefore been
welcome (e.{\ts}g.,  Phillips et al.~1996, Balcells et al.~2003, Fathi
\& Peletier 2003). Carollo et al.~(2002) provide the best statistics.  Their
Table 1 classifies central components as ``$r^{1/4}$-law'',
``exponential," or ``not fitted'' based on the $V$-band images.
Galaxies were ``not fitted'' when the brightness distribution was
badly affected by dust, young stars, or patchiness. After correcting the 
Hubble type of NGC 2344 from Sc (RC3) to Sb (UGC), 
the $V$-band statistics are as follows: $r^{1/4}$-law bulges,
exponential pseudobulges, and galaxies not fitted account for the following percentages of the Hubble types indicated. 
S0\ts$+$\ts{Sa}: 50\%, 10\%, 40\%; 
Sab: 60\%, 0\%, 40\%; 
Sb: 17\%, 11\%, 72\%; 
Sbc: 0\%, 28\%, 72\%; 
Sc: 0\%, 60\%, 40\%; and
Scd\ts{to}\ts{Sm}: 0\%, 50\%, 50\% of the galaxies.
When we classify the 45 galaxies that were not fitted in $V$-band
using the $H$-band images and $V-H$ images, we get 11 classical
bulges and 34 pseudobulges.   The statistics on classical and
pseudobulges~then~become as follows: S0\ts$+$\ts Sa: 50\%,
50\%; Sab: 60\%, 40\%; Sb: 44\%,
56\%; Sbc: 6\%; 94\%; and 
Sc\ts{to}\ts{Sm}: 0\%, 100\%.   The $V$\null$+$\ts$H$-band results are in
satisfactory agreement with the optical results.  The majority of early-type
galaxies have classical bulges; there is a sharp transition at
Hubble type Sb, and later-type galaxies mostly contain pseudobulges.

      Balcells (2001) reviews implications.  Andredakis (1988) comments~that
``The exponential bulges \dots~remain essentially unexplained; [his
results] suggest that they \dots~were probably formed, at least in part, by
different processes from those of early-type spirals.''  Even though we do not
understand quantitatively how inner exponentials are built, their close
association with other disky bulge phenomena supports our tentative conclusion
and that of many other authors that S\'ersic indices $n \sim 1$ are a signature
of secular formation.

\vfill\eject

\vsl\vsss
\ni {\bf 4.3.~Some ``Bulges'' Are As Flat As Disks}
\vsss

     In Section 4.1, we repeatedly noted that pseudobulges examples were very
flat, based on observed axial ratios or spiral structure.  Secular formation
out of disks does not require them all to be flat (Section 7.1), but it
appears that we are fortunate and that many are flat.  

     This is seen in the distribution of observed bulge ellipticities derived 
by Kent (1985, 1987a, 1988).  He decomposed major- and minor-axis profiles of
disk galaxies into $r^{1/4}$-law bulges and exponential disks.  The bulge and
disk ellipticities were fit parameters that were allowed to be different.
Figure 8 in Kormendy (1993) shows the following:

\vs
\nnhi 1.~A majority of bulges appear rounder than their associated disks.
         These include the well known classical bulges in M{\ts}31, M{\ts}81,
         NGC 2841, NGC 3115, and NGC 4594 (the Sombrero galaxy).

\vs
\nnhi 2.~Some bulges have apparent flattenings that are similar to those of
         their associated disks, as Kent noted.

\vs
\nnhi 3.~Some bulges appear more flattened than their associated disks; these
         may be nuclear bars (Section 4.4).

\vs
\nnhi 4.~The median ratio of bulge and disk ellipticities, 
         $\epsilon_{\rm bulge}/\epsilon_{\rm disk}$, is smallest for Sas and
         increases toward later Hubble types.  This agrees with other evidence
         that pseudobulges are more common in later-type galaxies.  

\vs
\nnhi 5.~However, the median $\epsilon_{\rm  bulge}/\epsilon_{\rm disk}$ for S0
         galaxies is similar to that for Scs, not Sas.  Kinematically disklike
         bulges also are more common in S0s than in Sas (Sections 4.6 and 4.7).
         Similar effects led van den Bergh (1976b) to develop his ``parallel 
         sequence'' classification. 

\vs

      Bulge-disk decompositions should be interpreted with caution.~The bulge
and disk parameters are strongly coupled.  Even when the bulge ellipticity is a
fit parameter, it assumed to be constant with radius; this is necessary for
computational stability.  But Figures 10\ts--\ts{12} and much other data show
that this is a serious oversimplification. Also, most decompositions in the
literature are not suitable.  Some have too little leverage on the bulge.
Non-parametric decompositions depend on the assumption that the bulge and disk
have different flattenings; they force the bulge to be rounder than the disk.
So we have few checks of the above results.  Those that are available are
consistent with points 1\ts--\ts5 but show a large dispersion in numbers.
Here are two examples:

      Fathi \& Peletier (2003) carry out bulge-disk decompositions for 35
\hbox{S0\ts--\ts{Sb}} and 35 Sbc\ts--\ts{Sm} galaxies based on {\it HST\/}
NICMOS $H$-band images.  The high spatial resolution provides good leverage
on small ``bulges''.  The results show that 
$\epsilon_{\rm bulge}/\epsilon_{\rm disk} > 0.9$ in 36\ts\% of  S0\ts--\ts{Sb} 
galaxies and 51\ts\% of Sbc\ts--\ts{Sm} galaxies.  This is consistent with
Kent's decompositions and confirms that flat ``bulges'' are more common in
late-type galaxies.  

      In contrast, M\"ollenhoff \& Heidt (2001) find that only 10\ts\% of their
decompositions imply $\epsilon_{\rm bulge}/\epsilon_{\rm disk} > 0.9$. These
are $K$-band measurements of a sample of S0{\ts}--\ts{Sc} galaxies weighted
toward later Hubble types.  The galaxies are relatively face-on; this
reduces sensitivity to the flattening.  However, the above results refer to the
39 galaxies that meet the selection criterion used for points 1\ts--\ts5,
$\epsilon_{\rm disk} < 0.14$.  So different authors get substantially different
distributions of bulge flattening.  On the other hand, Figures 10\ts--\ts{12}
clearly show that some pseudobulges are as flat as disks. 

\vsl\vsss
\ni {\bf 4.4.~Bars Within Bars}
\vsl

      Figure 14 shows galaxies that have a secondary bar interior to the main 
bar.  The inner bar is, in fact, the ``bulge'' -- its surface brightness 
increases rapidly toward the center, far above the inward extrapolation of the
disk brightness profile.  However, bars are disk phenomena.  Seeing a nuclear
bar is strong evidence that a galactic center is dominated by a pseudobulge.

      The nuclear bar in NGC 1291 was seen as long ago as Evans (1951). 
de Vaucouleurs (1975) saw nuclear bars in four of the six galaxies
illustrated in Figure~14: NGC 1291, NGC 1433 (see also Sandage \& Brucato 1979),
NGC 1543 (see de Vaucouleurs 1959), and NGC 3081.  

      Other early examples are NGC 1326 (de Vaucouleurs 1974b), NGC~2859, NGC
3945, NGC 7743 (Kormendy 1979b), NGC 1543 (Sandage \& Brucato 1979), NGC 1317
(Schweizer 1980), and NGC 2950 (Kormendy 1981,1982a, b).  Kormendy concluded:
{\it ``triaxial SB bulges and bars rotate rapidly and are therefore dynamically similar.  Both are different from elliptical galaxies, which rotate slowly.''}
We return to these points in Section 4.6.

      The number of known examples grew rapidly as work on barred galaxies
accelerated (Jarvis et al.~1988; Buta 1990; Buta \& Crocker 1993; Shaw et
al.~1993b, 1995; Wozniak et al.~1995; Friedli et al.~1996; Elmegreen et
al.~1996;
Jungwiert, Combes, \& Axon 1997; Mulchaey et al.~1997; Erwin \& Sparke 1999,
2003; M\'arquez et al.~1999; Martini \& Pogge 1999; Colina \& Wada 2000;
Greusard et al.~2000; Rest et al.~2001).  Erwin (2004) has compiled a catalog,
and Friedli (1996) and Erwin (2004) provide reviews.

      Recent studies focus on larger and more representative samples and
therefore yield better estimates of what fraction of SB galaxies contain nuclear
bars.  Erwin \& Sparke (2002) find nuclear bars in $26 \pm 7$\ts\% of their
sample of 38 SB galaxies.  They remark that the true fraction could be as large
as 40\ts\%; they could not detect nuclear bars in the (many) objects that have
central dust.  As in the previous section, pseudobulge features are surprisingly common.  The galaxies in the above survey are S0\ts--\ts{Sa}; these are the
Hubble types that are most likely to contain classical bulges.

      Laine et al.~(2002) analyze {\it HST\/} NICMOS $H$-band images of a
matched sample of Seyfert and non-Seyfert galaxies.  The sample is slightly 
biased toward early Hubble types but otherwise is representative.  They
find that 28 $\pm$ 5\ts\% of their barred galaxies have a nuclear bar.
They also find several indications that nuclear and main bars have a
different origin, most notably that main bar sizes are proportional to the
scale length of the disk while nuclear bar sizes are uncorrelated with the
size of the disk and almost always smaller than $\sim$ 1.6 kpc in radius.
Nuclear bars and nuclear star-forming rings have similar size distributions when
normalized by the galactic diameter $D_{25}$.  They argue plausibly that this
means that nuclear bar radii, like nuclear ring radii, are bounded approximately
by ILR (see also Pfenniger \& Norman 1990; Friedli \& Martinet 1993).  

      Observations like these support the cononical hypothesis that nuclear 
bars form when infalling disk gas builds up a central, cold, and disky system
that is sufficiently self-gravitating to become barred.  How this
happens is not known.  One possibility is that a cold nuclear disk suffers its
own bar instability, independent of that of the main bar (Friedli \& Martinet
1993; Combes 1994).  

      A good sign that we understand the essence of nuclear bar dynamics is the
observation (Figure 14) that inner bars are oriented randomly with
respect to main bars (Buta \& Crocker 1993; Friedli \& Martinet 1993; Shaw et al.~1995; Wozniak et al.~1995; Friedli et al.~1996; Erwin \& Sparke 2002). This
can be understood within the dynamical framework of \S\ts2.2.  At small
radii, $\Omega(r) - \kappa(r)/2$ reaches a high peak in galaxies that have such
high central mass concentrations. A bar's pattern speed $\Omega_p$ seeks
out approximately the local angular velocity $\Omega - \kappa/2$ at which closed
ILR orbits precess.  Therefore, the pattern speeds of inner bars are almost
certainly much higher than those of main bars\footnote{$^4$}
{Similarly, because $\Omega - \kappa/2$ decreases outward, the pattern speeds 
of spiral arms are likely to be slower than
those of bars (Sellwood 1985; Sparke \& Sellwood 1987; Sellwood \& Sparke 1988;
SW93).  This accounts for the comment in \S\ts2.1 that the spiral arms of SB(r)
galaxies ``often [begin] downstream from the ends of the bar'' (Sandage \& Bedke
1994).} (see Pfenniger \& Norman 1990, Friedli \& Martinet 1993, Buta \& Combes
1996, and Maciejewski \& Sparke 2000 for further discussion).
Kinematic decoupling of main and nuclear bars is observed by Emsellem et
al.~(2001) and Corsini et al.~(2003).

      Shlosman, Frank, \& Begelman (1989) suggest that bars within bars are a
primary way to transport gas farther inward than the gravitational torque of 
the main bar can achieve.  To fuel nuclear activity in galaxies, they envisage 
a hierarchy of bars within bars.  Triple bars have been seen (Friedli
1996 and Erwin \& Sparke 1999 provide reviews).

NGC 4736 is an example of a nuclear bar in an unbarred but oval galaxy (Block 
et al.~1994; M\"ollenhoff, Matthias, \& Gerhard 1995).  It emphasizes again the
similarity between bars and ovals as engines for secular evolution.

\vfill\eject

\cl{\null}

\vfill







\col{\includegraphics{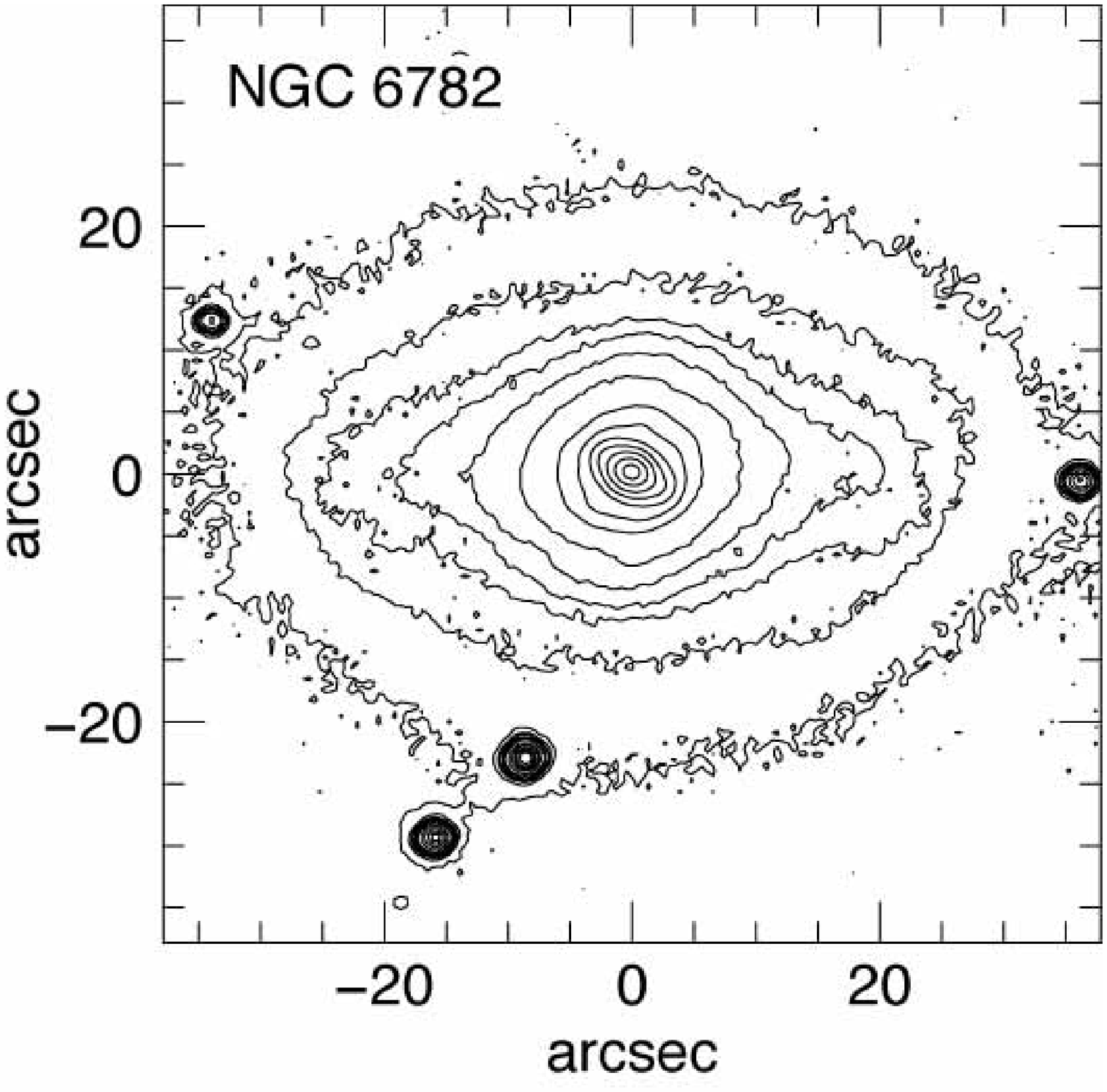}}

\col{\includegraphics{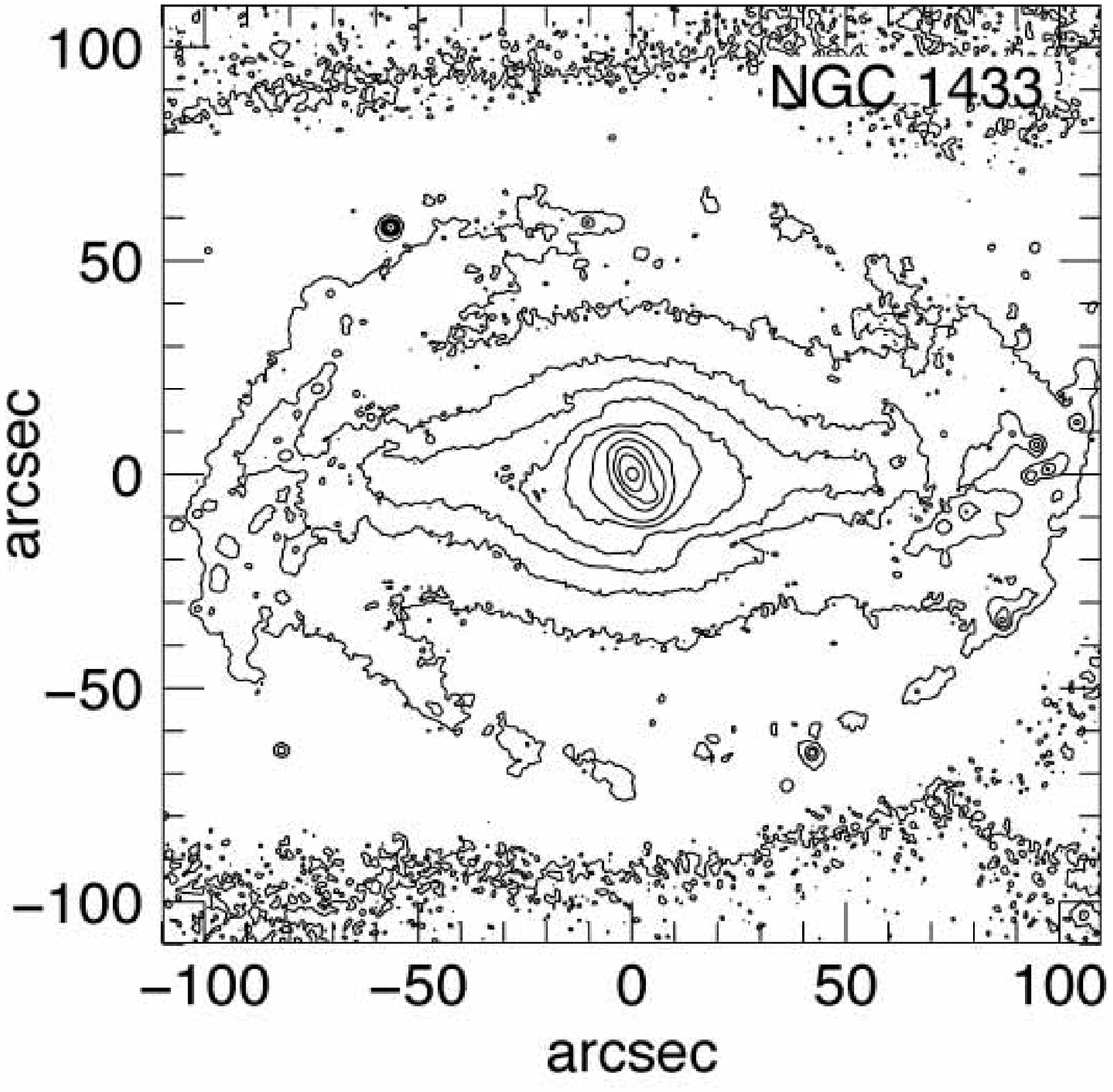}}

\col{\includegraphics{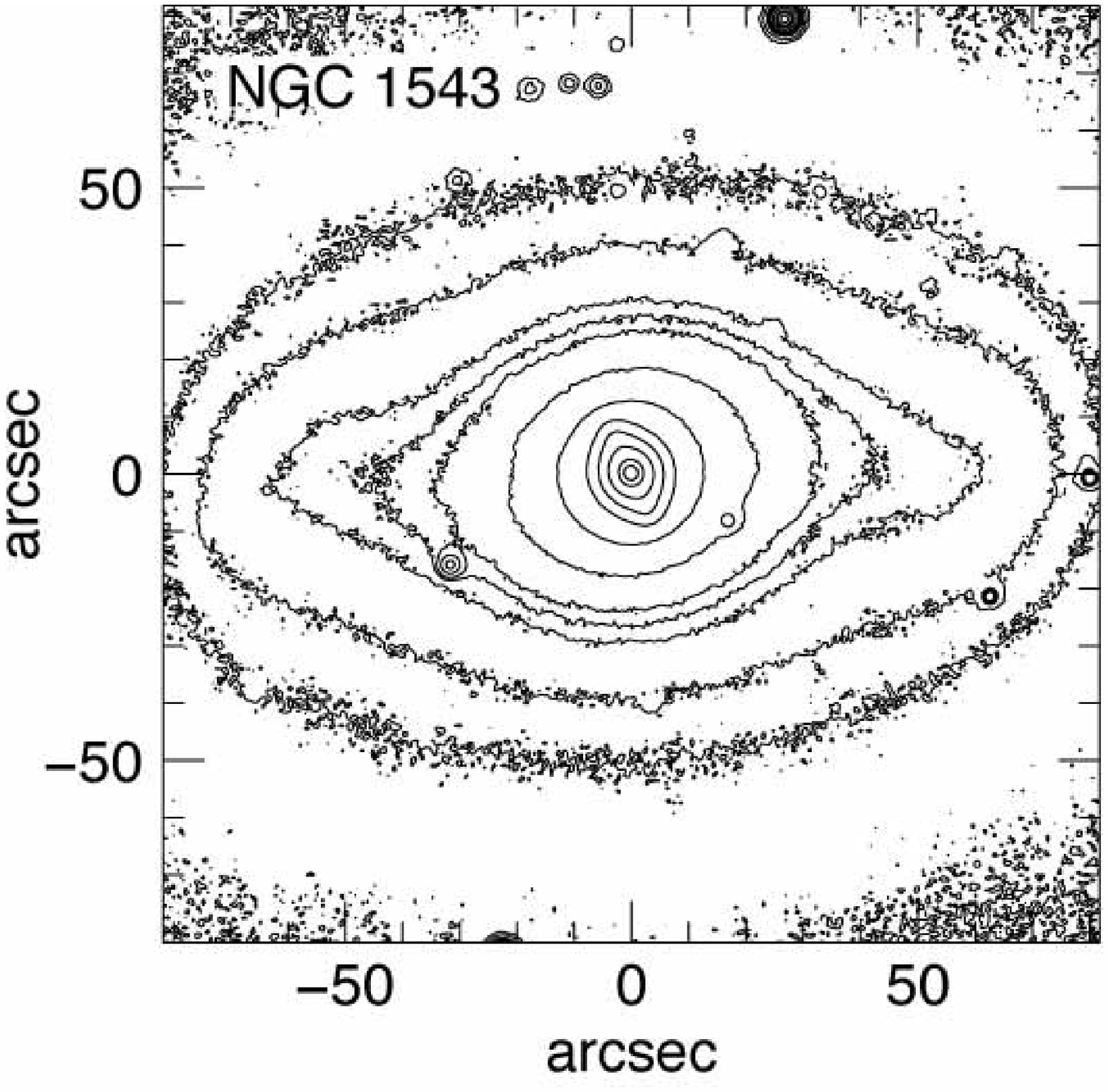}}

\col{\includegraphics{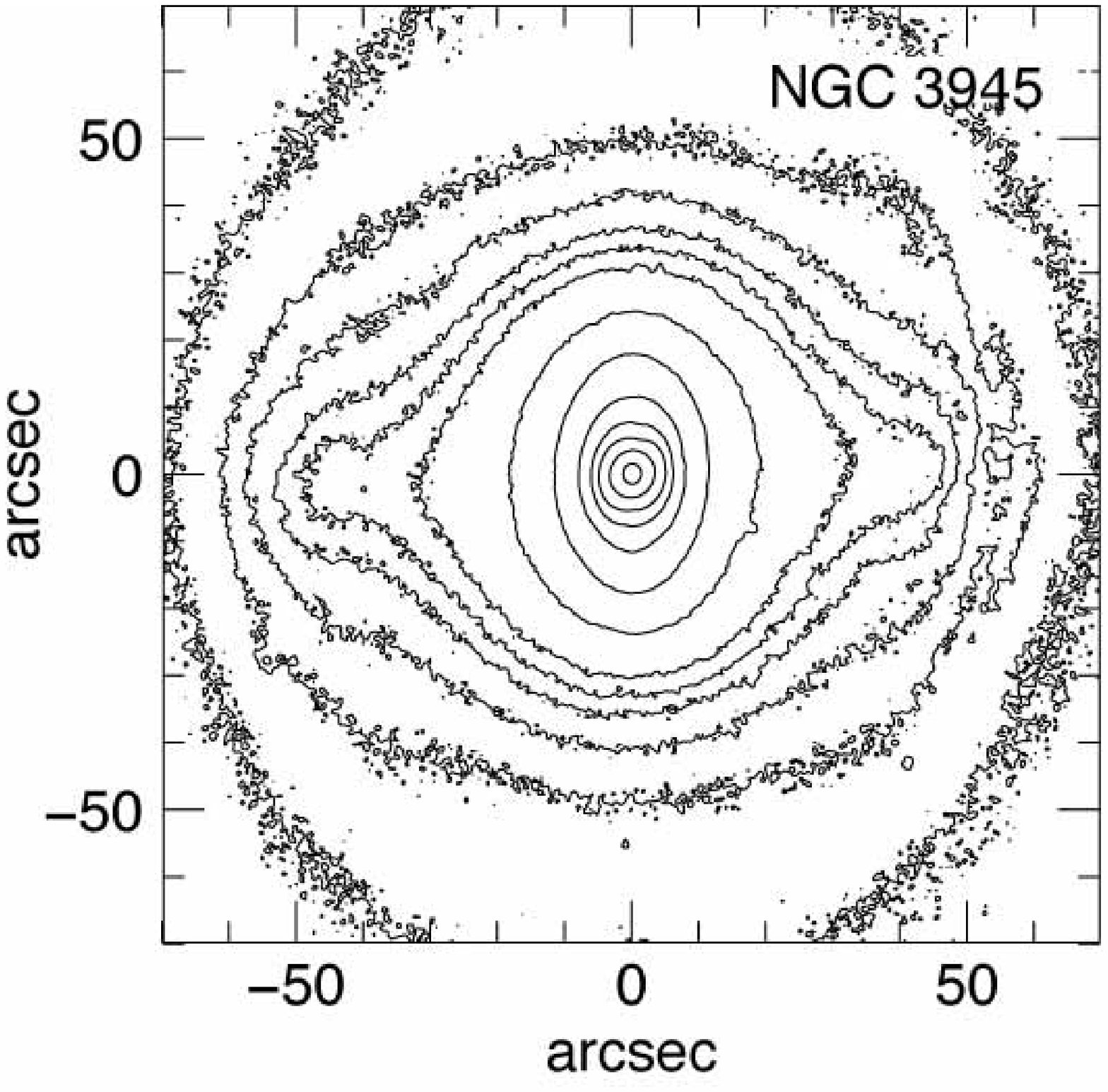}}

\col{\includegraphics{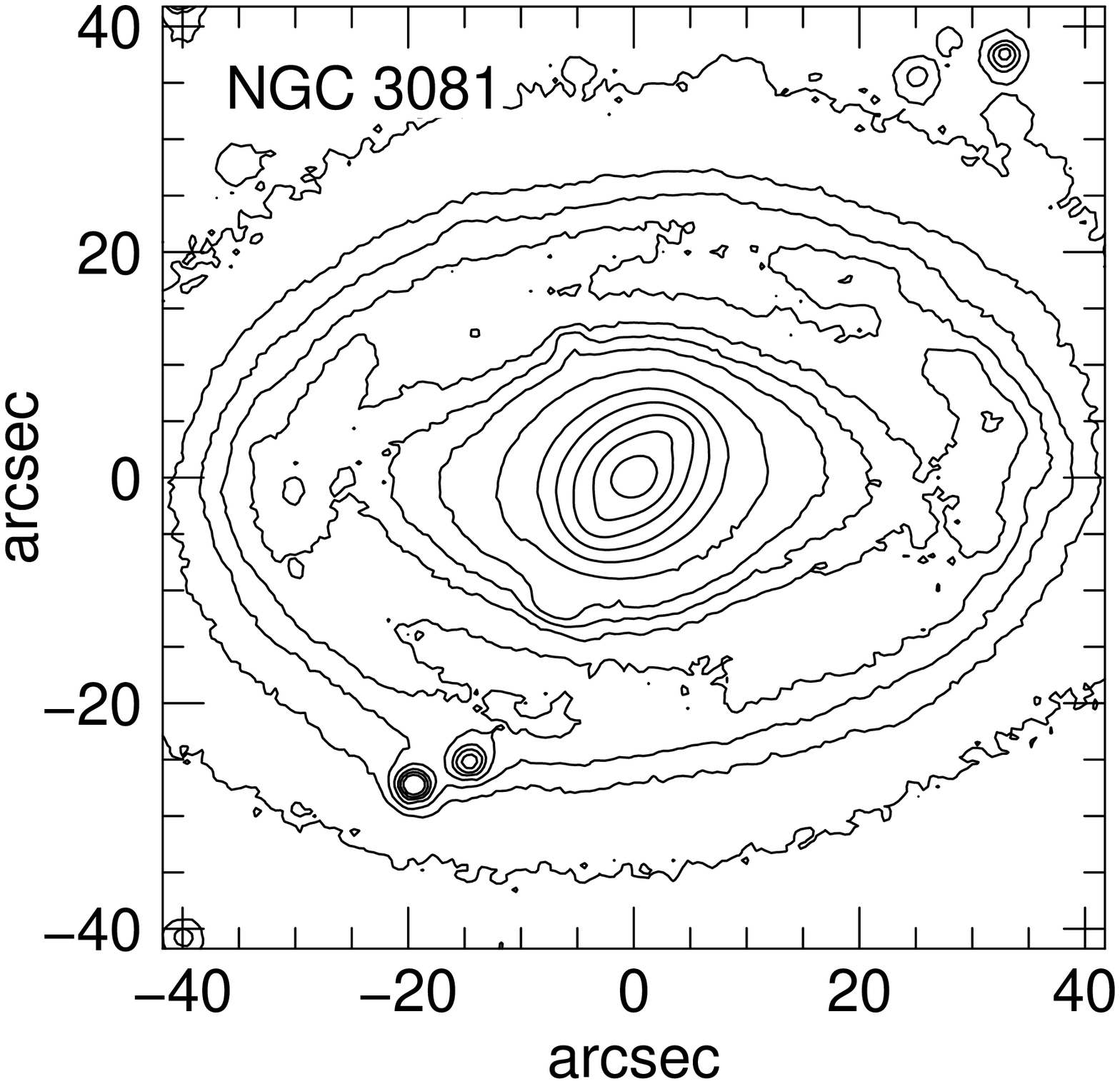}}

\col{\includegraphics{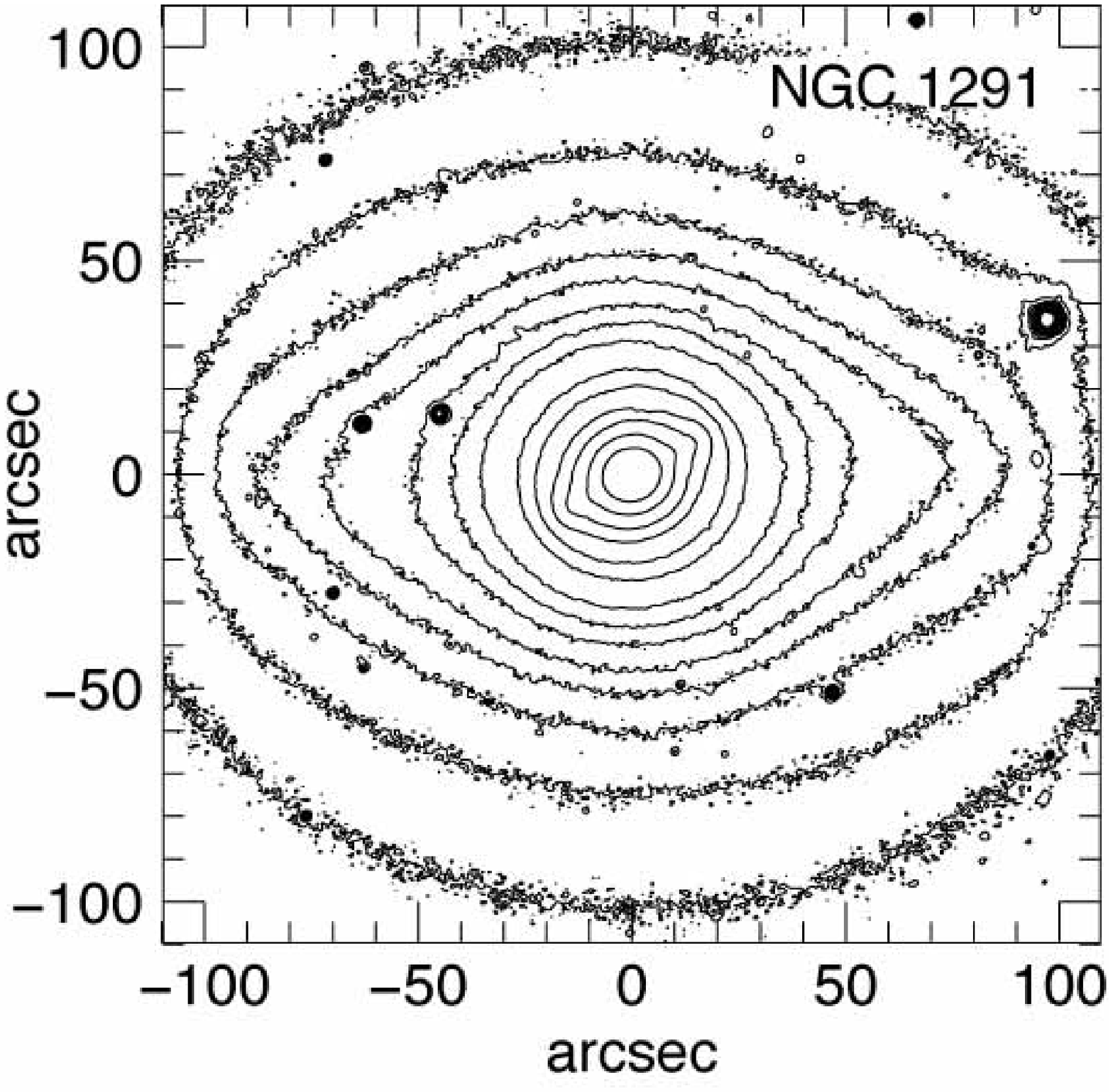}}

     {\it Figure 14}\quad Bars within bars.~Each galaxy image is
rotated so that the main bar is horizontal.  Contour levels are close 
together at large radii and widely spaced in the nuclear bars.  NGC 3081 and
NGC 1433 have inner rings.  NGC 1291 is also shown
in Figure 2, NGC 3081 and NGC 3945 in Figure 5, and NGC 6782 in Figure 8.
The images are courtesy Ron Buta.

\eject

\vsl\vsss
\ni {\bf 4.5.~Box-Shaped Bulges}
\vsl

     Bulges with box-shaped isophotes (Figure 15) are well known (Burbidge \&
Burbidge 1959; Sandage 1961; de Vaucouleurs 1974b).  Clear examples are
seen in at about one-fifth of edge-on galaxies (Jarvis 1986; Shaw 1987; de Souza
\& dos Anjos 1987; L\"utticke, Dettmar, \& Pohlen 2000a).  Numerical simulations
universally show that bars heat themselves in the vertical direction; they
suggest that box-shaped bulges are edge-on~bars.  If this is correct, then
observing box-shaped isophotes is a sufficient criterion for identifying a
pseudobulge.  Probably indepently of this, boxy bulges also present us with a
serious collision between simulations and observations. There are at least two
problems. (1) Observations imply that bars are flat in the edge-on galaxies in
which they can reliably be identified. (2) Bars and boxy bulges that are clearly
distinct from each other occur together in several galaxies. In these galaxies,
the major-axis radii of the boxy bulges are much shorter than the lengths of the
bars.  

      That bars heat themselves in the axial direction was an immediate result
of the first three-dimensional $n$-body simulations of unstable disks; it has
been a robust theoretical prediction ever since (see SW93~for~a~review). Combes
\& Sanders (1981) were the first to point out that $n$-body bars look
like boxy bulges (e.{\ts}g., NGC 7332) when seen end-on and like peanut-shaped 
bulges (e.{\ts}g., NGC 128) when seen side-on (both galaxies are illustrated in
Sandage 1961 and in Sandage \& Bedke 1994).  Edge-on $n$-body bars looked boxy
in some previous papers (e.{\ts}g., Miller \& Smith 1979), but these resulted
from the collapse of spherical stellar systems, so it was not clear
that their vertical structure was relevant to the evolution of disks.
The Combes \& Sanders (1981) results have been confirmed and extended by many
authors (e.{\ts}g., Combes et al.~1990; Pfenniger \& Friedli 1991; Berentzen
et al.~1998; Athanassoula \& Misiriotis 2002; Athanassoula 2003).  Early papers
concluded that the orbits that contribute most to the boxy structure are in
vertical ILR with the bar.  With two vertical oscillations for each revolution, 
it is easy to arrange that a star be at its maximum height above the disk plane
when it is near apocenter.  It then contributes naturally to a box-shaped
structure.~The importance of vertical resonant heating was emphasized by
Pfenniger (1984, 1985) and especially by Pfenniger \& Norman (1990).~From
``sticky particle'' simulations, Pfenniger \& Norman (1990) found both the mass
inflow discussed earlier and vertical heating that fed stars into a component
with the scale height of a bulge.  Timescales were short, on the order of
one-tenth of a Hubble time.

      In contrast, Raha et al.~(1991) showed that buckling instabilities thicken
bars in the axial direction.  These are collective phenomena, so they are
different from resonant heating.  Raha et al.~(1991) suggested that buckling
instabilities also occurred in the above simulations; Pfenniger \& Friedli
(1991) acknowledged this possibility.  Additional examples of buckling
instabilities are in Sellwood (1993b), Kalnajs (1996), and Griv \& Chiueh
(1998). Further discussion is provided by Toomre (1966); Merritt \& Sellwood
(1994); Pfenniger (1996a); and Merrifield (1996).  

      However the heating happens, all of the simulators agree that bars and
boxy bulges are connected. A few papers suggest only that disk stars are heated
vertically and fed into the bulge (pre-existing or not), giving it a box-shaped
appearance.  But most authors advocate a stronger conclusion, namely that boxy
bulges are nothing more nor less than bars seen edge-on.  

      What do the observations say?  Persuasive observations show that boxy
bulges occur in SB galaxies.  However, they also suggest that box bulges are 
not identical to edge-on bars.

      The obvious sanity check -- that boxy bulges are seen in edge-on
galaxies as frequently as well developed bars are seen in face-on galaxies -- 
is passed with flying colors.  References are in the first paragraph of this
subsection.

      A link between $n$-body bars and boxy bulges is the observation in
both of cylindrical rotation to substantial heights above the equatorial plane
(see
Bertola \& Capaccioli 1977;
Kormendy \& Illingworth 1982;
Jarvis 1990;
Shaw, Wilkinson, \& Carter 1993a;
Shaw 1993a;
Bettoni \& Galletta 1994;
Fisher, Illingworth, \& Franx 1994;
D'Onofrio et al.~1999; and
Falc\'on-Barroso et al.~2004 for the observations and
Combes et al.~1990;
Sellwood 1993a;
Athanassoula \& Misiriotis 2002 for simulations).
Classical bulges and ellipticals do not rotate cylindrically, as
evident from early long-slit spectroscopy (Illingworth \&  Schechter 
1982; Kormendy \& Illingworth 1982; Binney, Davies, \& Illingworth 1990) 
and now beautifully shown by integral-field spectroscopy (de Zeeuw
et al.~2002; Verolme et al.~2002, Bacon et al.~2002; Copin, Cretton, \& Emsellem
2004; Falc\'on-Barroso et al.~2004; Krajnovi\'c et al.~2004).

    Kuijken \& Merrifield (1995) and Merrifield (1996) suggest that a kinematic 
signature of edge-on bars is a splitting in the gas velocities just interior to
corotation because the gas there is depleted by radial transport.  They observe
such velocity splitting in NGC 5746 and NGC 5965 and argue that both galaxies
are barred.  Merrifield \& Kuijken (1999) and Bureau \& Freeman (1999) show 
additional examples.  NGC 5746 from the latter paper is shown in Figure 15. 
The ``figure 8'' pattern in the emission line is the bar signature. The rapidly
rotating gas is identified
with a nuclear disk of $x_2$ orbits, and the slowly rotating component shows the
line-of-sight velocities in the disk beyond the end of the bar.  The lobes of
the ``figure 8'' are empty because an annulus between the nuclear disk and the
end of the bar contains little gas.  The idea is that the missing gas has
been transported to the center or to an inner ring at the end of the bar.
This is an interpretation: an axisymmetric disk containing an annulus devoid
of gas would also show the ``figure 8''.  The connection~with~bars~is~indirect:
(1) in face-on galaxies, gasless annuli are seen only in
mature SB(r) galaxies, and
(2) [N{\ts}II] $\lambda$6584 \AA~emission is much stronger than H$\alpha$ in
the steep-rotation-curve central disk; this is a possible diagnostic of the
shocks expected in the inner parts of the bar (Bureau \& Freeman 1999).
On the other hand, we noted in \S\ts2.1 that mature SB(r) galaxies -- the ones
in which an annulus interior to the inner ring has been cleared of gas -- do 
not have the radial dust lanes that are characteristic of shocks.
Despite these uncertainties, the almost universal detection of figure-8-like
(or at least, X-shaped) line splitting in boxy bulges and -- equally important
-- the lack of such splitting in elliptical bulges argues that the former are
found in barred galaxies.

    A third observation that connects boxy bulges with bars is the detection 
in the disks of a few edge-on examples of density enhancements that plausibly
are inner rings (Aronica et al.~2003).

    Galaxy mergers probably create a minority of boxy bulges (Jarvis 1987).
Also, Patsis et al.~(2002) illustrate a simulation that makes a boxy-bulge-like
structure in the absence of a bar.  However, the conclusion that galaxies with
boxy bulges generally contain bars seems reasonably secure.  

    This is not a proof that they are the same things. There are two problems
with the simple, well motivated, and almost universally accepted notion that
boxy bulges are edge-on bars.

     First is the observation that at least some edge-on bars are flat.  The ``Rosetta stone'' object for this subject is NGC 4762.  It is studied in an
important paper by Wakamatsu \& Hamabe (1984) and is illustrated in Figure 16.

\cl{\null}

\vfill

%

%

\col{\includegraphics{n5746.ps}}

     {\it Figure 15}\quad (Top) NGC 5746 (Sb) has a prominently box-shaped bulge
(see also Sandage \& Bedke 1994).  (Bottom) Position-velocity diagram of the
[N\ts{II}] $\lambda$6584 \AA~emission line along the major axis registered in
position with the image.  The ``figure 8'' pattern is interpreted as the
signature of a barred galaxy by Bureau \& Freeman (1999, who kindly supplied
this figure) and by Kuijken \& Merrifield (1995).

\eject

\cl{\null}

\vfill

\col{\includegraphics{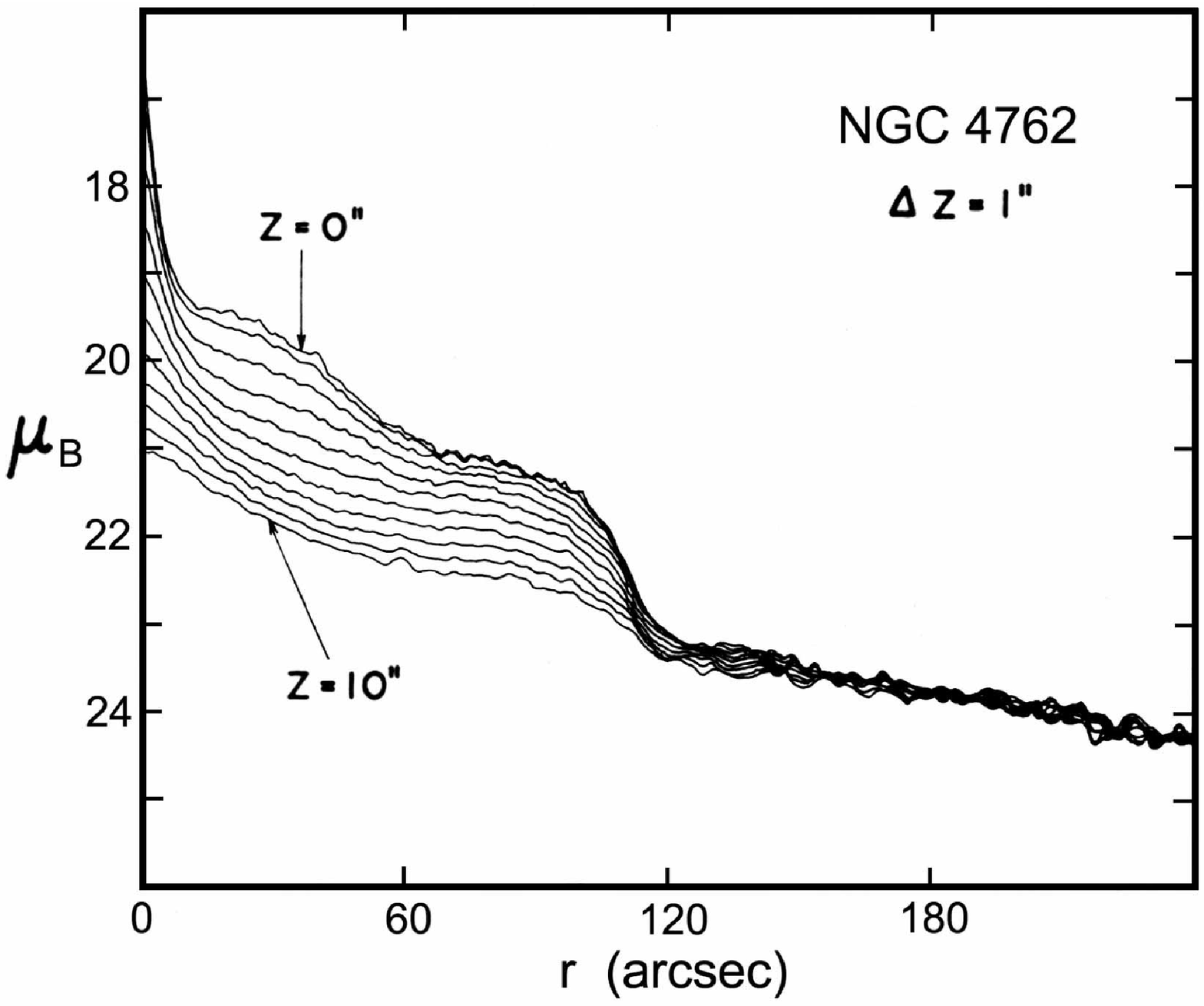}}

\col{\includegraphics{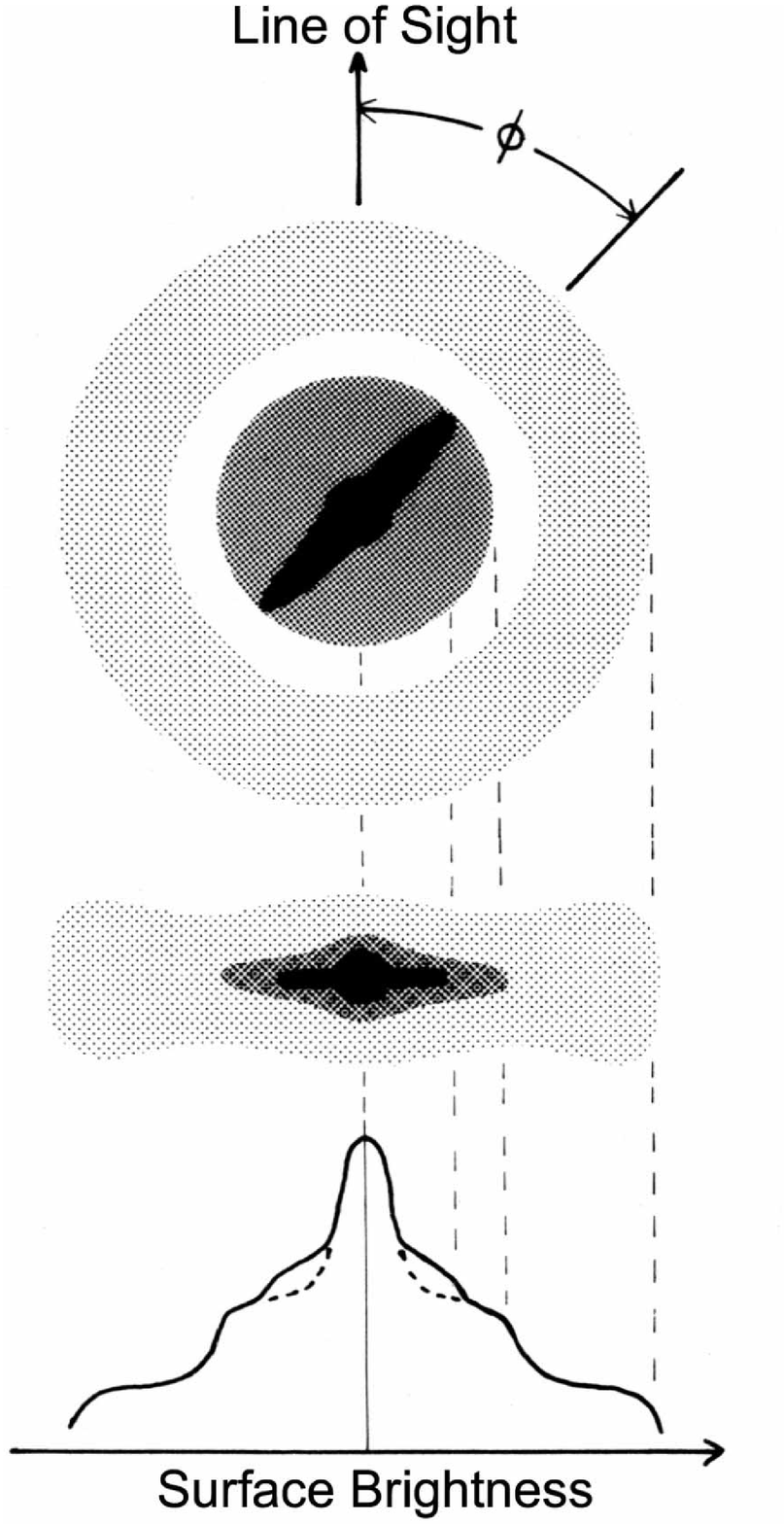}}

     {\it Figure 16}\quad From Wakamatsu \& Hamabe (1984), (top) brightness
cuts parallel to the major axis of NGC 4762 and displaced from it by 
${\Delta}{z}$ along the minor axis; (bottom) assumed viewing geometry: face-on
(upper diagram) and as seen by us (middle sketch and major-axis brightness cut).

\eject

      NGC 4762 is unique among edge-on galaxies studied so far because it has,
in addition to a bulge, three clearcut shelves in its major-axis brightness
distribution.  All three shelves are visible in the {\it Hubble Atlas} images
(Sandage 1961), which also show that the bulge is slightly boxy.  More face-on
galaxies show us that three shelves in the surface brightness profile are common
in early-type galaxies that contain a bar, a lens, and an outer ring (see NGC
1291 in Figure 2; NGC 3945 in Figure 5; NGC 2217 and NGC 2859 in the 
{\it Hubble Atlas}).  Lenses and outer rings have shallow brightness gradients
interior to a sharp outer edge; their two nested ovals are exactly analogous to
those in later-type oval galaxies (Figure~9).  Kormendy (1979b) emphasizes that
the bar almost always fills the lens in its longest dimension.  Since SB(lens)0 
galaxies are common and since they are the only S0s with three prominent shelves
in the brightness profile, interpreting NGC 4762 is reasonably straightforward.
Wakamatsu \& Hamabe (1984) suggest that the outer shelf is an outer ring, that
the middle shelf is a lens, and that the inner shelf is a bar.  Since the inner
shelf has a smaller radius than the middle shelf, the bar must be seen at a skew
orientation $\phi$ (Figure 16).  Wakamatsu and Hamabe point out that their 
interpretation is supported by four observations: (1) The deprojected profile of
the outer shelf is that of a ring: it has a minimum interior to an outer 
maximum.  (2) The radius of the outer shelf satisfies
the correlation between outer ring radii and galaxy luminosity; (3) the radius
of the inner shelf satisfies the correlation between lens radii and galaxy
luminosity; both correlations are from Kormendy (1979b).  (4) The ratio of
the radius of the outer shelf to the radius of the inner shelf is $2.4 \pm 0.2$,
consistent with the average ratio of outer ring to lens radii, $2.21 \pm 0.12$
(Kormendy 1979b; Buta \& Combes 1996 and references therein).

      We belabor these points because it is critically important to know that
the inner shelf is the bar.  The reason is illustrated in the top panel of
Figure 16.  Wakamatsu \& Hamabe (1984) show convincingly that the bar is flat.
In the series of brightness cuts parallel to the disk major axis and displaced
from it by $\Delta z = 0^{\prime\prime}$, $1^{\prime\prime}$, 
$2^{\prime\prime}$, \dots~$10^{\prime\prime}$, the bar disappears as a feature
distinct from the lens by $\Delta z \simeq 5^{\prime\prime}$.  That is, its
scale height is less than that of the lens and much less than that of the bulge.
The bar is the flattest component in the galaxy.

      Also, the bar and the bulge are photometrically distinct.~The boxy 
outer part of the bulge (which is not evident in the brightness cuts in
Figure 16) has a radius about half as big as the projected radius of the bar.
If the bar fills the lens, then this is about one-fifth of the true radius of the bar.

      Similar evidence for flat bars is presented in de Carvalho \& da Costa
(1987); L\"utticke, Dettmar, \&  Pohlen (2000b); and Quillen et al.~(1997).

      The second problem with the assumption that boxy bulges are edge-on bars
is the observation that both occur together but are distinct from each other
in NGC 7582 (Quillen et al.~1997).  We see this galaxy at an inclination
$i \simeq 65^\circ$ that is close enough to edge-on so that the boxy bulge is
visible in the infrared but
far enough from edge-on so that the bar can be recognized (Sandage \&
Bedke 1994).  In fact, the galaxy has the morphology of a typical oval disk
with the bar filling the inner oval along its apparent major axis.  Therefore
the bar is seen essentially side-on.  However, the bar is very flat, the boxy
bulge is clearly distinct from it, and the maximum radius of the boxy structure 
along the disk major axis is about one-third of the radius of the bar.  



      These observations suggest that boxy bulges and edge-on bars are
not exactly equivalent.  Interestingly:

   Observations and theory are consistent with the hypothesis that at least some
and possibly most box-shaped bulges are edge-on {\it nuclear} bars.  E.{\ts}g.,
the two nested triaxial components in our Galaxy proposed by Blitz \& Spergel
(1991, their Figure 1) are similar to the bar-within-bar structure in Section
4.4.  If the inner bar has a radius of 1\ts\--\ts2 kpc (Binney et al.~1991;
Blitz \&
Spergel 1991; Binney \& Gerhard 1993; Sellwood 1993b; cf.~Dwek et al.~1995), 
then it is more nearly the length of typical nuclear bars than of typical main
bars.  (Scaling our Galaxy to other Sbcs, a normal bar should be $\sim 3.5$ kpc
in radius.)  It is the inner bar that looks boxy in {\it COBE} images (Weiland
et al.~1994; Dwek et al.~1995).  We may live in a weakly barred or oval galaxy
with a boxy nuclear bar.  However, only one-quarter of strongly barred galaxies
contain nuclear bars.  There may be too few of them to account for all boxy
bulges.

      Another solution may be the indication in Figure 1.1{\ts}(b) of Shen \&
Sellwood (2004) that the boxy part of their $n$-body bar is smaller than the bar
as a whole.  We are indebted to Jerry Sellwood for pointing this out.
Athanassoula (private communication) emphasizes the same point.

The safest conclusion -- and one sufficient for our purposes -- is that boxy
bulges are connected with bars and owe their origin to them.  All mechanisms
under discussion build the box structure out of disk material.  We therefore
conclude that detection of boxy bulge isophotes is sufficient for the
identification of a pseudobulge.  However, the disagreement between the bar
simulations and the above observations needs attention.


\vfill\eject

\vsl\vsss
\ni {\bf 4.6.~``Bulges'' With The Dynamics of Disks: 
                The V/$\sigma$ -- $\epsilon$ Diagram}
\vsss

    Figure 17, the $V_{\rm max}/\sigma$ -- $\epsilon$ diagram (Illingworth
1977; Binney 1978a,{\ts}b), shows that pseudobulges (filled symbols) are more
rotation-dominated than classical bulges (open circles), which are more
rotation-dominated than giant elliptical galaxies (crosses).  This is disky
behavior.

\vfill


\col{\includegraphics{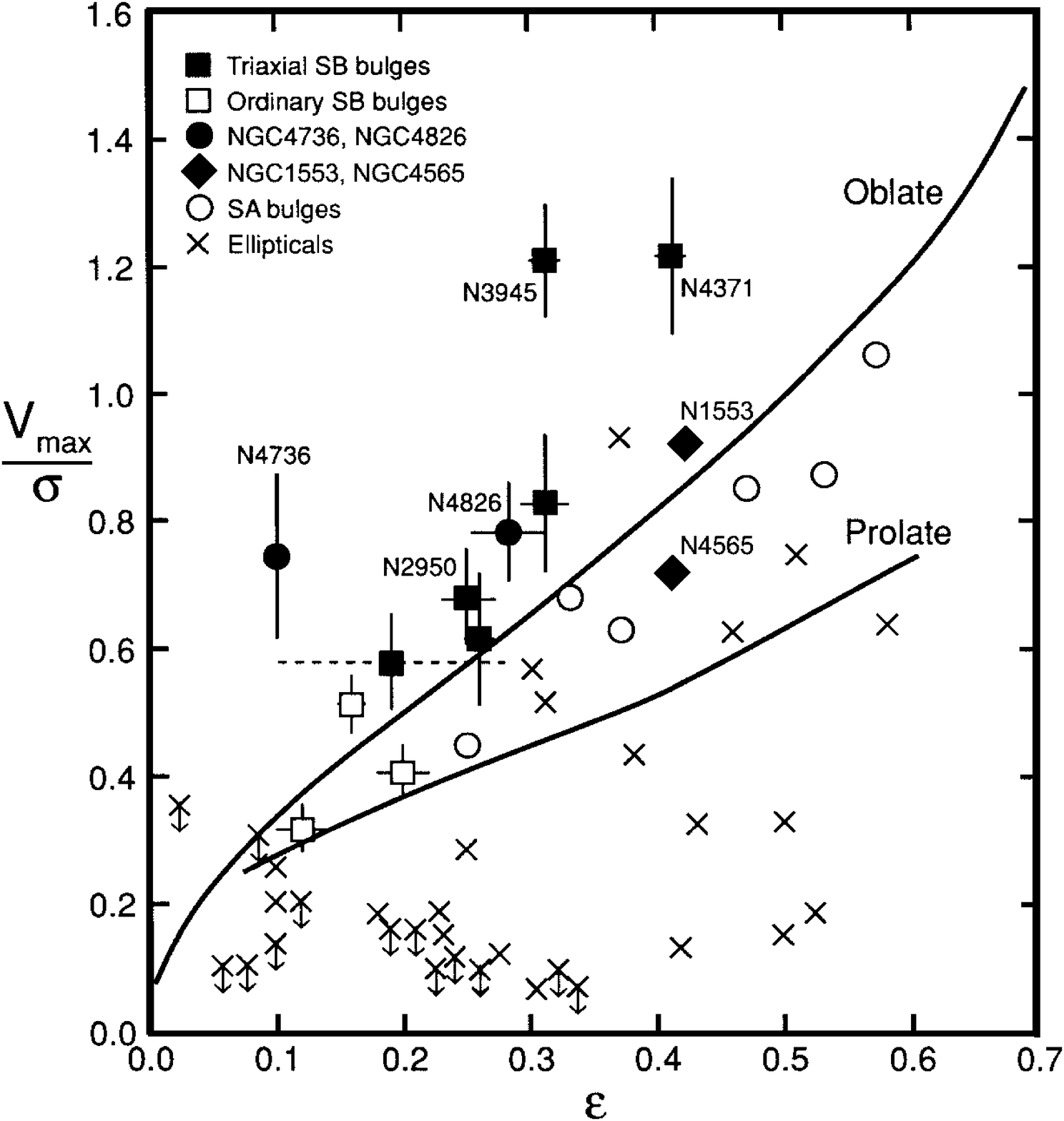}}

     {\it Figure 17}\quad The relative dynamical importance of rotation and
random motions as a function of observed ellipticity for various kinds of
stellar systems.  Here $V_{\rm max}/\sigma$ is the ratio of maximum rotation
velocity to mean velocity dispersion interior to the half-light radius and
$\epsilon$ = 1 $-$ axial ratio. The ``oblate'' line describes oblate-spheroidal
systems that have isotropic velocity dispersions and that are flattened only by
rotation; it is a consequence of the tensor virial theorem (Binney \& Tremaine
1987).  The ``prolate'' line is one example of how prolate spheroids can rotate
more slowly for a given $\epsilon$ because they are flattened partly by velocity
dispersion anisotropy.  This figure is updated from Kormendy (1993).

\eject

      The essential features of the  $V_{\rm max}/\sigma$ -- $\epsilon$ diagram
are as follows:

\vs
\nnhi 1.~The virial theorem relates the gravitational potential and kinetic
energy tensors; the former involves the shape of the stellar system; the
latter involves the balance between rotational and random
kinetic energies (Binney 1978a; Binney \& Tremaine 1987).  In Figure
17, $V_{\rm max}/\sigma$ is a surrogate for the (square root of) the ratio of
ordered to random kinetic energies and $\epsilon$ is the apparent flattening.

\vs
\nnhi 2.~If rotation is dynamically unimportant ($V_{\rm max}/\sigma \ll 1$) and
if the system is flattened, then it must be anisotropic (Binney 1976; 1878a, b;
1980; 1982).  Stars climb farthest out of their mutual gravitational potential
well in the direction in which the velocity dispersion is largest.

\vs
\nnhi 3.~Rotation adds extra flattening regardless of velocity anisotropy,
because rotation plus random motions allow stars to climb farther out of their
mutual gravitational potential well than do random motions alone.  Isotropic
systems that are flattened into spheroids by rotation have a simple relationship between flattening and $V_{\rm max}/\sigma$ that is shown by the ``oblate'' line
in Figure 17.  Binney (1978a) gives it implicitly; Fall (1981) provides an
explicit equation for the projected configurataion seen edge-on, and Kormend
(1982a) gives an approximation formula, 
$V_{\rm max}/\sigma \simeq [\epsilon/(1 -\epsilon)]^{1/2}$, 
that is good to 1\ts\% for $0 \leq \epsilon \leq 0.95$.

\vs
\nnhi 4.~For 0.1\ts\lapprox\ts$\epsilon$\ts\lapprox\ts0.5, projection moves an
isotropic oblate spheroid almost parallel to the oblate line.  If an isotropic
spheroid with $\epsilon = 0.5$ is seen at a skew orientation so that it looks
like an $\epsilon = 0.3$ system seen edge-on, then both sustems have
approximately the same value of $V_{\rm max}/\sigma$.  However, an edge-on disk
that is near the oblate line at $\epsilon \sim 0.9 \pm 0.1$ projects well above
the oblate line when it is seen other than edge-on.

\vs
\nnhi 5.~Observations show that most giant ellipticals have insignificant 
rotation and are dominated by velocity anisotropy (Bertola \& Capaccioli 1975;
Illingworth 1977; crosses in Figure 17).  Low-luminosity ellipticals are
more nearly isotropic and consistent with the oblate line (Davies et al.~1983).
And classical bulges are consistent with being isotropic oblate rotators
(Illingworth \& Schechter 1982; Kormendy \& Illingworth 1982; Kormendy 1982b;
open circles in Figure 17).  Classical bulges fall slightly to the right of
the oblate line, but the extra flattening is provided by the disk potential
(Jarvis \& Freeman 1985).   

\vs

      The above papers show that anisotropic giant ellipticals are triaxial.  
We emphasize that this triaxiality is different from that
of bars.  Ellipticals are triaxial because they have little angular momentum.
They are made largely out of ``box orbits'' that have no net angular momentum
(see Binney \& Tremaine 1987 for discussion).  Rotation is provided by
``$z$-axis tube orbits'' that encircle the $z$ = rotation axis; in a triaxial
elliptical, these are somewhat elongated in the direction of the longest axis.
Other orbits, including chaotic ones, are present as well.  But the essential
character of an elliptical is defined by its box orbits.

      In contrast, a barred galaxy is barred not because it has little angular
momentum but rather because it has too much  for the combination of
its velocity dispersion and its central concentration.  This is why the disk
made a bar.  Bars are not made of box orbits; they are made of $x_1$ orbits.
These are very elongated $z$-axis tubes that, in some cases, include baroque
decorations such as loops.  They have lots of angular momentum.  
\omit{
In Figure 17, bars have their highest  $V_{\rm max}/\sigma$ values -- 
ones that are well above the oblate line -- when seen nearly end-on.  Then the
shape that we see is not very elongated, but the component of rotation that is
along the line of sight is larger than the local circular velocity because we
see stars near the pericenters of very elongated orbits.  In contrast, bars
generally appear below the oblate line when they are seen side-on; now the
apparent flattening is maximum but the rotation component along the line of
sight is smaller than the local circular velocity.  Examples of both behaviors
are shown in Figure 17.
}
It is important to keep in mind the distinction between bars and ellipticals.
They are not different versions of each other, and they virtually never occur
together.  
\omit{
To our knowledge, they never occur together\footnote{$^5$}{\kern -3pt An 
extreme statement such as this
one invites the reader to find an exception.  In the context of what we know 
about galaxy formation, two kinds of exception can be imagined.  If a 
nonrotating, triaxial elliptical accretes enough gas to build a self-gravitating
disk near the center -- that is, one whose internal dynamics are controlled by
nuclear disk stars and not by the elliptical -- then such a disk could be
bar-unstable.  Alternatively, if a merger that involves two bulges or
ellipticals happens with just the right geometry to maximize the angular
momentum of stars near the center, then one might imagine a rather hot but
nevertheless rapidly rotating and triaxial center in the merger remnant.  This
could be qualitatively like a bar without the presence of a disk.  Only a few
objects with this morphology are known; possible examples include NGC 2699 and
NGC 4648 in Rest et al.~(2001).  Of course, we do not know how these objects
formed.  There is even a chance that they are misclassified SB0 galaxies.
Neither invalidates our basic point, which is that bars and ellipticals are
fundamentally different and that bars are very generally a disk phenomemon.}.
}
A bar is fundamentally a disk phenomenon.

       Contrast the behavior of the bulges that are plotted in Figure~17
as filled symbols.  They are above the oblate line and even more above the
distribution of classical bulges (open symbols).  Rotation is more
important in these objects than it is in classical bulges and ellipticals.  
Point 4, above, shows why this is disk-like behavior.  It indicates an 
admixture of stars that are flattened, dynamically cold, and rapidly revolving
around the galactic center -- that is, a disk contribution that would appear
near the oblate line if seen edge-on but that plots well above the oblate line
at the skew inclinations of these galaxies.  The filled symbols include barred
galaxies and the prototypical unbarred oval galaxy NGC 4736 (Figures 2 and 8). 
Another prominent example is NGC 3945; its rapidly rotating ``bulge'' is the
nuclear bar shown in Figure 14 (see also Erwin et al.~2003 for a detailed
discussion).  Another is NGC 2950, also a nuclear bar.  Thus the dynamical
evidence agrees with other evidence that these are pseudobulges.

      Two pseudobulges from Kormendy \& Illingworth (1982) deserve comment
(filled diamonds in Figure 17).  NGC 1553 contains a
prototypical lens in an unbarred galaxy (point 7
of Section 2.1).  Figure 17 shows that it has an unusually high value of
$(V_{\rm max}/\sigma)^*$ for an unbarred galaxy.  Consistent with the suggestion
 that the lens is a defunct bar (Kormendy 1979b, 1984) (Section 3.3),
the hint is that the galaxy grew a pseudobulge while it was still barred.
In contrast, the boxy bulge of NGC 4565 has $(V_{\rm max}/\sigma)^* = 
0.86 \pm 0.16$.  This is smaller than 
$(V_{\rm max}/\sigma)^*$ values for other pseudobulges.  However,
a box-shaped pseudobulge rotates cylindrically, so $(V_{\rm max}/\sigma)^*$
underestimates the dynamical importance of rotational kinetic energy
compared with ellipsoidal bulges.

      The difference between classical and pseudobulges need not always be
large.  It is entirely implausible
that secular evolution sometimes augments a classical bulge with new, disky
material.  We also point out in Section 7.1 that pseudobulges can
heat themselves in the vertical direction, thereby decreasing their dynamical
difference from classical bulges.  Therefore, large $(V_{\rm max}/\sigma)$ is
evidence for a pseudobulge, but values comparable to those on the oblate line
do not guarantee that the bulge is classical.
      
\vfill\eject

\vsl\vsss
\ni {\bf 4.7.~Velocity Dispersions and the Faber-Jackson Relation}
\vsss

    Figure 18 correlates velocity dispersion and absolute magnitude for
elliptical galaxies and for bulges of disk galaxies.  Most early-type galaxy
bulges are consistent with the well known correlation that bulge luminosity 
$L_B \propto \sigma^4$ (Faber \& Jackson 1976; see Minkowski 1962 for an early
hint).  But a few early-type bulges and a large majority of Sbc -- Sm
``bulges'' fall well below the correlation.  One possible interpretation is that
the bulges of late-type galaxies are low in velocity dispersion.  This is disky
behavior.  Alternatively, the late-type ``bulges'' may lie to the right of the
early-type systems because they are actively forming stars and therefore have
small mass-to-light ratios.  Figure 18 then implies that most late-type bulges
have young stellar population ages.  This means that the star formation must be
secular, not episodic.  In either case, Figure 18 suggests that most late-type
galaxies contain pseudobulges.  In contrast, most early-type bulges are similar
to elliptical galaxies.

\vfill


\col{\includegraphics{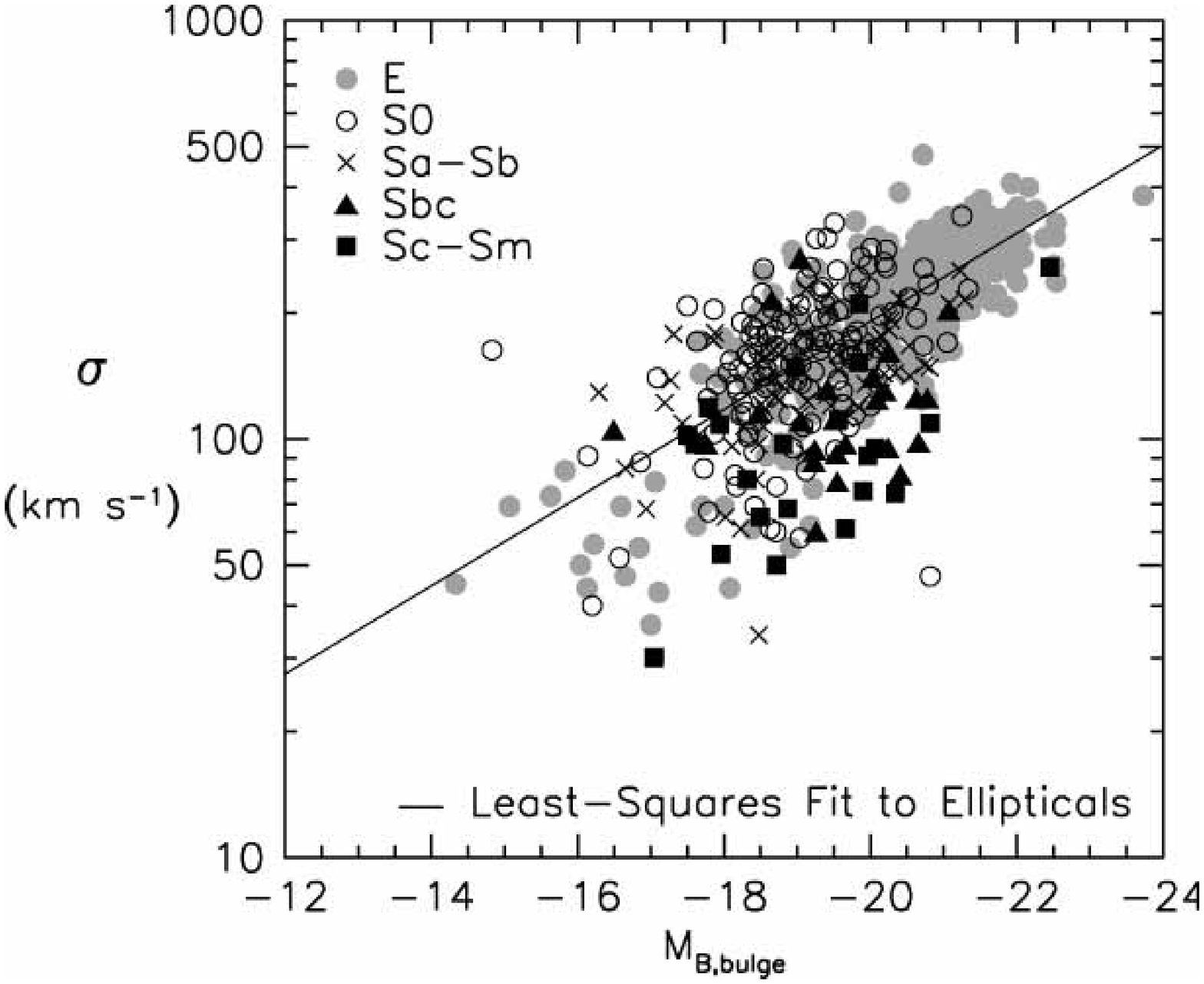}}

     {\it Figure 18}\quad Correlation between central velocity dispersion and
bulge absolute magnitude for all galaxies of the indicated Hubble types that
have velocity dispersions tabulated in Hypercat.  The straight line is a
least-squares fit to the ellipticals.  Updated from Kormendy \& Illingworth
(1983), this figure is from Kormendy \& Cornell (2004).

\eject

\vsl\vsss
\ni {\bf 4.8.~Pseudobulges and the Fundamental Plane Correlations}
\vsl

      If pseudobulges have S\'ersic indices $n$ $\sim$ 1 that are smaller
than~those of elliptical galaxies ($n$ $\sim$ 4), then this signals a breakdown
in the homology that, together with the Virial theorem, is the reason why 
classical bulges and ellipticals satisfy the Fundamental Plane (FP) correlations
(e.{\ts}g., Djorgovski \& Davis 1987; Dressler \etal 1987; Faber \etal 1987;
Lauer 1987; Djorgovski, de Carvalho, \& Han 1988; Kormendy \& Djorgovski 1989;
Bender, Burstein, \& Faber 1992, 1993; Burstein et al.~1997).  That is, bulges 
and ellipticals lie in an inclined plane,
\hbox{$ R \propto \sigma^{1.4 \pm 0.15} ~ I^{-0.9 \pm 0.1}$}
in the space of observed radius $R$, surface brightess $I$, and velocity
dispersion~$\sigma$.  The scatter of ellipticals around the FP
is small (see the above papers and Saglia, Bender, \& Dressler 1993). Therefore
deviations from the FP are a sensitive test of whether the structure of
a \hbox{bulge-like} component is or is not similar to that of an elliptical.  
Carollo (1999) finds that pseudobulges deviate from the FP of classical bulges
and elliptical galaxies in the direction of having lower densities (see also
Kent 1985; Andredakis, Peletier, \& Balcells 1995).  They lie
closer to the locus of disks than to that of hot stellar systems.  Similar
studies of larger samples may provide an additional, quantitative way
to recognize~pseudobulges.

\vsl\vsss
\ni {\bf 4.9.~Nuclei}
\vsl

      Nuclei are compact star clusters\footnote{$^6$}{\kern -3pt AGN light is
expressly excluded.  Caution:~Not all authors do this.} at galactic centers 
(see Kormendy~\& Djorgovski 1989 for a review).  They should not be confused
with steep density cusps that are the central parts of nearly featureless 
\hbox{power-law} profiles (Lauer et al.~1995).  E.{\ts}g., M{\ts}32 does not
have a nucleus (Lauer et al. 1992, 1998).  Rather, nuclei are clearly
differentiated from the surrounding (pseudo)bulge and disk in the sense that
they have much smaller effective radii $r_e$ and much higher effective surface
brightnesses $\mu(r_e)$ than their surroundings.  Figure 19 illustrates the
prototypical example in M{\ts}33 (Kormendy \& McClure 1993).  Tremaine \& 
Ostriker (1982) show that the nucleus and bulge of M{\ts}31 are dynamically 
independent; we assume the same for other nuclei.  Local Group examples occur in
M{\ts}31 (e.{\ts}g., Lauer et al.~1993, 1998), M{\ts}33, and NGC 205 (Jones et 
al.~1996).  Not much farther away are nuclei in IC 342 (B\"oker et al.~1997,
1999) and in NGC 7793 (D\'\i az et al.~2002).  The literature on nuclei is
extensive; we review only the part of it that is relevant to this paper.  

      Late-type disk galaxies usually contain nuclei.  Many of these have young
stellar populations.  They imply episodic star formation over long periods of
time and so are consistent with secular growth in a manner that is similar to
the proposed formation of pseudobulges.  But the ``smoking gun'' that is most
compelling is not the one that points at secular evolution.  Rather, the
observations are screaming that there is another physical process taking place
that we do not understand.
Because we will see that nuclei do not appear to be the low-luminosity limit of
pseudobulges.  Also, nuclei and pseudobulges often occur in the same galaxy. 
If they form similarly, why are they so different?  Our review highlights a
remarkable list of enigmas.

      Surveys of late-type galaxies and detailed studies of individual objects
provide the following list of properties: 

\vs
\nnhi 1.~Nuclei very common in late-type spirals.  B\"oker et al.~(2002)
find them in 75\ts\% of 77 Scd\ts--\ts{Sm} galaxies in an $I$-band {\it HST\/}
survey.  Carollo et al.~(2002) find nuclei in 30\ts\% of S0\ts--\ts{Sa}
galaxies, 59\ts\% of Sab\ts--\ts{Sb} galaxies, and 77\ts\% of Sbc\ts--\ts{Sm}
galaxies.  Nuclei are also common in spheroidal galaxies (Binggeli et al.~1984,
1985; van den Bergh 1986).

\vs
\nnhi 2.~Nuclei are rare in irregulars (van den Bergh 1995);~this is not
         understood.

\vs
\nnhi 3.~Nuclei are fairly homogeneous in their properties. Typical luminosities
are $10^6$ to $10^7$ $L_\odot$ 
(Kormendy \& McClure 1993;
Matthews et al.~1999a;
Carollo et al.~2001;
Matthews \& Gallagher 2002;
B\"oker et al.~2002).
Effective radii are typically $10^{1 \pm 0.5}$ pc (Carollo et al. 1999; B\"oker et al.~2003b).
Observed central densities are high and limited by the spatial resolution of
the images: $10^4$ to $10^5$ $L_\odot$ pc$^{-3}$ in examples in Matthews et al.~(1999a) and at least and $10^{7}$ $L_\odot$ pc$^{-3}$
in M{\ts}33 (Lauer et al.~1998).  These values are enormously higher than the
surrounding disk densities.

\vs
\nnhi 4.~Most nuclei are at the centers of their galaxies to within measurement
         errors (B\"oker et al.~2002).  Exceptions are rare (Matthews et
         al.~1999a; Binggeli, Barazza, \& Jerjen
         2000; Carollo et al.~2002).  This is hard to understand.  In their
         absence, the center does not look like a special place.  The
         gravitational potential gradient of the visible matter is shallow
         (Kormendy \& McClure 1993; Matthews et al.~1999a; B\"oker et al. 2002).
         Why does a nucleus form at the center and why is its scale
         length so short compared to that of the rest of the galaxy
         (Figure 19; B\"oker, Stanek, \& van der Marel 2003)?
         Could the reason be that cold dark matter halos are cuspier than the
         baryons (Navarro, Frenk, \& White 1996, 1997), or, if they are not so
         now (Moore 1994), could they have been so in the past, before baryonic
         physics intervened (Navarro, Eke, \& Frenk 1996)?  Could nuclei be 
         compact not because the galactic center is a special place
         but rather because galaxies know how to make compact clusters
         and they can sink to the center by dynamical friction (Tremaine, 
         Ostriker, \& Spitzer 1975)?  Carollo (1999) argues that the timescale
         for dynamical friction against dark matter is interestingly short.
         This may explain the ``seeds'' of the observed nuclei, but it remains
         remarkable that subsequent star formation keeps them so compact.

\vfill\eject

\vs
\nnhi 5.~Stellar populations often imply young ages for the stars that
         contribute most of the light.  The M{\ts}33 nucleus is typical. 
         It has a composite, late-A to early-F spectrum dominated by younger
         stars at bluer wavelengths (e.{\ts}g., van den Bergh 1976a, 1991;
         Gallagher, Goad, \& Mould 1982; O'Connell 1983; Schmidt, Bica, \&
         Alloin 1990; Gordon et al.~1999).  Population synthesis by Long,
         Charles, \& Dubus (2002) gives a best fit to the spectrum between
         1150\ts\AA~and 5700\ts\AA~for two starbursts, one with a mass of 9000
         $M_\odot$, 40 Myr ago and the other with a mass of 76,000 $M_\odot$,
         1 Gyr ago.  (The total nuclear mass is $2 \times 10^6$ $M_\odot$;
         Kormendy \& McClure 1993). The spectra are insensitive
         to still older starbursts. 

\vs
\hhhi Additional examples of nuclei with blue colors indicative of young stars
      are discussed in 
      D\'\i az et al.~(1982);
      Bica, Alloin, \& Schmidt (1990);
      Shields \& Filippenko (1992);
      B\"oker et al.~(1997, 1999)
      Matthews et al.~(1999a); 
      Davidge \& Courteau (2002);
      B\"oker et al.~(2001, 2003b);
      see also Ho, Filippenko, \& Sargent (1997).
      Carollo et al.~(2001) observe colors that are consistent with a range of
      ages; there is some tendency for bluer nuclei to be associated with bluer
      surrounding disks.  In fact, ``brighter nuclei ($M_V$ \lapprox \ts$-12$)
      are typically found \dots~in the centers of galaxies with circumnuclear 
      rings/arms of star formation or dust and an active, i.{\ts}e., H\ts{II} 
      or AGN-type \dots~spectrum'' (Carollo 1999).  Since many nuclei contain
      young stars, star formation does not happen over only a small
      fraction of the life of the cluster but rather is secular.  

\vs
\nnhi 6.~Nuclei are not more common in barred galaxies (Carollo et al.~2002;
         B\"oker et al.~2003b).  Evidently, supplying gas to feed their star
         formation does not require a bar.

\vs
\nnhi 7.~In the Fundamental Plane parameter correlations, nuclei are more 
         similar to large Galactic globular clusters and to compact young
         clusters in interacting and merging galaxies than they are to
         (pseudo)bulges (Carollo 1999; Geha, Guhathakurta, \& van der Marel
         2002; B\"oker et~al. 2003b).  There is no sign that nuclei form the
         faint end of the sequence of (pseudo)bulge properties (see also
         \S\ts4.10). \vs

\vs
\nnhi 8.~The luminosities of nuclei correlate with the luminosities and
         central surface brighnesses of their host galaxies (B\"oker et
         al.~2003b). \vs

      What does all this mean?  Point (5) -- the prevalence of young
      stars in nuclei and their correlation with the surrounding star
formation -- is the strongest evidence that nuclei are built by secular
processes like those that we suggest make pseudobulges.  Point (8) also
seems consistent.  So is observation (1) that nuclei are more common in
later-type galaxies; they are approximately as common as pseudobulges.
However, observations (2), (3), (4), (6), and (7) either are major
puzzles or suggest that nuclei and pseudobulges are fundamentally different.  

      The prudent conclusions at present are these:~Nuclei are not a problem
for our picture of pseudobulge formation by secular inward transport of gas.
In fact, many authors quoted above -- including one author of this paper -- 
have argued that this is how they grow.  But nuclei are not a secure argument
for secular evolution, either.  We find it compelling that nuclei and
pseudobulges are very different in their parameters but occur together in the
same galaxies.  Nuclei appear to be related to globular clusters and to young
clusters in merger starbursts.  {\it Several mysteries would be easier to
understand if they got their start as such clusters and then sank to the center
by dynamical friction.}  In particular, our problem with observation (3, 4) 
that nuclei are tiny and dense compared to pseudobulges and disks would 
vanish.  This does not mean that we understand how star-forming clusters get 
so compact.  But galaxies clearly know how to make them.  Also, they are common
in late-type disks, so it is easy to engineer a correlation with pseudobulges.
Further work is needed to clarify how much nuclei then grow by secular 
evolution. 

\vfill\eject

\vsl\vsss
\ni {\bf 4.10.~In Which Pseudobulges Fade Out Into Disks}
\vsl

      B\"oker, Stanek, \& van der Marel (2003) investigate how pseudobulges fade
out into disks.  Their most subtle examples of pseudobulges also highlight
fundamental uncertainties about the meaning of profile decomposition.  
M{\ts}33 (Figure 19) provides an example.  The disk has somewhat 
irregular but global-pattern spiral structure (Sandage 1961; Sandage \&
Bedke 1994).  There is no sign of an ILR, i.{\ts}e., the spiral arms become
radial near the center and pass through it.  
At the center, a very distinct nucleus (Kormendy \& McClure 1993) is
representative of the ones discussed in \S\ts4.9.  Figure 19 shows that
the disk surface brightness profile has a subtly two-component look; it is
well fitted by the sum of two exponentials.  The figure shows a decomposition
into a S\'ersic function plus an outer exponential; it confirms that $n = 1.09
\pm 0.18$ for the inner component.   This inner component is essentially the
``bulge'' discussed by Minniti, Olszewski, \& Rieke (1993).  It is also visible
as a subtle upturn in surface brightness at $r < 150^{\prime\prime}$ in the
$JHK$ profiles posted at the 2MASS web site.  What does this mean?  Does
M{\ts}33 contain a pseudobulge?

\vfill


\col{\includegraphics{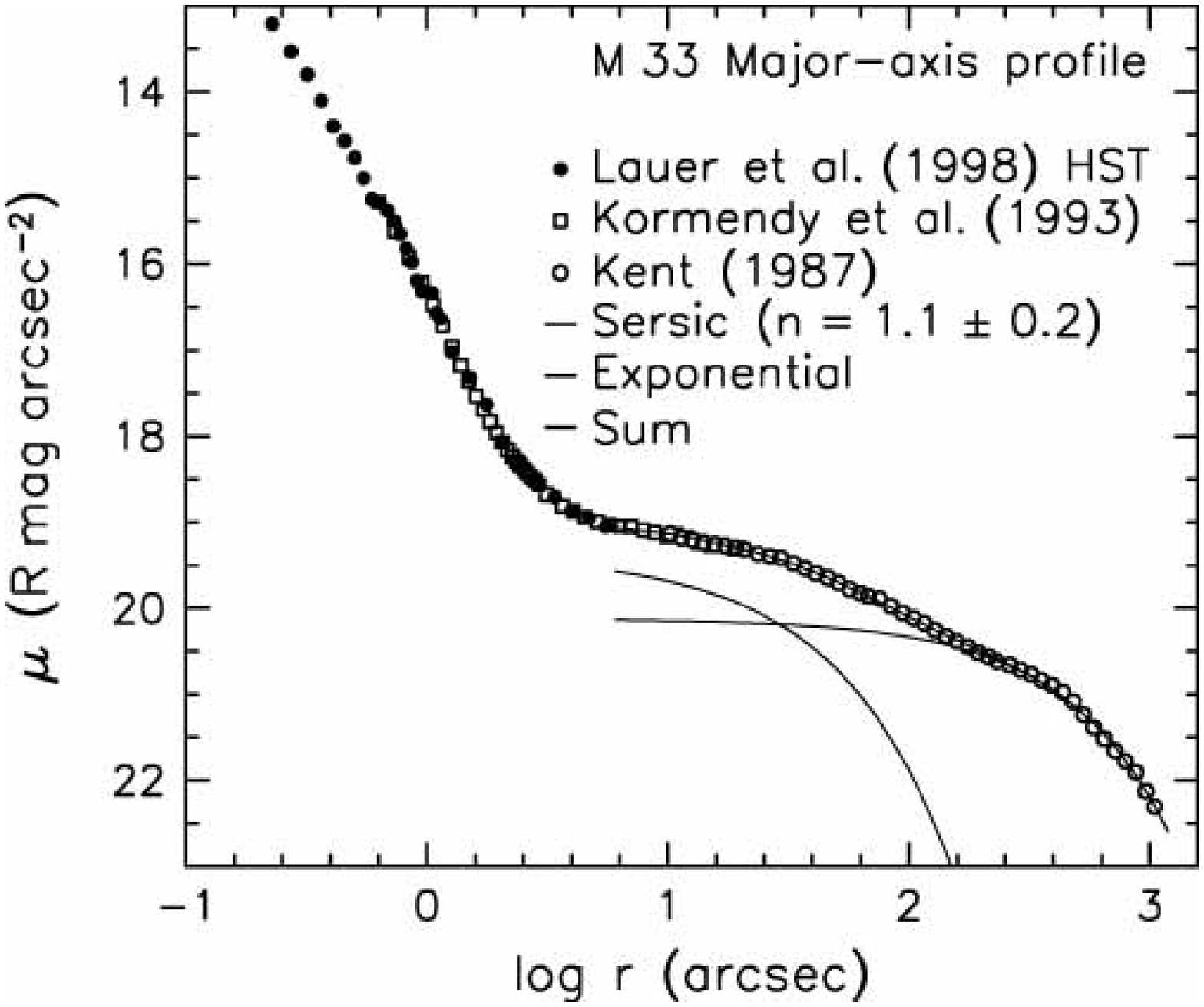}}

     {\it Figure 19}\quad Major-axis surface brightness profile of
M{\ts}33.  The steep rise in surface brightness near the center is the nucleus.
The rest of the profile has been decomposed into an inner S\'ersic function plus an exponential.  We thank S.~Faber and the Nuker team for suggesting
a discussion of M{\ts}33.

\eject

      The inner component is consistent with, although on the low-density side 
of, the parameter distribution for ``exponential bulges'' given in Carollo
(1999).  The effective radius of the inner component is $r_e = 0.31 \pm 0.05$
kpc, and the mean surface brightness within $r_e$ is 20.7 $\pm$ 0.2 $R$ mag
arcsec$^{-2}$.  (All errors quoted take account of parameter coupling in the
decomposition.)  If one applies profile decomposition in the canonical way, 
then one could reasonably conclude that this is a pseudobulge not unlike the
fluffiest ones discussed in the literature.

      On the other hand, we are uneasy about the decomposition in Figure~19.
The distinction between the components is subtle.  The inner one is at 
most a factor of 2 brighter than the outer one, and it is so only at $r < 10^{\prime\prime}$ where the outer exponential has already been extrapolated 
far inward.  The stars in both components are presumably in nearly circular
orbits; stars that define the outer exponential do not, by and large, visit the
inner exponential and vice versa.  Does it really make sense to say that half of
the disk stars at $r \simeq 25^{\prime\prime}$ belong to the main disk and half
belong to a pseudobulge?  If we could observe each star, how would we decide
which ones belong to the bulge and which to the
disk?  It is difficult to believe, given substantial spread in kinematic and
composition properties, that there would be such a clean separation into two
components that we could label each star correctly.  Another way to put it is
this: Since the fitting functions used for each component are not physically
motivated or explained, is there any reason to believe that each one
extrapolates without change into the part of the galaxy that is dominated by the
other?  And still another way: No theory of the formation of exponential disks
explains why an exponential is so magic and so required that modest departures
from it cry out for explanation.  We already accept Freeman (1970a) ``Type II''
profiles as canonical disk behavior, even though we can explain it in only a
few cases (e.{\ts}g., Talbot, Jensen, \& Dufour 1979).  We accept outer cutoffs
(e.{\ts}g., van der Kruit \& Searle 1981a, b; 1982).  Oval disks are only
piecewise exponential (\S\ts3.2).  Would it be a surprise if disks also knew 
how to deviate above an inward extrapolation to small radii?

      We believe that it is not possible, given the available information, to
distinguish between the following possibilities:  It is entirely plausible that
the inner exponential is a protopseudobulge.  Alternatively, disk profiles can
have a variety of wiggles and this is one of them.  Interestingly, the inner
component is significant out to a radius that is comparable to the width of the
spiral arms.  Spiral arms pass through the center when there is no ILR.  This
means that the profile near the center measures light only from a spiral arm 
crest, while the profile farther out is an average over arm crests and arm
troughs.  This could be part of the explanation of the ``inner component''.

      This discussion highlights a fundamental conceptual uncertaintly with
the blind application of profile decompostion.  In the case of a
big, classical bulge plus an obvious disk, we can be confident that a
decomposition has physical meaning.  We know this from edge-on galaxies.
The bulge in the Sombrero galaxy clearly extends into and beyond the radii
where we would see the disk if the galaxy were face-on; at these radii, we
would see bulge stars in front of and behind the disk.  Bulge formation via
mergers guarantees that this be so, because some stars in a merger
always splash out to large radii.  But secular evolution involves the slow
transport of gas that is never far from dynamical equilibrium.  No substantial
splashing occurs.  The concept that a pseudobulge coexists, at some radius of
interest, with a disk that was already in place has much less physical
meaning than it does in the case of a classical bulge.  It is not clear to
us that decomposition has any meaning at all in distinguishing pseudobulges 
from disks.  It may be more meaningful to fit the profile piecewise.

      Decomposition remains a useful way to derive diagnostic parameters.
We should not overinterpret the results.  Further work is needed to define the
boundaries between what deserves interpretation and what does not.

      What are the implications?  We believe that the conclusions about well
developed pseudobulges -- the ones whose profiles rise well above the inward
extrapolation of the disk profile, as in Figures 10 to 12 -- are unchanged.  
The Fundamental Plane correlations then tell us that the dividing line between
pseudobulges and disks is just as fuzzy as the one between classical bulges and
pseudobulges.  The reason is not that our machinery is inadequate.  The reason
is that there is a physical continuum between disks and pseudobulges, as
suggested by Kormendy (1993) and by B\"oker, Stanek, \& van der Marel (2003c).
The sequence of pseudobulges fades out not where they become tiny, like nuclei,
but where they become large, low in surface brightness, and indistinguishable
from disks.

\omit{  MAKE SURE THESE IDEAS ARE IN SECTION 5:
      Finally, we comment on ongoing star formation near the centers of
late-type galaxies.  Recall Sandage \& Bedke's (1994) important comment about
classical bulges: ``the central bulge is~\dots~nearly always devoid of recently
formed stars.''  This is consistent with a merger origin: mergers are episodic
events that trigger star formation, but between mergers, few stars form and
the ones that are present have plenty of time to age.  On the other hand, the
morphology of late-type galaxies discussed in the above papers (e.{\ts}g.,
Figures 10 and 13), the results of \S\ts4.3, and the star formation indicators
reviewed in \S\ts5 imply that star formation is exceedingly common in late-type
``bulges''.  If star formation is ubiquitous, then it cannot be caused mostly by
mergers.  Also, the dynamical effects of a recent merger could not be hidden in
such a large fraction of late-type disk galaxies.  Instead, the star formation
must be ongoing over long periods of time.  This implies secular growth in the
stellar density.  
}

\def\sfr{M$_\odot$~yr$^{-1}$} 
\def\arcsec{$^{\prime\prime}$}

\hfuzz = 100pt

\vsl\vsss\vsss\vsss
\ni {\bf 5.~\hbox{CENTRAL~STAR~FORMATION~\&~PSEUDOBULGE~GROWTH}}
\vsl

\hfuzz = 10pt

      In this section, we estimate the current rate at which star formation
is building stellar mass density in pseudobulges.  We wish to see whether the
picture of secular evolution in Sections 2 and 3 is consistent via plausible 
formation timescales with the properties of pseudobulges in Section 4.

      Figure 8 showed central gas disks in barred and oval galaxies that have
radii and masses comparable to those of pseudobulges and that are intensely
forming stars.  They are a window on pseudobulge formation.  We begin this section by discussing well-studied systems in which star formation rates (SFRs),
gas masses, stellar mass deposition rates, and hence evolution timescales can 
be constrained accurately.  We then review the broader body of observations of
circumnuclear gas disks and SFRs.  Star formation is ubiquitous in late-type
galaxies.  This means that it cannot be driven by episodic events such as
mergers.  It must be secular.

\vsl
\ni {\bf 5.1.~Case Studies:~NGC~1326, NGC~1512, NGC~4314, NGC~5248}
\vsl

  These galaxies are excellent prototypes for studying circumnuclear disks,
because each has been studied in depth using a combination of {\it HST\/} and
ground-based imaging (Buta et al.\ 2000, Maoz et al.\ 2001, Benedict et al.\
2002, Jogee et al.\ 2002).  Also, the four objects span a representative range
of host galaxy properties.  They are all barred, and they all have similar
luminosities ($-19.0 \ge M{{_B}{^0}} \ge -20.3$), but they cover a wide range
of morphological types (SB0/a -- SBbc) and environments.  NGC~1326 is in the
Fornax cluster; NGC~1512 is in an interacting pair with NGC~1510; and NGC~4314
and NGC~5248 are relatively isolated field galaxies located in loose groups.
Nearly all ($>$\ts80\ts\%) of the star formation in  NGC~1326 and NGC~4314 is
contained in the circumnuclear rings, while NGC~1512 and NGC~5248 have actively 
star-forming outer disks, with less than 40\ts\%  of the total SFR near the
center.  The diversity in galaxy properties and environments already suggests
that internal structure (e.{\ts}g., bars) is more important than external
influences in feeding the central star formation.

      The central star-forming rings of NGC 1512, NGC 1326, and NGC 4314 are
illustrated in Figure 8.  At high spatial resolution, the rings of HII
regions and young star clusters often are revealed to be pairs of tightly-wound
spiral arms.  This is shown for NGC 1512 in Figure 3.  The spiral structure is
seen most clearly in red continuum images, where networks of dust features
spiraling toward the center from within the star-forming rings can be seen.
The continuum images also reveal large numbers of bright stellar knots ($>$\ts70
in NGC~4314; 500 -- 1000 in the others). The luminosities and dereddened colors
of these knots indicate that they are not single stars but instead are luminous
associations or star clusters.  The brightest of these have stellar masses of
order $10^5$ $M_\odot$, placing them in the class of populous blue clusters
observed in the Magellanic Clouds and other gas-rich galaxies.  Some may be 
young progenitors of globular clusters.  Many of the clusters are coincident
with HII regions, but most are free of surrounding nebulosity, and these
probably are older than the 5 -- 10 Myr lifetimes of typical HII regions.

      Current SFRs in these regions can be estimated from H$\alpha$ or
Pa$\alpha$ measurements converted using the SFR calibrations of Kennicutt
(1998a).  The largest uncertainties come from heavy and patchy dust obscuration.
When both H$\alpha$ and Pa$\alpha$ data are available, the flux ratio
of the two lines can be used to infer the extinction $A$.  Typically, 
$A_{H\alpha}$ = 1 to 3 mag across these regions, larger than normal values
of $\sim$\ts1 mag in spiral disks (Kennicutt 1983; Kewley et al.~2002), but low 
enough so that the extinction-corrected emission-line fluxes should provide
reasonable measures of the SFRs.  Star formation rates estimated independently
from extinction-corrected ultraviolet or far-infrared photometry of the regions 
(when available) are in general agreement with the above results.  This 
increases our confidence in the SFR measurements.  The resulting SFRs range 
from $\sim$\ts0.13 \sfr\ in NGC~4314 (Benedict et al.\ 2002) to 1 \sfr\ in
NGC~1326 and NGC~1512 (Buta et al.\ 1999, Maoz et al.\ 2001) and $\sim$\ts2
\sfr\ in NGC~5248 (Maoz et al.\ 2002 corrected to a distance of 15 Mpc). 
These values are probably accurate to within $\pm$\ts50\ts\%, given
uncertainties in the amounts and patchiness of the extinction and in the 
assumed distances to the galaxies.  This is sufficient to characterize the
evolutionary properties and physical conditions in these regions.

      SFRs of 0.1{\ts}--{\ts}2 \sfr\ are modest compared to the total SFRs in
giant spiral galaxies, which typically range from 0.1{\ts}--{\ts}1 \sfr\ in
normal Sa galaxies to 1{\ts}--{\ts}10 \sfr\ in Sb--Sc galaxies (Kennicutt 1983,
1998a). However, they are quite exceptional considering the physical compactness
of the star-forming regions. The star-forming rings have radii of 500 -- 700 pc,
so the surface densities of star formation are 0.1 -- 1 \sfr~kpc$^{-2}$.  This
is \hbox{1 -- 3} orders of magnitude larger than the typical disk-averaged SFR
densities in normal galaxies.  It approaches the SFR densities
seen in some infrared starburst galaxies (Kennicutt 1998a, b).  If these SFRs
were to persist over a Hubble time, they would produce ``bulges'' with
stellar masses of $10^9 - 10^{10}$ $M_\odot$.  Thus, while the total
amounts of star formation in these regions are not unusual by galactic
standards, the character of the star formation is quite distinct.  

   The distinctive character of the star formation is underscored by
large populations of luminous young star clusters.  Their extinction-corrected
absolute magnitudes range from $M{{_V}{^0}} = -13$ to $M{{_V}{^0}} \sim -8$.
Fainter than this, {\it HST\/} photometry becomes very incomplete.  The
corresponding masses, corrected as discussed below for the ages of the clusters,
are $\sim 10^3$ to $10^5$ $M_\odot$.  These are similar to the masses of giant
OB associations such as those in supergiant HII regions like 30 Doradus in the
LMC and to the masses of the populous blue star clusters found in the LMC and
other gas-rich galaxies (e.{\ts}g., Kennicutt \& Chu 1988).  The luminosity
functions of the knots are well fitted by a power law with slope $dN/dm \sim 
-2$.  They are consistent with the luminosity functions of HII regions
and their embedded OB associations (e.{\ts}g., Kennicutt et al.\ 1989a; Bresolin
\& Kennicutt 1997). They are also similar to the young star cluster populations
in merger remnants such as the Antennae (e.{\ts}g., Zhang \& Fall 1999).
However, no examples are found in these galaxies of the so-called
``super star clusters'' (SSCs) with $M_V < -15$ that are often seen in merger
remnants and luminous starburst galaxies.  This may be a reflection of
the lower total amounts of star formation in these rings rather than any sign
of a different cluster mass spectrum.  Even if the power-law cluster mass
spectra extend to the realm of the SSCs in these objects, the number of SSCs
that we expect to observe at any one moment is less than one, based on
the total size of the populations observed.  We need to observe more central
star-forming rings to determine whether they can form SSCs.

      The star clusters can be age-dated using multi-color photometry and
synthesis models (e.{\ts}g., Leitherer et al.\ 1999).  They provide a powerful
probe of the star formation histories in these circumnuclear regions.  In all
four of these galaxies, multiband {\it HST\/} imaging in different combinations
of $U$, $B$, $V$, $I$, $H$, and H$\alpha$ have been used to derive
reddening-corrected colors, luminosities, and hence age distributions.  The
galaxies all show a spread in cluster ages from zero to 200{\ts}--{\ts}300 Myr.
The age distributions are heavily weighted toward younger clusters, but this is
readily accounted for by dimming with age and by dynamical disruption effects.
When corrections are applied for these processes, the age distributions are
generally consistent with a roughly constant cluster formation rate over the
past 200{\ts}--{\ts}300 Myr (Maoz et al.\ 2001).  However, more sporadic
histories cannot be ruled out.

      If the rings have been forming stars at the current rate for 0.2 -- 0.3
Gyr, then the total mass of stars formed is (2 -- 6) $\times 10^8$ $M_\odot$ in
NGC~1326, NGC~1512, and NGC~5248 and about 2 $\times 10^7$  $M_\odot$ in
NGC~4314.  We can check the consistency of these results by comparing the SFRs
with the masses of the circumnuclear gas disks.  Millimeter measurements of CO
emission in the centers of NGC~1326, NGC~4314, and NGC~5248 have been reported
by Garcia-Barreto et al.\ (1991), Combes et al.\ (1992), Benedict et al.\
(1996), Sakamoto et al.\ (1999), Jogee et al.\ (2002), and Helfer et al.\
(2003).  The corresponding molecular gas masses range from 0.7 $\times 10^8$
$M_\odot$ in NGC~4314 to (5 -- 12) $\times 10^8$ $M_\odot$ in NGC~1326 and
NGC~5248.  These values assume a ``standard'' Galactic conversion factor between
CO intensity and H$_2$ column density.  Several authors (e.{\ts}g., Wilson 1995;
Paglione et al.\ 2001; Regan 2000) have advocated using a lower conversion
factor for these metal-rich environments; doing so would reduce the above 
masses by factors of up to 2 -- 3.  The gas masses are comparable to the masses
of stars already formed in the central disks during the current star formation 
burst, as one would expect if one typically observes these systems at random
times during the burst.  Combining the gas masses with the SFRs also shows that
there is sufficient fuel to feed the current circumnuclear SFRs for another 
0.2 -- 1 Gyr.  By the time the gas is exhausted, central stellar disks with
masses of $10^8$ to $10^9$ $M_\odot$ will have formed.  Of course, the masses
will be even larger if gas from the galaxies' bars continues to feed the star
formation.  In the cases of NGC~1326, NGC~1512, and NGC~5248, the above masses
are factors of several higher than the mass in stars, $\sim 1 \times 10^8$
$M_\odot$ (see below), formed in the main exponential disks if the parameters of
these disks are extrapolated to the center.  In fact, the stellar disks
being formed by the star formation rings have characteristic masses and sizes 
that are comparable to those of pseudobulges.  In these four galaxies, we almost
certainly are observing the formation of pseudobulges, or the continued growth
of pre-existing pseudobulges.

\vsl
\ni{\bf 5.2.~General Properties of Circumnuclear Regions}
\vsl


      The circumnuclear star-forming rings discussed above are not extreme
cases; even higher SFRs are observed at the centers of NGC~1097 and some other
nearby galaxies.  Nevertheless, before we attempt to
characterize the global rates of star formation in these objects, it is
important to review the properties of circumnuclear star-forming rings and disks
in general.  This subject was reviewed by Kennicutt (1998a), with emphasis on
the most luminous starburst galaxies.  These are nearly always associated
with major mergers of gas-rich galaxies that are forming high-mass bulges and
elliptical galaxies (Sanders \& Mirabel 1996; Kennicutt, Schweizer, \& Barnes
1998, and references therein).  Here, we focus exclusively on the central
regions of normal spiral galaxies, where the circumnuclear activity is fed by
the kinds of secular processes discussed in this paper.

   The frequency of occurrence of dense central gas disks and vigorous star 
formation can be estimated from two independent lines of evidence,
surveys of central star formation in the ultraviolet, visible, or mid-infrared,
and CO surveys of central molecular gas.  Prominent circumnuclear rings like
those in Figure 8 are easily identified.  Maoz et al.\ (1996) estimate that
approximately 10\ts\%\ of Sc and earlier-type spirals contain such strong
circumnuclear star-forming regions, based on an ultraviolet imaging survey of
110 nearby spirals.  This is roughly consistent with the frequency of
circumnuclear ``hotspot'' galaxies in the survey of S\'ersic (1973), which was
based on blue photographic plates.  Most of these galaxies are barred.  This
includes all of galaxies that have ultraviolet-bright rings identified by Maoz 
et al.\ (1996), 88\ts\%\ of the hotspot galaxies in the S\'ersic (1973)
compilation, and 81\ts\% of all galaxies with peculiar nuclei identified by
S\'ersic.  Galaxies with strong circumnuclear star formation tend to have
Hubble types between Sa and Sbc (Devereux 1987; Pogge 1989; Ho et al.\ 1997), 
although there are earlier- and later-type exceptions.  Putting all this
together, the frequency of circumnuclear
rings among the core population of massive, intermediate-type barred galaxies
is of order 20\ts\%.

      Quantifying the star formation statistics in less spectacular star-forming
galaxy centers is more difficult.  In early-type galaxies, the typical levels of
extended disk star formation are relatively low (Kennicutt 1998a and references
therein).  Any nuclear star formation stands out.  However, in the gas-rich,
late-type spirals that dominate the total star formation in the local universe,
it can be difficult to distinguish central star formation that might be 
associated with pseudobulge growth from the central extrapolation of the 
general disk star formation.  

      CO interferometer surveys give a clearer picture.  Some of the more
comprehensive aperture synthesis CO surveys include a study of 20 spirals by
Sakamoto et al.\ (1999), the BIMA Survey of Nearby Galaxies (SONG), which
observed 44 nearby spirals (Regan et al.\ 2001; Sheth et al.\ 2002; Helfer et
al.\ 2003; Jogee 1998; Jogee et al.~2004).  Notable studies
of individual galaxies include Kenney et al. (1992); Garcia-Burillo et al.\ 
(1998, 1999, 2000); Sofue et al.\ (1999); Sakamoto et al.\ (2000); Jogee et 
al.~(2002); and Schinnerer et al.\ (2002, 2003).  Larger samples of galaxies
have been observed in the $^{12}$CO (1--0) and (2--1) rotational lines with
single-dish telescopes, with typical beam diameters of 11{\ts}--{\ts}50\arcsec.
Such large surveys include Young \& Devereux (1991); Braine et al.\  (1993);
and B\"oker et al. (2003a).  A survey in HCN including a large subsample of
normal galaxies is given in Gao \& Solomon (2004).  

      These surveys show that central molecular disks are common but not
universal. For example, 45\ts\% of the spiral
galaxies in the BIMA SONG show central gas concentrations that exceed the
highest peak column densities anywhere else in the disks.  This fraction was
even higher (75\ts\%) in the Sakamoto et al.\ (1999) CO survey of 20 spiral
galaxies, but in this case the sample was partly selected on the basis of 
strong CO emission.  The presence and masses of these disks are strongly
enhanced in barred galaxies (Sakamoto et al.\ 1999, Helfer et al.\ 2003), as
is the case for central star formation.
Although a few objects show centrally peaked distributions that one might be
tempted to associate with the exponential profiles of stellar disks or
pseudobulges, the predominant structures are barlike distributions, bipolar
``twin-peak" distributions (Kenney et al.\ 1992), circumnuclear rings, spiral
arms, or combinations of these structures.  In many systems, the gas is
unlikely to be in steady-state equilibrium, and the interpretation is
complicated by the likely presence of spatial variations in gas temperature
that will modulate the distribution of CO emission.  We can conclude only that
the gas disks have radii that are characteristic of central bars, bulges, and
pseudobulges.

      The first CO observations that spatially resolved galaxies showed that the
distribution of molecular gas often follows the starlight (e.{\ts}g., Young \&
Scoville 1991).  Recent observations confirm this result (Regan et al.\ 2001;
B\"oker et al.\ 2003a).  Even when the stellar brightness rises steeply toward
the center above the inward extrapolation of the outer exponential -- this is
what we would conventionally call a ``bulge'' -- the CO emission often does so
also.  Five examples are shown in Figure 20.  All are excellent examples of
objects in which a bar (top row), oval (middle row), or global spiral structure
that reaches the center (NGC 4321 in the bottom row) provides an engine for 
inward gas transport. Consistent with this, the molecular gas is very centrally
concentrated.  Since star formation rates increase faster than linearly with gas
density, the observation that the molecular gas density follows the starlight
guarantees that star formation will further enhance the density contrast between
the (pseudo)bulge and the outer disk.  We discussed several of these
objects as typical pseudobulges.  The exception in Figure 20 is NGC 7331, a
galaxy that contains a probable classical bulge.  Other galaxies in Regan 
et al.\ (2001) behave similarly.

\vfill\eject

\cl{\null}

\vfill

\col{\includegraphics{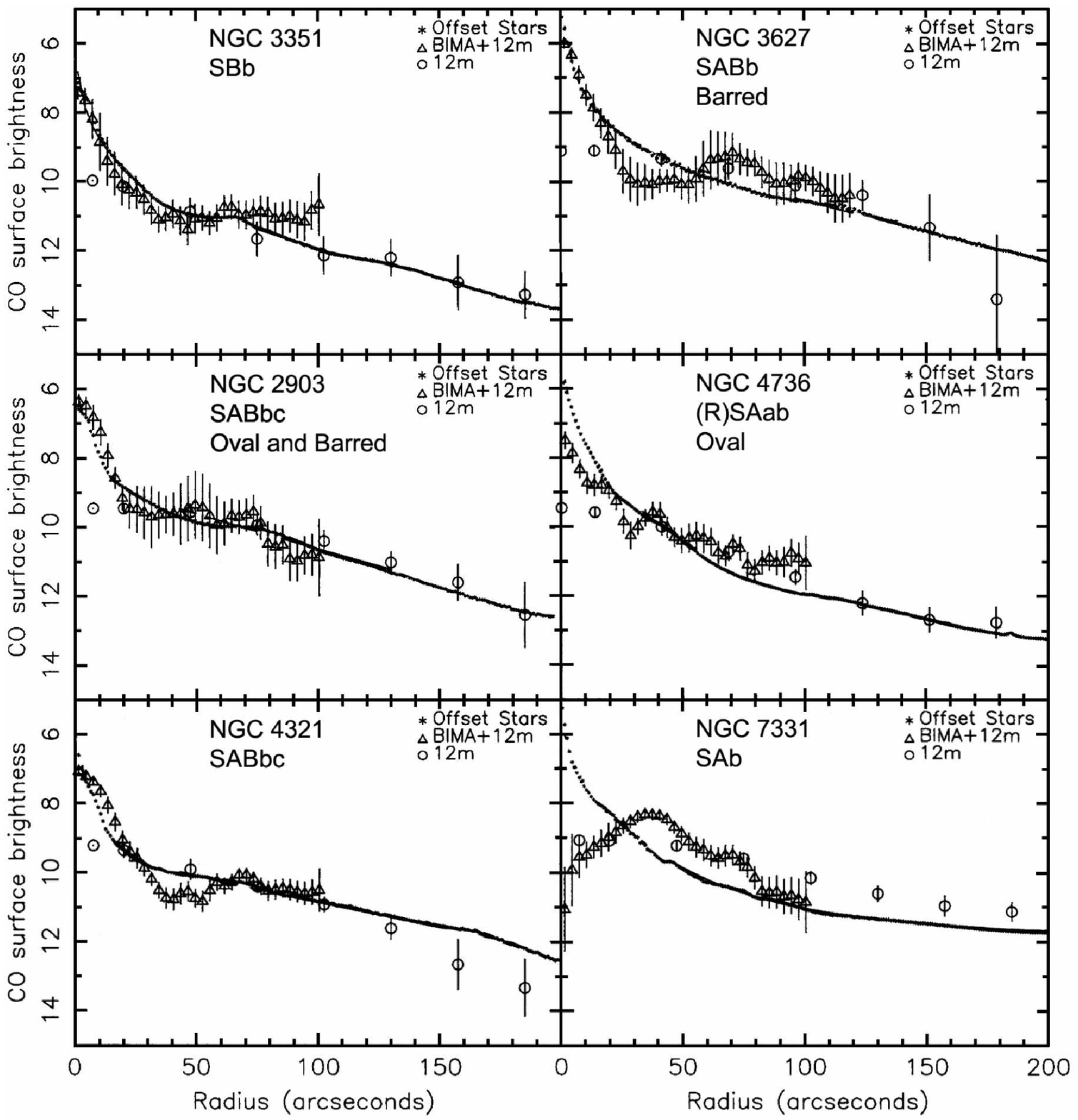}}

  {\it Figure 20}\quad Radial profiles of CO and stellar $K$-band
surface brightness from the BIMA SONG (adapted from Regan et
al.~2001).  CO surface brightness is in magnitudes of Jy km s$^{-1}$
arcsec$^{-2}$ with zeropoint at 1000 Jy km s$^{-1}$ arcsec$^{-2}$. The
stellar surface brightness profiles have been shifted vertically to the
CO profiles. Morphological types are from~the~RC3.  NGC 2903 and 3627
are clearly barred in the $K$-band images shown in Regan et
al.~(2001).  NGC 2903 and 4736 are oval galaxies (Section 3.2).  NGC
4736 contains a prototypical pseudobulge; it is also
illustrated here in Figures 2, 8, and 17.  NGC 4321 is an unbarred
galaxy with no ILR; it was discussed in Section 3.4.  All galaxies in this
figure except NGC 7331 have structures that are expected to cause gas to 
flow toward the center.  NGC 7331 is included to show the very different 
CO profile in a galaxy with a probable classical bulge.
  
\cl{\null} \cl{\null} \eject

      When one combines the data from the above aperture synthesis surveys
with small-beam, single dish measurements (Braine et al.\
1993, B\"oker et al.\ 2003), the resulting gas masses show a large
range, from $\sim$10$^6$ $M_\odot$ to 2 $\times 10^9$ $M_\odot$.  
These values assume a ``standard'' CO--to--H$_2$ conversion factor,
$X_{CO} = 2.8 \times 10^{20}$~cm$^{-2}$~(K~km~s$^{-1})^{-1}$ 
(Bloemen et al.\ 1986).  If instead we use a variable, metallicity-dependent
conversion factor (e.{\ts}g., Wilson 1995; Paglione et al.\ 2001; Boselli et
al.\ 2002), then this range narrows to $\sim$\ts10$^7 - 10^9$ $M_\odot$ (B\"oker
et al.\ 2003).  Pseudobulges are expected to grow to at least these masses.
More massive pseudobulges would be result if gas continues to be added to the
observed nuclear disks.

Studies of the SFRs in individual systems are too numerous to be listed here,
but some of the most extensive and notable studies include 
Kennicutt et al.~(1989b), 
Pogge (1989), 
Phillips (1993), 
Maoz et al.~(1996, 2001), 
Elmegreen et al.~(1997, 1998, 1999, 2002),
Usui et al.~(1998, 2001),  
Buta et al.~(2000), 
Colina \& Wada (2000),
Inoue et al.~(2000), 
Alonso-Herrero \& Knapen (2001), 
Ryder et al.~(2001), 
Benedict et al.~(2002), and 
Knapen et al.~(2002).  
A variety of star formation tracers have been used, including measurements of
ultraviolet and infrared continua, and H$\alpha$, P$\alpha$, Br$\gamma$, and
other hydrogen recombination lines (see Kennicutt 1998a).  The Spitzer Space
Observatory will have a major impact on this subject by providing
spatially-resolved maps of the far-infrared dust emission in these regions.

      SFRs measured by different authors are generally consistent at the 
factor-of-two level; this is comparable to the uncertainties that are typically
quoted for these highly dust-attenuated regions.  Although this limits the
reliability of SFRs for any individual object, good measurements are available
for about 40 galaxies, and this is sufficient to characterize the range of star
formation properties.  The absolute SFRs within circumnuclear rings and disks
range over a factor of about a thousand, from 0.01 to 10 \sfr.  This brackets
the range of SFRs observed in our four case studies and is comparable to the
range observed in the {\it integrated} SFRs of normal spiral galaxies 
(Kennicutt 1998a and references therein).  The central star formation accounts
for 10{\ts}--{\ts}100\% of the total SFR of spirals galaxies.  The highest
fractions occur in early-type galaxies, which typically have low SFRs in their
outer disks (Kennicutt 1983, 1998a).  

\vsl
\ni{\bf 5.3.~Constraining Evolution Timescales and Pseudobulge Growth}
\vsl

      We can combine the data on SFRs and gas contents of the central
regions of galaxies to constrain the evolutionary timescales
and formation rates of pseudobulges.  We first consider the prominent
circumnuclear star-forming rings, which represent only the high-luminosity
exttreme of this activity, but for which we can derive relatively hard
constraints.  A search of the literature reveals 20 galaxies with
circumnuclear star formation and reliable data on the SFRs, central
gas masses, and sizes of the star-forming regions.
For each galaxy, we derived the mean molecular gas surface density (using 
standard CO -- H$_2$ conversion factors) and the mean SFR surface density
within the circumnuclear regions.  These are plotted as filled squares 
with error bars in Figure 21.  The SFRs were derived using a combination
of methods, extinction-corrected H$\alpha$ and Pa$\alpha$ fluxes or
far-infrared fluxes.  The large error bars reflect considerable uncertainties
in the SFRs due to dust extinction and possible AGN contamination and
uncertainties in the CO -- H$_2$ conversions that provide the gas masses.
Ignoring atomic gas introduces another uncertainty, but this is expected
to be of order 10\ts\% or less (e.{\ts}g., Young \& Scoville 1991, Sanders 
\& Mirabel 1996).  For comparison, Figure 21 also shows the disk-averaged
SFR and total gas densities for 61 spiral galaxies (solid circles), and
the same data for the centers of these galaxies when spatially resolved
data are available (open circles).

\vfill




\col{\includegraphics{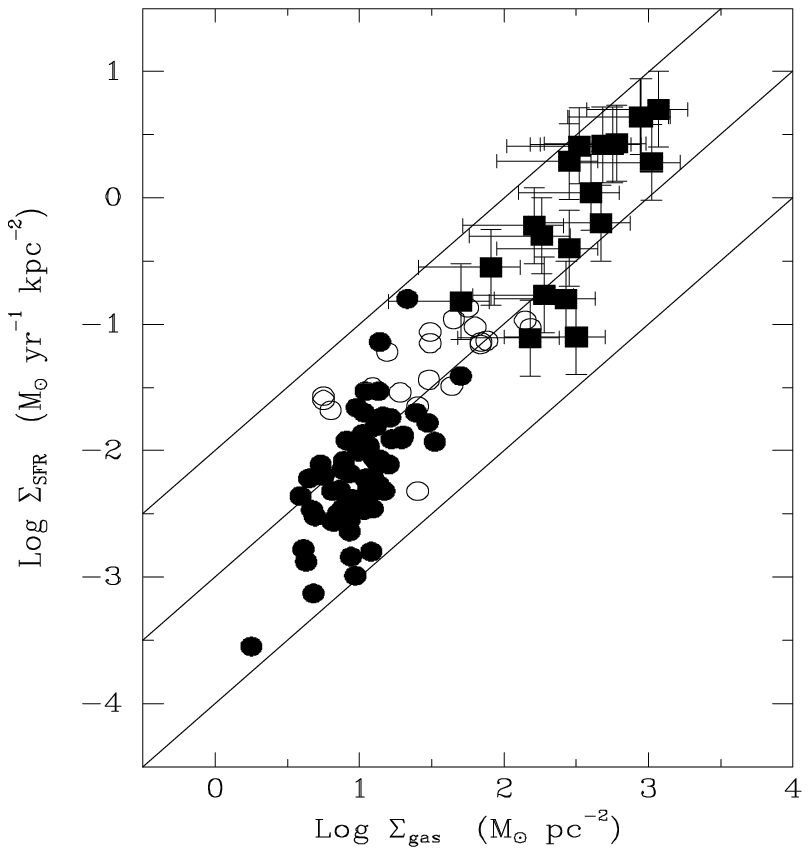}}

  {\it Figure 21}\quad Correlation between SFR surface density and 
  total gas surface density for 20 circumnuclear star-forming rings
  (filled squares with error bars) compared to disk-averaged values 
  for 61 spiral galaxies (filled circles) and the centers of these galaxies
  (open circles).  The circumnuclear data are compiled in this paper, and the
  comparison data are from Kennicutt (1998b).  The solid diagonal lines show
  constant gas consumption time scales (increasing downward) of 0.1, 1, and 
  10 Gyr.
  
\eject

      Figure 21 clearly shows that the circumnuclear rings populate a unique
regime of molecular gas density and SFR.  They extend  the Schmidt SFR power 
law that is seen in the other galaxies (Kennicutt 1998b).  For the sake of
consistency, we adopted the same standard CO--H$_2$ conversion factor for all
of the points; adopting a lower conversion factor for the centers would move
the filled squares and open circles to the left by up to 0.3 -- 0.5 dex.
This would increase the best-fitting Schmidt-law slope from $N \sim 1.4$ 
to $N \sim 1.5$.

      The same diagram can be used to constrain the timescale on which
the circumnuclear gas disks turn into stellar disks.  Figure 21 shows lines
of constant gas consumption times of 0.1, 1, and 10 Gyr.  The outer
star-forming disks of these galaxies are characterized by mean
star formation efficiencies of approximately 5\ts\% per $10^8$ yr
and gas depletion times of $\sim$2 Gyr on average.  However
the star formation efficiencies in most the circumnuclear disks are
much higher, of order 10 -- 50\ts\% per $10^8$ yr.  Gas consumption
timescales are 0.2 -- 1 Gyr.  If we assume that we observe the
average disk at the midpoint of a gas accretion and starburst
episode, this means that the typical formation timescales for 
these pseudobulges is approximately 0.4 -- 2 Gyr.  This is reasonably
consistent with the (luminosity-averaged!)~star cluster ages of 
0.0 -- 0.3 Gyr inferred in the HST studies cited earlier.  Note, however,
that if the standard CO--H$_2$ conversion factor overestimates
the molecular masses of the disks, this would lower the inferred
timescales, probably by a comparable amount.

      We can now put together a rough picture of the growth rates of
pseudobulges in present-day spirals.  We infer a typical formation 
timescale of $\sim$ 1 Gyr
for the central disks from the arguments given above.  When
we combine this with a typical SFR range of 0.01 -- 10 \sfr, we
expect these gas accretion episodes to form pseudobulges with
masses of order $10^7 - 10^{10}$ $M_\odot$.  Typical systems like
the examples in Figure 8 fall in the $10^8$ to (2 -- 3) $\times 10^9$ \msun\
range.  Furthermore, circumnuclear star-forming rings of this
type are seen in $\sim$ 10\ts\% of intermediate-type spiral galaxies
(S\'ersic 1973, Maoz et al.\ 1996).  As discussed earlier, lookback
studies suggest that strong bars first formed at least 5 Gyr
ago.  Combining these numbers suggests that approximately half of
unbarred spirals and nearly all barred spirals may have formed a pseudobulge
in this mass range.    Of course this rough calculation is subject to a
chain of possible systematic errors.  However, it demonstrates that a 
scenario in which pseudobulges are a common or even ubiquitous constituent 
of intermediate-Hubble-type, massive spiral galaxies is plausible.

      So far, our results are based solely on the occurrence
of the most prominent circumnuclear star-forming rings in barred
galaxies.  Is there independent evidence based on the statistics
of central molecular gas disks for lower levels of
star formation in central pseudobulge disks?  To make such
an estimate, we used the BIMA SONG survey (Helfer et al.\ 2003) to
derive the median central molecular gas surface density in their
objectively selected sample of 44 nearby spiral galaxies.  This is
$\sim$ 200 {\msun}pc$^{-2}$, for a standard CO--H$_2$ conversion
factor.  Interestingly. this gas density already exceeds the
typical stellar surface density in local spiral disks.  A typical, late-type
disk has a mass-to-light ratio $M/L_B \simeq 1$ and a central surface brightness
$\mu_B = 21.7$ mag arcsec$^{-2}$ (Freeman 1970). The corresponding stellar 
density is $\sim$ 130 {\msun}pc$^{-2}$.  Consequently, if the current
gas disks are converted into stars, the central surface brightness of the
disk will more than double.  If the gas infall continues for a few Gyr, 
a proportionally brighter stellar component will be formed.

We can repeat this calculation by using Figure 21 to estimate
the typical SFR densities in the centers of these barred galaxies,
and combine this with the typical star formation timescales derived
earlier to estimate the total surface density of stars formed.
This calculation is not entirely independent, because the star
formation timescales are partly derived from the measured molecular
gas densities.  However, there are independent constraints on the
star formation timescales from HST studies of the star clusters
in circumnuclear starbursts and from photometric constraints from
integrated light.  For a typical SFR density of 0.1--1 
{\msun}yr$^{-1}$~kpc$^{-2}$ (Figure 21), we expect to build up
central stellar densities of 50 -- 500 {\msun}pc$^{-2}$ for star
formation lifetimes of 0.5 Gyr, and 500 -- 5000 {\msun}pc$^{-2}$ if
the feeding of gas from the bar persists for 5 Gyr.  This compares
to ``Freeman disk" central densities of $\sim$ 100 -- 250 {\msun}pc$^{-2}$
for $M/L_B$ = 1 -- 2.  The total masses in these components are
of the same order as the observed molecular gas disks in the centers
of these galaxies, $10^7 - 10^9$ $M_\odot$, if there is no continued
feeding of the nuclear disks, and up to 5 times larger if
the gas feeding persists for 5 Gyr at a rate that is sufficient
to replace the mass lost from star formation.

      We reiterate that there are large uncertainties in these numbers.
The most important uncertainties are the total duration of the inward gas
transport in the bars, the CO--H$_2$ conversion factors used to 
estimate the molecular gas masses, and uncertainty in separating
``pseudobulge" star formation from steady-state disk and/or 
nuclear star formation.  However, our estimates demonstrate that in a 
typical barred spiral, the total central star formation that results
from secular gas inflow can easily exceed that in the underlying disk.
By the same token, even the high end of the mass ranges described here 
falls 1 -- 2 orders of magnitude short of the massive bulges that are 
typical of giant S0 -- Sab and elliptical galaxies.

\vfill\eject

\hfuzz = 100pt

\vsl\vsss\vsss\vsss
\ni {\bf 6.~\hbox{COSMOLOGICAL IMPLICATIONS OF SECULAR EVOLUTION}}
\vsl

\hfuzz = 10pt

\ni {\bf 6.1.~Evolution Along the Hubble Sequence}
\vsl

      The qualitative arguments in Sections 2 -- 4 and the star formation
measures in Section 5 imply that secular evolution increases bulge-to-total
luminosity ratios $B/T$.  How much evolution along the Hubble sequence
(e.{\ts}g., Pfenniger 1996b) is plausible?

      This question is too important to be postponed, but we warn readers that
the results of this section are very uncertain.  To address the question, 
we compare predicted $B/T$ ratios with the distribution of values observed
by Simien \& de Vaucouleurs (1986).  They decomposed the
$B$-band surface brightness profiles of 98 galaxies to measure $B/T$ as a
function of RC2 type.  They added up all of the central light in excess of the
inward extrapolation of exponentials fitted to the outer disks, so $B/T$ 
measures the sum of bulge and pseudobulge light.  They found that $B/T$ is
typically 2\ts\% in Sd galaxies, 9\ts\% in Sc galaxies, 16\ts\% in Sbc galaxies,
24\ts\% in Sb galaxies, and 41\ts\% in Sa galaxies.  The scatter around these
values is large.  In Sections 5.1 and 5.2, we estimated that circumnuclear
star-forming rings grow pseudobulges with masses of $\sim 10^9$ $M_\odot$.  The
plausible range of these masses is also large, from $\sim 10^7$ to $10^{10}$
$M_\odot$.  The total stellar masses of these galaxies are of order
(1{\ts}--{\ts}5) $\times 10^{10}$ $M_\odot$.  Therefore secular evolution can
reasonably be expected to have produced pseudobulges with masses ranging from
0\ts\% to $>$\ts10\ts\% of the total stellar masses of the systems.  This is
comparable to the $B/T$ value in Sc galaxies, consistent with our conclusion
that Scs contain pseudobulges.  Evolution of one Hubble stage -- e.{\ts}g., 
from Sd to Sc -- is plausible at the late end of the Hubble sequence.  Evolution
from Sc to Sbc is also plausible. 

      However, it is less easy for secular processes to form
the more massive bulges of S0{\ts}--{\ts}Sb galaxies.  The $B/T$ ratio in
these galaxies is large, and the galaxies themselves tend to be very massive.
Total bulge masses are at least $10^{10} - 10^{11}$ $M_\odot$.  The evidence 
from stellar populations (Section 8.1) is that the stars in these bulges formed
quickly and long ago.  We conclude that the stars in these bulges formed mostly
during hierarchical clustering.  That is, S0{\ts}--{\ts}Sb galaxies mainly
contain classical bulges.  Secular processes can contribute modestly to the
growth of classical bulges, but evolution by as much as half of a Hubble stage
is expected to be unusual. Based on present star formation rates, Sab galaxies 
like NGC 4736 that have dominant pseudobulges should be rare.

      In fact, they are not rare, and even some S0s have pseudobulges.  
The above estimates are lower limits for at least two reasons.  First, typical
disk galaxies presumably contained more gas in the past.  Second, some secular
processes, such as buckling instabilities, do not depend on concurrent star
formation.  They elevate pre-existing disk stars into the pseudobulge.

\vsl\vsss
\ni {\bf 6.2.~Merger-Induced Versus Secular Star Formation in Bulges}
\vsl

      Finally, we make a preliminary comparison of the relative
importance of secular and merger-induced star formation in the present
universe.  As shown above, it is plausible that most of the stars in
Sc{\ts}--{\ts}Sm (pseudo)bulges and a significant fraction of the stars in
Sb{\ts}--{\ts}Sbc (pseudo)bulges formed as a result of secular evolution. Given
the relative numbers of early- and late-type galaxies, classical bulges and
pseudobulges are not very different in number.  However, the masses of (mostly)
classical bulges in early-type galaxies are at least 1{\ts}--{\ts}2 orders of
magnitude larger than the  masses of pseudobulges in late-type spirals.
Integrated over the history of the universe, star formation caused by
secular processes has contributed at most a few percent of bulge stars. 
The vast majority of bulge stars are believed to have formed in collapse and
merger events.  

      However, most of these events occurred in the distant past.  Observations
show that merger rates increase dramatically with increasing cosmic lookback
time (Patton et al.~2002; Conselice et al.~2003).  
The fractional contribution of galaxy interactions and mergers to the total 
present-day SFR in the universe has been estimated by numerous workers,
starting with a classic study by Larson \& Tinsley (1978).   Kennicutt et
al.~(1987) estimate that 6\ts\% $\pm$ 3\ts\% of current star formation is
induced by galaxy-galaxy interactions.  This contains two, partly
compensating uncertainties.  It underestimates dust-extincted star formation in
bulges, but it overestimates bulge star formation because it includes a
contribution from distant tidal interactions as well as mergers.  How does
this value compare to the contribution from secular evolution?  In Section 5.3, 
we noted that $\sim$ 10\ts\% of intermediate-type spirals contain circumnuclear
disks or rings and that their star formation accounts for
10{\ts}\%--{\ts}80\ts\% of the current SFR in those galaxies.  These same
intermediate-type spirals dominate the current total cosmic SFR (Brinchmann et
al.\ 2003).  Combining these numbers suggests that a few percent of 
present-day star formation is attributable to secular processes.  The hint 
is that galaxy mergers and secular evolution produce comparable star formation
in the present universe.   Both contributions are small compared to the
dominant source of star formation at $z = 0$, i.{\ts}e., namely the quiescent
star formation in the disks of spiral and irregular galaxies.  Nevertheless, 
as argued in the Introduction, it appears that we live approximately at 
the epoch of transition when secular processes are overtaking mergers as the
primary mechanism that forms stars in the central parts of galaxies.

      We emphasize that all of the estimates in Sections 6.1 and 6.2 are
very uncertain.  We include them primarily to stimulate further work. 

\vfill\eject 

\vsl\vsss\vsss\vsss
\ni {\bf 7.~COMPLICATIONS}
\vsl

      This section discusses issues that complicate the identification of
pseudobulges.  They do not threaten the conclusion that pseudobulges form
by secular evolution of disks; on the contrary, they make it likely that such
evolution has operated even in situations where this is not obvious.  They
highlight areas that need further work.

\vsl\vsss
\ni {\bf 7.1.~Pseudobulges Do Not Have To Be Flat}
\vsl

    Several dynamical heating processes are expected to puff pseudobulges
up in the axial direction.  We are fortunate that some pseudobulges are disky
enough so that we can detect the ``smoking gun'' that points to a secular
origin.  It is plausible that others are so similar to classical bulges that
they cannot easily be recognized.

      One heating mechanism that we have already discussed is bar buckling
(\S\ts4.5).  If the density profile along the ridge line of the bar is not too
steep, then buckled bars produces box-shaped structures that can easily be
recognized when they are seen edge-on.

      Resonant vertical heating by the bar may also be important (Pfenniger
1984, 1985; Pfenniger \& Norman 1990; Friedli 1999).  It is not limited to the
relatively few stars that are in resonance at any one time, because the radii
of all resonances change as the central concentration increases.

      Another heating mechanism involves in-the-plane instabilities that result
if the nuclear disk gets too dense for its velocity dispersion.  Secular gas
inflow is slow compared to the disk's rotation velocity and even the disk's 
radial velocity dispersion $\sigma_r$.  If, as a result, star formation builds
up the disk surface density $\mu$ without much changing $\sigma_r$, then the
disk gets less stable.  Toomre (1964) showed that violent instability sets in
when $Q \equiv \sigma_r/\sigma_{\rm crit} \rightarrow 1$, where 
$\sigma_{\rm crit} = 3.36 G \mu / \kappa$.  Gas inflow increases both the
density and the epicyclic frequency, but density wins and $\sigma_{\rm crit}$ 
increases as the evolution proceeds.  As $Q$ drops toward 1, instabilities
should form and heat the growing pseudobulge in the disk plane.  Toomre (1966)
showed further that buckling instabilities heat the disk vertically if
$\sigma_r$ gets bigger than about 3.3 times the vertical velocity dispersion
even if there is no bar.  So heating in the plane results in heating 
perpendicular to the plane.  Because the density increases rapidly toward the
center, it is unlikely that the result will look box-shaped.  Rather, the
thickness of the pseudobulge is likely to be larger at smaller radii, much like
in a classical bulge.  

      Finally, at radii that are smaller than the disk thickness,
it is no longer relevant that the incoming gas comes from a disk.  There is no
reason why the innermost parts of a pseudobulge should be flattened at all.

\vfill\eject

\vsl\vsss
\ni {\bf 7.2.~Pseudobulges Do Not Have To Be Young}
\vsl

      Most of this paper emphasizes evolution in progress, because this is the
easiest way to see that evolution is happening at all.  However, we do not mean
to create the mistaken impression that pseudobulges must be young or that they
must be made of young stars. 

      Secular evolution is by definition slower than nonequilibrium processes
such as mergers, but it can have timescales that are much shorter than a Hubble
time.~Rates are uncertain, but evolution is thought to be possible on
timescales as short as $\sim$ 5 galactic rotations.  Bars started to be 
abundant at least 5 Gyr ago (Abraham et al.~1999; van den Bergh
et al.~2002; van den Bergh 2002).  Therefore it is possible that secular
evolution built some pseudobulges quickly \gapprox \ts5 Gyr ago and then
stopped.  

      Also, heating by bars can elevate disk stars to scale heights
characteristic of pseudobulges.  These stars can be
as old as the oldest disks.

      So we expect that pseudobulges have a range of stellar population ages
from nearly zero to at least 5 Gyr (e.{\ts}g., Bouwens, Cay\'on, \& Silk (1999).
Much older stellar populations are not out of the question; how much older is
plausible is not known.  The signs
from population studies are mixed.  Some support a continuity from classical
bulges in early-type galaxies to pseudobulges in late-type galaxies.  Others
point to a possible collision with the secular evolution picture.  We defer a
review to Section 8.1.

\vsl\vsss
\ni {\bf 7.3.~Demise of Bars.~II.~Is Pseudobulge Formation Self-Limiting?}
\vsl

      If building a central mass concentration of 5\ts\% -- 10\ts\% of the
disk mass destroys the bar, does secular evolution then stop?  Do we already
know the maximum bulge-to-disk ratio $B/D$ that secular evolution can produce?

      The answer is probably ``no''.  (1) Many SB0 and SBa galaxies have
bulge-to-disk ratios of $\sim 1$.  So bars can coexist with surprisingly
large central mass concentrations.  (2) Once bars grow non-linear, simple
heuristic arguments about how to make a successful bar lose force.
In particular, it may no longer be necessary that the radius of ILR be small.
(3) The simulations that demonstrate bar suicide do not take into
account enough physics.  Many do not include gas.  Almost none allow the bar to
interact with all of the components that we see~in~galaxies.  The competition
between angular momentum sinks that help to strengthen bars and the damaging
effects of pseudobulge building may be weighted more in favor of the angular
momentum sinks than current simulations suggest.  (4) Gravitational tickling
of an unbarred galaxy with an encounter (but not a merger) may re-excite a bar
(Noguchi 1988; Gerin, Combes, \& Athanassoula 1990; Barnes \& Hernquist 1991).

      Pseudobulge building may be self-limiting, but it does not
necessarily stop at the small bulge-to-disk ratios quoted above.

\vfill\eject

\vsl\vsss\vsss\vsss
\ni {\bf 8.~CAVEATS}

\vsl
\ni {\bf 8.1.~Stellar Populations in Classical Bulges and Pseudobulges}
\vsl

      Published studies of stellar populations and metallicities promote quite a
different view of bulges than the one discussed in this paper.  They emphasize
that stellar populations are generally old, metal-rich, and similar to those of
elliptical galaxies.  On the basis of such evidence, many papers suggest that
observations of stellar population are inconsistent with secular evolution and
instead point to early and rapid formation.  Do these results reveal a problem
with the present picture?

     The themes of this section are: (1) Most stellar population
studies concentrate on early-type galaxies; their results are consistent with
our conclusion that these generally contain classical bulges. (2) In contrast,
many Sbc and later-type bulges, plus a few early-type objects with
independent evidence for pseudobulges, do contain young stars.  However, (3)
some galaxies with clearcut evidence for pseudobulges certainly have old stellar
populations.  Point (2) provides some ``wiggle room''; tentatively, we see no
compelling collision between stellar population and secular evolution studies.
But the situation is not clearcut, and a collision is possible.  More than
any other subject discussed in this review, the stellar populations of late-type
(pseudo)bulges need further work to make sure that there is no fundamental
problem or else to uncover it.

      Stellar populations in galaxies are a giant subject; we do not have space
to review it in detail.  Excellent recent reviews are available, including 
Sandage (1986);
Wyse, Gilmore, \& Franx (1997), 
papers in Carollo, Ferguson, \& Wyse (1999),
particularly Renzini (1999), 
and may others.  We concentrate on a few seminal papers that illustrate the
above points.

      Peletier et al.~(1999) studied the $B - I$ versus $I - H$ color-color
diagram of bulges using {\it HST} NICMOS.  High spatial resolution is
important: they show convincingly that dust absorption dominates the central
colors in many galaxies and that colors at the effective radius $r_e$ of the
bulge are much less affected by dust. Their conclusions are as follows.
 
      For 13 bulges of S0{\ts}--{\ts}Sab galaxies, colors at $r_e$ mostly show
little scatter and are consistent with the colors of Coma cluster ellipticals.
The authors conclude that the age spread of these bulges is small, at most 2
Gyr, and that the stars formed \gapprox \ts12 Gyr ago.  Two early-type galaxies
have blue colors indicative of younger ages.  NGC 5854 is classified Sa in
Sandage \& Bedke (1994), who note, ``The inner spiral pattern [of two] is in 
the bulge (it forms the bulge) and has the form and the sense of the opening 
of a stubby S.''  This observation is more consistent with a pseudobulge than
with a classical bulge.  Spiral structure requires a dynamically cold system,
and indeed, the weighted mean of the best velocity dispersion measurements is
$102 \pm 4$ km s$^{-1}$ (Simien \& Prugneil 1997; Vega Beltr\'an et al.~2001;
Falc\'on-Barroso, Peletier, \& Balcells 2002).  
The second ``young'' bulge is in the S0 galaxy NGC 7457; Kormendy (1993) showed that it has an
unusually small velocity dispersion indicative of a pseudobulge.  

      Of four Sb bulges, one (NGC 5443) is young; it is also barred.  The other
three are made of old stars.  It is important to note that these include the
boxy pseudobulge NGC 5746 (Figure 15).  

      Peletier et al.~(1999) note that their three Sbc bulges are ``considerably
bluer, have lower surface brightness, show patchy dust and star formation
together, and are rather different from the rest of the galaxies.''  So the
sample is small, but the results in Peletier et al.~(1999) are consistent
with the present picture that secular evolution dominates late-type bulges but
that early-type bulges, on the whole, are like ellipticals.  One important new 
result is that at least one pseudobulge is made of old stars.
Generally similar results are found in Bica \& Alloin (1987).  

      One stellar population result that is consistent with and even suggestive
of secular evolution is an observed correlation between bulge color and the
color of the adjacent part of the disk (Peletier \& Balcells 1996; Gadotti \&
dos Anjos 2001).  Bulges and disks both show large ranges in colors, but 
``bulges are more like their disk than they are like each other'' (Wyse et
al.~1997). Also, some bulge colors found in the above studies are indicative of
young ages, especially for Sc{\ts}--{\ts}Sm galaxies (de Jong 1996c).  Bulge and
disk scale lengths and surface brightnesses also correlate (see the above
papers; Courteau, de Jong, \& Broeils 1996; Courteau 1996b). Courteau and
collaborators interpret these correlations as products of ``secular dynamical
evolution \dots~via angular momentum transfer and viscous [gas] transport.''  

      Finally, Trager (2004) reviews recent work which shows that bulges of
S0/a{\ts}--{\ts}Sbc spirals span a large range of ages and [$\alpha$-element/Fe]
abundance enhancements.  The latter are a particularly important indicator
of star formation history because $\alpha$-elements are ejected by massive
stars when they explode as supernovae of type II; their abundances are diluted 
with Fe when type I supernovae become important $\sim 1$ Gyr after a starburst.
After that, [$\alpha$-element/Fe] can never again be much greater than solar.
So overabundances of the $\alpha$-elements indicate that almost all of the star
formation occurred quickly (e.{\ts}g., Terndrup 1993; Bender \& Paquet 1995;
Thomas, Greggio, \& Bender 1999; Thomas, Maraston, \& Bender 2002; cf.~Worthey,
Faber, \& Gonzalez 1992). Trager (2004) reviews evidence (e.{\ts}g.,
Jablonka, Martin, \& Arimoto 1996; Proctor \& Sansom 2002; Thomas et al.~2002;
Mehlert et al.~2003) that bulges with high luminosities or velocity dispersions
show $\alpha$-element enhancements but those with low luminosities or velocity
dispersions do not.  For example, many S0 bulges, which tend to be high in
luminosity, tend to have $\alpha$-element enhancements indicative of
rapid formation (e.{\ts}g., Bender \& Paquet 1995; Fisher, Franx, \& Illingworth
1996), although even they show a large range in $\alpha$-element enhancements
and ages (Kuntschner \& Davies 1998; Kuntschner 2000; Kuntschner et al. 2002;
Mehlert et al.~2003).  These results do not ring alarm bells, but they would be
a more decisive test of secular evolution if they included more late-type
galaxies.

      It is well known that a few S0 bulges have post-starburst spectra; these
are more likely to be the result of galaxy accretions than processes that form
either classical or pseudo bulges (e.{\ts}g., 
NGC 4150: Emsellem et al.~2002;
NGC 5102: van den Bergh 1976c; 
          Pritchet 1979; 
          Rocca-Volmerange \& Guideroni 1987; 
          Burstein et al.~1988; 
          Deharveng et al.~1997).

      In summary, stellar population data appear reasonably consistent with
the conclusion of previous sections that S0{\ts}--{\ts}Sb galaxies tend to
have classical bulges and that Sbc{\ts}--{\ts}Sm galaxies usually have 
pseudobulges.  However, (1) the galaxy samples studied are too heavily
weighed toward early-type galaxies to be a decisive test, and (2) there 
certainly exist galaxies that are difficult to understand.  For example,
while the boxy bulge of the S0 galaxy NGC 7332 is likely to be younger than
its disk (Bender \& Paquet 1995), the even more boxy bulge of NGC 5746 appears
to be old (Peletier et al.~1999).

      Finally, the best-studied boxy bulge is the one in our Galaxy.  The
low-absorption field that has received the most attention is Baade's window.
At a Galactic latitude of $-4^\circ$, it is well up into the boxy part of the
bulge revealed by COBE (see Figure 1 in Wyse, Gilmore, \& Franx 1997).  Still,
it is almost along the minor axis, so there is a small chance that the stars 
that define the boxiness are not completely the same as the ones in Baade's 
window.  In any case, the evidence is overwhelmingly that bulge stars far from
the Galactic plane are old (Terndrup 1993; Ortolani et al.~1995; Feltzing \& Gilmore 2000; Kuijken \& Rich 2002; Zoccali et al.~2003, all of which review 
earlier work; see Sandage 1986; Wyse, Gilmore, \& Franx 1997; Renzini 1999; and
Rich 1999 for further review).  The absolute age is uncertain but is
approximately 11 -- 13
Gyr.  Moreover, moderate $\alpha$-element overabundances with respect to iron
(McWilliam \& Rich 1994; Barbuy et al.~1999) again imply rapid star formation
over a period of \lapprox \ts1 Gyr.  Finally, the observed correlation that
more metal-poor bulge/halo stars have more eccentric, plunging orbits continues
to point to an early collapse with accompanying self-enrichment (Eggen,
Lynden-Bell, \& Sandage 1962; Sandage 1986, 1990).  The Galactic bulge is
clearly older than a secular evolution picture can easily accommodate. If we
try to solve this problem by postulating that the boxy structure was made by
heating a pre-existing disk of old stars, then the fact that the bulge and the
metal-poor halo are similar in age becomes a coincidence.  
Also, the Galactic center is currently forming stars at a rate that, if
sustained for an appreciable fraction of a Hubble time, adds up to much of the
stellar density observed there (Rich 1999).  We have argued in this paper for
the importance of secular evolution, but we would be the last to suggest that
the above results are easily understood.  On the other hand, the Galactic bulge
is clearly boxy.  At present, the only model that we have for its origin is via
secular processes. 

\vsl\vsss
\ni {\bf 8.2.~Can Minor Accretion Events Mimic Pseudobulge Growth?}
\vsl

      The answer is surely ``yes''.  Kannappan, Jansen, \& Barton (2004)
find a correlation between blue-centered, star-forming bulges and evidence of 
tidal encounters with neighboring galaxies.  Well known examples are M{\ts}82
and NGC 3077, which are connected to M{\ts}81 by H{\ts}I tidal bridges (Yun, 
Ho, \& Lo 1994).  Counterrotating gas and even stellar components in some
galaxies also imply accretion.  An example is NGC 4826 (Braun et al.~1994;
Rubin 1994; Walterbos et al.~1994; Rix et al.~1995; Garcia-Burillo et al.~2003).
Note, however, that NGC 4826's pseudobulge signature -- a very low stellar
velocity dispersion (Kormendy 1993) -- is a property of corotating stars and 
therefore predates the accretion of counterrotating material and has not yet
been affected by it.  Examples of galaxies with
pseudobulge characteristics that may instead be caused by gas accretions
include NGC 5102 (Section 8.1) and NGC 7457 (Kormendy \& Illingworth 1983;
Kormendy 1993; Peletier et al.~1999).

      There are three reasons why we suggest that secular evolution
accounts for more pseudobulges than do accretion events.  (1) Many of the most 
recognizable pseudobulges occur in strongly barred and oval galaxies,
especially in ones in which radial dust lanes imply that gas infall is ongoing
now.  (2) If galaxies approach each other closely enough to transfer gas, then
their dark matter halos are likely already to overlap and they are likely to
merge after a 
few more orbits.  A configuration like the M{\ts}81 -- M{\ts}82 -- NGC 3077
encounter can last for a billion years but not for a significant fraction of
a Hubble time and not at all without being recognizable.  Most pseudobulge
galaxies show no signs of tidal interactions in progress.  (3) Inhaling a tiny,
gas-rich dwarf does no damage to an existing disk, but a major merger heats
a thin disk too much to be consistent with flat, edge-on galaxies. 

      Nevertheless, the relative importance of internal and externally-driven
secular evolution is not known and needs further study.  It is likely that
accretions create more than an occasional quasipseudobulge.

\vfill\eject

\vsl\vsss
\ni {\bf 8.3.~Disky Distortions in Elliptical Galaxies}
\vsl

      Some ellipticals contain central disky distortions
(Bender, D\"obereiner, \& M\"ollenhoff 1988; see Bender et al.~1989;
Kormendy \& Djorgovski 1989; Bender 1990b; Kormendy \& Bender 1996; Rest et
al.~2001 for reviews).  In the more extreme cases, the line-of-sight velocity 
distributions show asymmetries or even a two-component structure indicative of
a cold, rotating nuclear disk embedded in a more slowly rotating elliptical host
(e.{\ts}g., Franx \& Illingworth 1988; Bender 1990a; Bender, Saglia, \& Gerhard
1994; Scorza \& Bender 1995).  Because these disks are not self-gravitating,
the processes discussed in this paper cannot operate.  Therefore there must be
some embedded nuclear disks in the earliest-type galaxies that are not related
to the themes of this paper.  

      How are they produced?  Minor accretion events in which an elliptical
swallows a gas-rich dwarf almost certainly produce some of them, along
with the central dust disks commonly observed in ellipticals (see Kormendy \&
Djorgovski 1989 for a review and, Jaffe et al. 1994; van Dokkum \&
Franx 1995; Martini et al.~2003a for {\it HST\/} images).  There is evidence
that dust disks gradually turn into small stellar disks (Kormendy et al.~1994, 
2004).  Alternatively, gas shed by dying stars in the elliptical may, in some cases, cool and fall to the center.  These processes are quite different from
secular evolution driven by nonaxisymmetries in dominant, self-gravitating
disks.  

      We do not know whether the tiny nuclear disks seen, for example, in
NGC 3115 (Kormendy \& Richstone 1992; Lauer et al. 1995; Scorza \& Bender 1995;
Kormendy et al.~1996b; Emsellem, Dejonghe, \& Bacon 1999) and NGC 4594 
(Burkhead 1986, 1991; Kormendy 1988b; Wagner, Dettmar, \& Bender 1989;
Crane et al.~1993; Emsellem et al.~1996; Seifert \& Scorza 1996; Kormendy
et al.~1996a) are more nearly related to pseudobulges or to the disky
distortions discussed above.  The dividing line between 
the above processes and those that make pseudobulges deserves further
investigation.  However, we emphasize that this uncertainty affects only a
minority of disky components embedded in the largest, earliest-type classical
bulges.  It is not a problem for the identification of most pseudobulges in Sb
and later-type galaxies.

\vfill\eject

\vsl\vsss
\ni {\bf 9.~CONCLUSION}
\vsl

\ni {\bf 9.1.~A Preliminary Prescription for Recognizing Pseudobulges}
\vsl

      Any prescription must cope with the expected continuum from pure classical
bulges built by mergers and rapid collapse through objects with some E-like and 
some disk-like characteristics to pseudobulges built completely by secular
processes.  Uncertainties are inevitable when we deal with transition objects.
Keeping them in mind, a preliminary list of pseudobulge characteristics
suggested by the previous sections include: \vs

      1 -- The candidate pseudobulge is seen to be a disk in images, e.{\ts}g.,
its apparent flattening is similar to that of the outer disk. \vs

      2 -- It is or it contains a nuclear bar (in relatively face-on 
galaxies). \vs

      3 -- It is box-shaped (in edge-on galaxies). \vs

      4 -- It has S\'ersic index $n \simeq 1$ to 2. \vs

      5 -- It is more rotation-dominated than are the classical bulges in the
$V_{\rm max}/\sigma$ -- $\epsilon$ diagram; e.{\ts}g., 
$(V_{\rm max}/\sigma)^* > 1$. \vs

      6 -- It is a low-$\sigma$ outlier in the Faber-Jackson (1976) correlation 
between (pseudo)bulge luminosity and velocity dispersion.  \vs

      7 -- It is dominated by Population I material (young stars, gas, and
dust), but there is no sign of a merger in progress. \vs

      If any of these characteristics are extreme or very well developed,
it seems safe to identify the central component as a pseudobulge.  The more
of 1 -- 7 apply, the safer the classification becomes.  If several of 1 -- 7
apply but all are relatively subtle, then the central component may be a
pseudobulge or it may be a transition object.

      Small bulge-to-total luminosity ratios $B/T$ do not guarantee that the
galaxy in question contains a pseudobulge, but if $B/T$ \gapprox \ts1/3 to 1/2,
it seems safe to conclude that the galaxy contains a classical bulge.

      Based on these criteria, galaxies with classical bulges include M{\ts}31, 
M{\ts}81, NGC 2841, NGC 3115, and NGC 4594.  Galaxies with prototypical pseudobulges
include 
NGC 1291 (Figures 2, 14),
NGC 1512 (Figures 3, 8),
NGC 1353 (Figure 10), 
NGC 1365 (Figure 7),
NGC 3945 (Figures 5, 14, 17),
NGC 4371 (Figure 17),
NGC 4736 (Figures 2, 8, 17, 20), and
NGC 5377 (Figure 11).
The classification of the bulge of our Galaxy is ambiguous; the observation that
it is box shaped strongly favors a pseudobulge, but the stellar population age 
and $\alpha$-element overabundance are most easily understood if the bulge is 
classical (Section 8.1).

\vfill\eject

\vsl\vsss
\ni {\bf 9.2.~Perspective}
\vsl

      Secular evolution provides a new collection of physical processes that 
we need to take into account when we try to understand galaxies.  Doing so has
already led to significant progress. Thirty years ago, Hubble classification
was in active and successful use.  However, we also knew about a long list of
commonly observed, regular features in disk galaxies, including lenses, boxy
bulges, nuclear bars, and nuclear star clusters, that were not understood and
that were not included in the classification schemes. In addition, we knew 
about uniquely peculiar galaxies (e,{\ts}g., Arp 1966) that were completely
outside the classification process.  Now, almost all of the common features 
and peculiar galaxies have candidate explanations within one of two
paradigms of galaxy evolution that originated in the late 1970s.  The peculiar
objects have turned out mostly to be interacting and merging galaxies.  And many
of the previously unexplained but common features of disk galaxies now are
fundamental to our growing realization that galaxies continue to evolve
secularly after the spectacular fireworks of galaxy mergers, dissipative
collapse, and their attendant nuclear activity have died down.

\vsl\vsss
\ni {ACKNOWLEDGMENTS}
\vsl

      We are indebted to
Ron Buta and
Jerry Sellwood, 
to scientific editors Geoffrey Burbidge and Allan Sandage,
and to the
Nuker team, especially
Sandy Faber and
Scott Tremaine,
for penetrating comments on the draft that resulted in important improvements
in the final paper.  We thank
Ralf Bender,
Martin Bureau,
Andi Burkert,
Leo Blitz,
Ken Freeman,
Sheila Kannappan,
Mike Rich, and
Guy Worthey
for helpful discussions or for permission to quote results before publication. 
We are exceedingly grateful to Ron Buta and Marcello Carollo for making
available many of the digital images used in the construction of figures.
Additional figures were kindly made available by 
Lia Athanassoula,
Martin Bureau, 
Ortwin Gerhard, and
Linda Sparke.
The color image of NGC 1326 in Figure 8 was constructed for this paper by
Zolt Levay of STScI from $UBVRI$ and $H\alpha$, {\it HST\/} PC images 
provided by Ron Buta.  We are most grateful to Mary Kormendy for extensive
editorial help.  Also, we thank Mark Cornell for technical support and for
permission to quote results before publication.
This paper has made extensive use of the NASA/IPAC Extragalactic Database 
(NED), which is operated by JPL and Caltech under contract with NASA.

      Approximately 50 references were removed from the published version 
because of editorial pressure to shorten this paper.  We regret this, and
we include these references here.

\eject

\magnification = \magstep0
\nopagenumbers



\hsize=11.3truecm  \hoffset=3.0truecm  \vsize=18.6truecm  \voffset=2.5truecm

\def\dblbaselines{\baselineskip=8.1pt    \lineskip=0pt   \lineskiplimit=0pt}

\font\sc=cmr8
\font\it=cmti8
\def\t#1{#1} 
\def\t#1{\empty}
  \def\s{\null}  
\def\vsl{\vskip\baselineskip}           \def\vs{\vskip 10pt}
\parskip = 0pt 
\def\ts{\thinspace} \def\cl{\centerline}
\def\ni{\noindent}  \def\nhi{\noindent \hangindent=0.3cm}

\def\makeheadline{\vbox to 0pt{\vskip-30pt\line{\vbox to8.5pt{}\the
                               \headline}\vss}\nointerlineskip}
\def\makeheadline{\vbox to 0pt{\vskip-30pt\fullline{\vbox to8.5pt{}\the
                               \headline}\vss}\nointerlineskip}
\def\toppageno{\headline={\hss\tenrm\folio\hss}}
\def\footnoterule{\kern-3pt \hrule width \hsize \kern 2.6pt \vskip 3pt}
\output={\plainoutput}
\pretolerance=9000   \tolerance=9000  \linepenalty=10000 
\def\sup1{$^{\rm 1}$} \def\sup2{$^{\rm 2}$}
\def\r0{$\rho_0$}  
\def\00{$\phantom{000000}$} \def\0{\phantom{0}} 
\def\1{\phantom{1}}         
\def\etal{{\it et~al.~}}
\def\gapprox{$_>\atop{^\sim}$} \def\lapprox{$_<\atop{^\sim}$}
          
\newdimen\sa  \def\sd{\sa=.1em \ifmmode $\rlap{.}$''$\kern -\sa$
                               \else \rlap{.}$''$\kern -\sa\fi}
\def\ss{\ifmmode ^{\prime\prime}$\kern-\sa$ \else $^{\prime\prime}$\kern-\sa\fi}
\def\mm{\ifmmode ^{\prime}$\kern-\sa$ \else $^{\prime}$\kern-\sa \fi}
\def\msun {M$_{\odot}$~} 
  
\def\m31{M{\ts}31}


\sc
\dblbaselines


\newdimen\fullhsize
\fullhsize=11.0 cm 
\hsize=5.25 cm
\def\fullline{\hbox to \fullhsize}
\let\lr=L \newbox\leftcolumn
\output={\if L\lr
         \global\setbox\leftcolumn=\columnbox \global \let\lr=R
         \else  \doubleformat \global \let\lr=L\fi
         \ifnum\outputpenalty>-2000 \else \dosupereject\fi}
\def\doubleformat{\shipout\vbox{\makeheadline
         \fullline{\box\leftcolumn\hfil\columnbox}
         \makefootline}
\advancepageno}
\def\columnbox{\leftline{\pagebody}}


\ni {LITERATURE CITED}
\vsl

\frenchspacing

\hfuzz = 5pt


\nhi Abraham RG, Merrifield MR, Ellis RS, Tanvir NR, Brinchmann J. 1999.
     {\it MNRAS} 308:569 


\nhi Aguerri JAL, Debattista VP, Corsini EM. 2003. {\it MNRAS} 338:465 


\nhi Alonso-Herrero A, Knapan JH. 2001. {\it Astron. J.} 122:1350

\nhi Alonso-Herrero A, Ryder SD, Knapen JH. 2001. {\it MNRAS} 322:757


\nhi Andersen DR, Bershady MA, Sparke LS, Gallagher JS, 
     Wilcots EM. 2001. {\it Ap.~J.} 551:L131

\nhi Andredakis YC. 1998. {\it MNRAS} 295:725

\nhi Andredakis YC, Peletier RF, Balcells M. 1995. {\it MNRAS} 275:874

\nhi Andredakis YC, Sanders RH. 1994. {\it MNRAS} 267:283

\nhi Aronica G, Athanassoula E, Bureau M, Bosma A, Dettmar R-J, et al. 2003.
     {\it Astrophys. Space Sci.} 284:753

\nhi Arp H. 1966. {\it Atlas of Peculiar Galaxies}.  Pasadena: California
     Inst.~of Technology

\nhi Arsenault R, Boulesteix J, Georgelin Y, Roy J-R. 1988.
      {\it Astron. Astrophys.} 200:29

\nhi Athanassoula E. 1992a. {\it MNRAS} 259:328 

\nhi Athanassoula E. 1992b. {\it MNRAS} 259:345 

\nhi Athanassoula E. 2002. {\it Astrophys. Space Sci.} 281:39


\nhi Athanassoula E. 2003. {\it MNRAS} 341:1179

\nhi Athanassoula E, Bosma A, Cr\'ez\'e M, Schwarz MP. 1982. {\it Astron. 
     Astrophys.} 107:101


\nhi Athanassoula E, Misiriotis A. 2002. {\it MNRAS} 330:35

\nhi Bacon R, Bureau M, Cappellari M, Copin Y, Davies R, et al. 2002. In
     {\it Galaxies: The Third Dimension}, ed. M Rosado, L Binette, L Arias,
     p. 179.  San Francisco: Astron. Soc. Pac. (astro-ph/0204129)

\nhi Balcells{\ts\ts\ts}M.{\ts\ts\ts}2001. In {\it The Central Kiloparsec of
     Starbursts and AGN: The La Palma Connection}, ed. JH Knapen, JE Beckman,
     I Schlosman,  TJ Mahoney, p. 140.  San Francisco: ASP

\nhi Balcells M. 2003. In {\it JENAM 2002 Galaxy Dynamics Workshop}. ESA.
     (astro-ph/0301647)

\nhi Balcells M, Graham AW, Dom\'\i nguez-Palmero L, Peletier RF. 2003. {\it Ap. J.}
     582:L79



\nhi Barbuy B, Renzini A, Ortolani S, Bica E, Guarnieri MD. 1999. 
     {\it Astron. Astrophys.} 341:539

\nhi Barnes JE, Hernquist LE. 1991. {\it Ap. J.} 370:L65



\nhi Bender R. 1990a. {\it Astron. Astrophys.} 229:441

\nhi Bender R. 1990b. In {\it Dynamics and Interactions of Galaxies}, 
     ed. R Wielen, p. 232. Berlin: Springer

\nhi Bender R, Burstein D, Faber SM. 1992. {\it Ap. J.} 399:462

\nhi Bender R, Burstein D, Faber SM. 1993. {\it Ap. J.} 411:153

\nhi Bender R, D\"obereiner S, M\"ollenhoff~C. 1988. 
     {\it Astron. Astrophys. Suppl.} 74:385

\nhi Bender R, Paquet A. 1995. In {\it IAU Symposium 164, Stellar Populations},
     ed. PC van der Kruit, G Gilmore, p. 259.  Dordrecht: Kluwer

\nhi Bender R, Saglia RP, Gerhard OE. 1994. {\it MNRAS} 269:785

\nhi Bender R, Surma P, D\"obereiner S, M\"ollenhoff C, Madejsky R. 1989.
     {\it Astron. Astrophys.} 217:35

\nhi Benedict GF, Higdon JL, Jefferys WH, Duncombe R, 
     Hemenway PD, et al. 1993. {\it Astron. J.} 105:1369

\nhi Benedict GF, Higdon JL, Tollestrup EV,
     Hahn JM, Harvey PM. 1992. {\it Astron. J.} 103:757

\nhi Benedict GF, Howell DA, J\o rgensen I, Kenney JDP, Smith BJ. 2002.
     {\it Astron. J.} 123:1411

\nhi Benedict GF, Smith BJ, Kenney JDP. 1996. {\it Astron. J.} 111:1861

\nhi Berentzen I, Heller CH, Shlosman I, Fricke KJ. 1998. {\it MNRAS} 300:49

\nhi Bertin G. 1983. {\it Astron. Astrophys.} 127:145



\nhi Bertola F, Capaccioli M. 1975. {\it Ap.~J.} 200:439

\nhi Bertola F, Capaccioli M. 1977. {\it Ap.~J.} 211:697


\nhi Bettoni D, Galletta G. 1994. {\it Astron. Astrophys.} 281:1


\nhi Bica E, Alloin D. 1987. {\it Asr.~Ap.~Suppl.} 70:281

\nhi Bica E, Alloin D, Schmidt AA. 1990. {\it Astron. Astrophys.} 228:23

\nhi Binggeli B, Barazza F, Jerjen H. 2000. {\it Astron. Astrophys.} 359:447

\nhi Binggeli B, Jerjen H. 1998. {\it Astron. Astrophys.} 333:17

\nhi Binggeli B, Sandage A, Tammann GA. 1985. {\it Astron.~J.} 90:1681 
 
\nhi Binggeli B, Sandage A, Tarenghi M. 1984. {\it Astron.~J.} 89:64 

\nhi Binney J. 1976. {\it MNRAS} 177:19
 
\nhi Binney J. 1978a. {\it MNRAS} 183:501
 
\nhi Binney J. 1978b. {\it Comments Ap.} 8:27
 
\nhi Binney J. 1980. {\it MNRAS} 190: 873 

\nhi Binney J. 1982. In {\it Morphology and Dynamics of Galaxies, Twelfth 
     Advanced Course of the Swiss Society of Astronomy and Astrophysics}, 
     ed. L Martinet, M Mayor, p.~1. Geneva Observatory: Sauverny

\nhi Binney J, Gerhard O. 1993. In {\it Back to the Galaxy}, ed. SS Holt,
     F Verter, p. 87.  Ney York: AIP

\nhi Binney J, Gerhard OE, Stark AA, Bally J, Uchida KI. 1991. {\it MNRAS}
     252:210



\nhi Binney J, Tremaine S. 1987. {\it Galactic Dynamics}. Princeton:
     Princeton Univ.~Press

\nhi Binney JJ, Davies RL, Illingworth GD. 1990. {\it Ap. J.} 361:78

\nhi Blitz L, Spergel DN. 1991. {\it Ap. J.} 379:631

\nhi Block DL, Bertin G, Stockton A, Grosb\o l P, Moorwood AFM, Peletier RF.
     1994. {\it Astron. Astrophys.} 288:365

\nhi Block DL, Buta R, Puerari I, Knapen JH, Elmegreen BG,
     et al. 2002. In {\it The Dynamics, Structure and History of Galaxies},
     ed. GS Da Costa, H Jerjen, p.~97.  San Francisco: Astron. Soc. Pac.

\nhi Block DL, Puerari I, Knapen JH, Elmegreen BG, Buta R, et al. 2001.
     {\it Astron. Astrophys.} 375:761

\nhi Block DL, Wainscoat RJ. 1991. {\it Nature} 353:48


\nhi B\"oker T, F\"orster-Schreiber NM, Genzel R. 1997. {\it Astron. J.}
      114:1883  

\nhi B\"oker T, Laine S, van der Marel RP, Sarzi M, Rix H-W, et al. 2002.
      {\it Astron.~J.} 123:1389

\nhi B\"oker T, Lisenfeld U, Schinnerer E. 2003a. {\it Astron.~Astrophys.}
      406:87  

\nhi B\"oker T, Sarzi M, McLaughlin DE, van der Marel RP, Rix H-W, et al.
      2003b. {\it Astron. J.} 127:105  

\nhi B\"oker T, Stanek R, van der Marel RP. 2003c. {\it Astron. J.}
      125:1073  

\nhi B\"oker T, van der Marel RP, Mazzuca J, Rix, H-W, Rudnick G, et al. 2001.
      {\it Astron. J.} 121:1473

\nhi B\"oker T, van der Marel RP, Vacca WD. 1999. {\it Astron. J.}
      118:831 

\nhi Boroson T. 1981. {\it Ap.~J.~Suppl.} 46:177

\nhi Boselli A, Lequeux J, Gavazzi G. 2002. {\it Astron. Astrophys.} 384:33

\nhi Bosma A. 1981a. {\it Astron. J.} 86:1791

\nhi Bosma A. 1981b. {\it Astron. J.} 86:1825


\nhi Bosma A. 1992. In {\it Morphological and Physical Classification of 
     Galaxies}, ed. G Longo, M Capaccioli, G Busarello, p.~207.
     Dordrecht: Kluwer

\nhi Bosma A, Ekers RD, Lequeux J. 1977a. {\it Astron.~Astrophys.} 57:97

\nhi Bosma A, van der Hulst JM, Sullivan WT. 1977b. 
     {\it Astron.~Astrophys.} 57:373

\nhi Bouwens R, Cay\'on L, Silk J. 1999. {\it Ap. J.} 516:77

\nhi Braine J, Combes F, Casoli F, Dupraz C, G\'erin M, et al. 1993.
     {\it Astron. Astrophys. Suppl.} 97:887

\nhi Braun R, Walterbos RAM, Kennicutt RC, Tacconi LJ. 1994. {\it Ap. J.} 
     420:558

\nhi Bresolin F, Kennicutt RC. 1997. {\it Ap.J.} 113:975

\nhi Brinchmann J, Charlot S, White SDM, Tremonti C, Kauffmann G, et al. 2003.
     {\it MNRAS}, in press (astro-ph/0311060)

\nhi Burbidge EM, Burbidge GR. 1959. {\it Ap. J.} 130:20

\nhi Burbidge EM, Burbidge GR. 1960. {\it Ap. J.} 132:30

\nhi Burbidge EM, Burbidge GR. 1962. {\it Ap. J.} 135:366


\nhi Bureau M, Freeman KC. 1999. {\it Astron. J.} 118:126

\nhi Burkhead MS. 1986. {\it Astron. J.} 91:777

\nhi Burkhead MS. 1991. {\it Astron. J.} 102:893


\nhi Burstein D, Bender R, Faber SM, Nolthenius R. 1997. 
     {\it Astron. J.} 114:1365

\nhi Burstein D, Bertola F, Buson LM, Faber SM, Lauer TR. 1988. {\it Ap. J.}
     328:440


\nhi Buta R. 1986a. {\it Ap. J. Suppl.} 61:609

\nhi Buta R. 1986b. {\it Ap. J. Suppl.} 61:631

\nhi Buta R. 1988. {\it Ap. J. Suppl.} 66:233

\nhi Buta R. 1990. {\it Ap.~J.} 351:62

\nhi Buta R. 1995. {\it Ap.~J.~Suppl.} 96:39

\nhi Buta R. 1999. {\it Astrophys. Space Sci.} 269-270:79

\nhi Buta R. 2000. In {\it Toward a New Millennium in Galaxy Morphology},
     ed. DL Block, I Puerari, A Stockton, D Ferreira, p.~79.
     Dordrecht: Kluwer

\nhi Buta R, Block DL. 2001. {\it Ap. J.} 550:243

\nhi Buta R, Combes F. 1996. {\it Fund. Cosmic Phys.} 17:95

\nhi Buta R, Corwin HG, Odewahn SC. 2003.  {\it The de Vaucouleurs
     Atlas of Galaxies}, in preparation. Cambridge: Cambridge Univ.~Press

\nhi Buta R, Crocker DA. 1991. {\it Astron. J.} 102:1715

\nhi Buta R, Crocker DA. 1993. {\it Astron. J.} 105:1344

\nhi Buta R, Crocker DA, Byrd GG. 1999. {\it Astron. J.} 118:2071 

\nhi Buta R, Crocker DA, Elmegreen BG, ed. 1996. {\it IAU Colloquium
     157, Barred Galaxies\/}.  San Francisco: ASP

\nhi Buta R, Purcell GB. 1998. {\it Astron. J.} 115:484

\nhi Buta R, Purcell GB, Cobb ML, Crocker DA, Rautiainen P,
     Salo H. 1999. {\it Astron. J.} 117:778  

\nhi Buta R, Ryder SD, Madsen GJ, Wesson K, Crocker DA, 
     Combes F. 2001. {\it Astron. J.} 121:225 

\nhi Buta R, Treuthardt PM, Byrd GG, Crocker DA. 2000.
     {\it Astron. J.} 120:1289 

\nhi Byrd G, Rautiainen P, Salo H, Buta R, Crocker DA. 1994.
     {\it Astron. J.} 108:476

\nhi Caon N, Capaccioli M, D'Onofrio M. 1993. {\it MNRAS} 265:1013



\nhi Carlberg RG. 1987. In {\it Nearly Normal Galaxies: From the Planck Time
     to the Present}, ed. SM Faber, p. 131. Berlin: Springer

\nhi Carollo CM. 1999. {\it Ap.~J.} 523:566

\nhi Carollo CM. 2003. In {\it Carnegie Observatories Astrophysics Series,
     Vol. 1: Coevolution of Black Holes and Galaxies}, ed. LC Ho, In press.
     Cambridge: Cambridge Univ. Press (astro-ph/0306021)

\nhi Carollo CM, Ferguson HC, Wyse RFG, ed. 1999. {\it The Formation
     of Galactic Bulges}. Cambridge: Cambridge Univ.~Press

\nhi Carollo CM, Stiavelli M. 1998. {\it Astron. J.} 115:2306 

\nhi Carollo CM, Stiavelli M, de Zeeuw PT, Mack J. 1997 
     {\it Astron. J.} 114:2366

\nhi Carollo CM, Stiavelli M, de Zeeuw PT, Seigar M,
     Dejonghe H. 2001. {\it Ap. J.} 546:216

\nhi Carollo CM, Stiavelli M, Mack J. 1998. {\it Astron. J.} 116:68

\nhi Carollo CM, Stiavelli M, Seigar M, de Zeeuw PT,
     Dejonghe H. 2002. {\it Astron. J.} 123:159

\nhi Cole S, Lacey C. 1996. {\it MNRAS}, 281:716

\nhi Colina L, Garc\'\i a Vargas ML, Mas-Hesse JM, Alberdi A, 
     Krabbe A. 1997. {\it Ap. J.} 484:L41

\nhi Colina L, Wada K. 2000. {\it Ap.~J.} 529:845

\nhi Combes F. 1991. In {\it IAU Symposium 146, Dynamics of Galaxies
      and Their Molecular Cloud Distributions}, ed. F Combes, F Casoli, p.~255
      Dordrecht: Kluwer

\nhi Combes F. 1994. In {\it Mass-Transfer Induced Activity in Galaxies},
     ed. I Shlosman, p. 170. Cambridge: Cambridge Univ. Press

\nhi Combes F. 1998. In {\it IAU Symposium 184, The Central Regions of the 
      Galaxy and Galaxies}, ed. Y Sofue, p.~257.  Dordrecht: Kluwer

\nhi Combes F. 2000. In {\it Dynamics of Galaxies: From the Early Universe
     to the Present}, ed. F Combes, GA Mamon, V Charmandaris, p.~15.  
     San Francisco: Astron.~Soc.~Pacific

\nhi Combes F. 2001. In {\it Galaxy Disks and Disk Galaxies}, ed. JG Funes,
      EM.~Corsini, p.~213. (San Francisco: Astron. Soc. Pac.)


\nhi Combes F, Debbasch F, Friedli D, Pfenniger D. 1990.
      {\it Astron. Astrophys.} 233:82

\nhi Combes F, Garc\'\i a-Burillo S, Boone F, Hunt LK, Baker AJ., et al. 2003. 
     (astro-ph/0310652)

\nhi Combes F, Gerin M. 1985. {\it Astron. Astrophys.} 150:327

\nhi Combes F, Gerin M., Nakai N, Kawabe R, Shaw MA. 1992. 
     {\it Astron. Astrophys.} 259:L27

\nhi Combes F, Sanders RH. 1981. {\it Astron. Astrophys.} 96, 164

\nhi Conselice CJ, Bershady MA, Dickinson M, Papovich C. 2003. 
     {\it Astron. J.} 126:1183

\nhi Contini T, Wozniak H, Consid\`ere S, Davoust E. 1997.
     {\it Astron. Astrophys.} 324:41

\nhi Contopoulos G. 1980. {\it Astron. Astrophys.} 81:198


\nhi Contopoulos G, Mertzanides C. 1977. {\it Astron.~Astrophys.} 61:477

\nhi Copin Y, Cretton N, Emsellem E. 2004. {\it Astron. Astrophys.} 415:889

\nhi Corsini EM, Debattista VP, Aguerri JAL. 2003. {\it Ap. J.} 599:L29

\nhi Corsini EM, Aguerri JAL, Debattista VP. 2003. In {\it IAU Symposium 220,
     Dark Matter in Galaxies}, ed. S Ryder, DJ Pisano, M Walker, KC Freeman,
     in press. (astro-ph/0311042)

\nhi Courteau S. 1996a. {\it Ap. J. Suppl.} 103:363

\nhi Courteau S. 1996b. In {New Extragalactic Perspectives in the New South
     Africa}, ed. DL Block, JM Greenberg, p. 255.  Dordrecht: Kluwer

\nhi Courteau S, de Jong RS, Broeils AH. 1996. {\it Ap. J.} 457:L73

\nhi Crane P, Stiavelli M, King IR, Deharveng JM, Albrecht R, et al. 1993.
     {\it Astron. J.} 106:1371

\nhi Crosthwaite LP, Turner JL, Buchholz L, Ho PTP, Martin RN, 2002.
     {\it Astron. J.} 123:1892

\nhi Curran SJ, Polatidis AG, Aalto S, Booth RS. 2001a.
     {Astron. Astrophys.} 368:824

\nhi Curran SJ, Polatidis AG, Aalto S, Booth RS. 2001b.
     {Astron. Astrophys.} 373:459

\nhi Dalcanton JJ, Bernstein RA. 2002. {\it Astron. J.} 124:1328


\nhi Davidge TJ, Courteau S. 2002. {\it Astron. J.}  123:1438


\nhi Davies RL, Efstathiou G, Fall SM, Illingworth G,
      Schechter PL. 1983 {\it Ap.~J.} 266:41

\nhi Debattista VP, Corsini EM, Aguerri JAL. 2002. {\it MNRAS} 332:65

\nhi Debattista VP, Williams TB. 2001. In {\it Galaxy Disks and Disk Galaxies},
     ed. JG Funes, EM Corsini, p. 553. San Francisco: ASP 

\nhi de Carvalho RR, da Costa LN. 1987. {\it Astron. Astrophys.} 171:66

\nhi Deharveng J-M, Jedrzejewski R, Crane P, Disney MJ, Rocca-Volmerange B.
     1997. {\it Astron. Astrophys.} 326:528

\nhi de Jong RS. 1996a. {\it Astron. Astrophys. Suppl.} 118:557

\nhi de Jong RS. 1996b. {\it Astron. Astrophys.} 313:45

\nhi de Jong RS. 1996c. {\it Astron. Astrophys.} 313:377

\nhi de Jong RS, van der Kruit PC. 1994. {\it Astron. Astrophys. Suppl.} 106:451

\nhi de Souza RE, dos Anjos S. 1987. {\it Astron. Astrophys. Suppl.} 70:465

\nhi de Vaucouleurs G. 1948. {\it Ann.~Astrophys.} 11:247


\nhi de Vaucouleurs G. 1959. {\it Handbuch der Physik} 53:275


\nhi de Vaucouleurs G. 1963. {\it Ap.~J.~Suppl.} 8:31

\nhi de Vaucouleurs G. 1974a. In {\it IAU Symposium 58, The Formation and 
     Dynamics of Galaxies}, ed. JR Shakeshaft, p. 1. Dordrecht: Reidel

\nhi de Vaucouleurs G. 1974b. In {\it IAU Symposium 58, The Formation and 
     Dynamics of Galaxies}, ed. JR Shakeshaft, p. 335. Dordrecht: Reidel

\nhi de Vaucouleurs G. 1975. {\it Ap. J. Suppl.} 29:193



\nhi de Vaucouleurs G, de Vaucouleurs A, Corwin HG, Buta RJ,
     Paturel G, Fouqu\'e P. 1991. {\it Third Reference Catalogue of Bright
     Galaxies}. Berlin: Springer

\nhi de Vaucouleurs G, Freeman KC. 1972. {\it Vistas Astron.} 14:163 


\nhi Devereux N. 1987. {\it Ap. J.} 323:91

\nhi Devereux NA, Kenney JDP, Young JS. 1992. {\it Astron. J.} 103:784



\nhi de Zeeuw PT, Bureau M, Emsellem E, Bacon R, Carollo CM, et al. 2002. 
     {\it MNRAS} 329:513


\nhi D\'\i az RJ, Dottori H, Vera-Villamizar N, Carranza G. 2002.
     In {\it Disks of Galaxies: Kinematics, Dynamics and Perturbations},
     ed. E Athanassoula, A Bosma, R Mujica, p.~278 (San Francisco: Astron.
     Soc. Pac.

\nhi D\'\i az RJ, Pagel BEJ, Edmunds MG, Phillips MM. 1982. {\it MNRAS} 201:49P

\nhi Djorgovski S, Davis M. 1987. {\it Ap. J.} 313: 59

\nhi Djorgovski S, de Carvalho R, Han M-S. 1988. In {\it The Extragalactic
      Distance Scale}, ed. S van den Bergh, CJ Pritchet, p. 329. San 
      Francisco: Astron. Soc. Pac.

\nhi D'Onofrio M, Capaccioli M, Caon N. 1994. {\it MNRAS} 271:523

\nhi D'Onofrio M, Capaccioli M, Merluzzi P, Zaggia S, Boulesteix J. 1999. 
     {\it Astron. Astrophys. Suppl.} 134:437


\nhi Dressler A, Lynden-Bell D, Burstein D, Davies RL, Faber SM, et al. 1987.
     {\it Ap. J.} 313:42


\nhi Duus A, Freeman KC. 1975. In {\it La Dynamique des Galaxies Spirales},
     ed. L Weliachew p.~419.  Paris: CNRS

 

\nhi Dwek E, Arendt RG, Hauser MG, Kelsall T, Lisse CM, et al. 1995.
     {\it Ap. J.} 445:716

\nhi Eggen OJ, Lynden-Bell D, Sandage AR. 1962. {\it Ap. J.} 136:748

\nhi Elmegreen BG. 1996. In {\it Barred Galaxies}, ed. R Buta, DA Crocker, 
     BG Elmegreen, p. 197. San Francisco: ASP

\nhi Elmegreen BG, Elmegreen DM. 1985. {\it Ap. J.} 288:438

\nhi Elmegreen BG, Elmegreen DM, Brinks E, Yuan C, Kaufman M, et al. 1998.
     {\it Ap. J.} 503:L119

\nhi Elmegreen BG, Wilcots E, Pisano DJ. 1998. {\it Ap. J.} 494:L37

\nhi Elmegreen D, Chromey F, McGrath EJ, Ostenson JM. 2002. {\it Astron. J.} 
     123:1381

\nhi Elmegreen D, Chromey F, Santos M, Marshall D. 1997. {\it Astron. J.} 
     114:1850

\nhi Elmegreen D, Chromey FR, Sawyer JE, Reinfeld EL. 1999. {\it Astron. J.} 
     118:777


\nhi Elmegreen DM, Elmegreen BG, Chromey FR, Hasselbacher DA, Bissell BA. 1996.
     {\it Astron. J.} 111:1880

\nhi Emsellem E, Bacon R, Monnet G, Poulain P. 1996. {\it Astron. Astrophys.}
     312:777

\nhi Emsellem E, Davies R, McDermid R, Kuntschner H, Peletier R, et al. 2002.
     In {\it Galaxies: The Third Dimension}, ed. M Rosado, L Binette, L Arias,
     p. 189. San Francisco: ASP

\nhi Emsellem E, Dejonghe H, Bacon R. 1999. {\it MNRAS} 303:495

\nhi Emsellem E, Greusard D, Combes F, Friedli D, Leon S, P\'econtal E, 
     Wozniak H. 2001. {\it Astron. Astrophys.} 368:52


\nhi Englmaier P, Gerhard O. 1997. {\it MNRAS} 287:57

\nhi Erwin P. 2004. {\it Astron. Astrophys.} 415:941 

\nhi Erwin P, Sparke LS. 1999. {\it Ap. J.} 521:L37

\nhi Erwin P, Sparke LS. 2002. {\it Astron. J.} 124:65

\nhi Erwin P, Sparke LS. 2003. {\it Ap. J. Suppl.} 146:299

\nhi Erwin P, Vega Beltr\'an JC, Graham AW, Beckman JE. 2003. {\it Ap. J.}
     597:929

\nhi Eskridge PB, Frogel JA, Pogge RW, Quillen AC, Berlind AA. 2002. 
     {\it Ap. J. Suppl.} 143:73

\nhi Eskridge PB, Frogel JA, Pogge RW, Quillen AC, Davies RL, et al. 2000.
     {\it Astron. J.} 119:536

\nhi Eskridge PB, Frogel JA, Taylor VA, Windhorst RA, 
     Odewahn SC, et al. 2003. {\it Ap. J.} 586:923

\nhi Evans DS. 1951. {\it MNRAS} 111:526


\nhi Faber SM, Dressler A, Davies RL, Burstein D, Lynden-Bell D, et al. 1987.
     In {\it Nearly Normal Galaxies: From the Planck Time to the Present}, 
     ed. SM Faber, p. 175. Berlin: Springer

\nhi Faber SM, Jackson RE. 1976. {\it Ap.~J.} 204:668
 


\nhi Falc\'on-Barroso J, Bacon R, Bureau M, Cappellari M, Davies RL, et al.
     2004. {\it Astron. Nachr.} 325:92 (astro-ph/0311078)

\nhi Falc\'on-Barroso J, Peletier RF, Balcells M. 2002. {\it MNRAS} 335:741

\nhi Fall SM. 1981. See Kormendy 1982a, p. 230.

\nhi Fathi K, Peletier RF. 2003. {\it Astron. Astrophys.} 407:61

\nhi Feltzing S, Gilmore G. 2000. {\it Astron. Astrophys.} 355:949

\nhi Fisher D, Franx M, Illingworth G. 1996. {\it Ap. J.} 459:110

\nhi Fisher D, Illingworth G, Franx M. 1994. {\it Astron. J.} 107:160

\nhi Forbes DA, Kotilainen JK, Moorwood AFM. 1994b. {\it Ap. J.} 
     433:L13

\nhi Forbes DA, Norris RP, Williger GM, Smith RC. 1994a.
     {\it Astron. J.} 107:984

\nhi Franx M, de Zeeuw T. 1992. {\it Ap.~J.} 392:L47

\nhi Franx M, Illingworth GD. 1988. {\it Ap. J.} 327:L55

\nhi Freeman KC. 1970a. {\it Ap.~J.} 160:811

\nhi Freeman KC. 1970b. In {\it IAU Symposium 38, The Spiral Structure of
     Our Galaxy}, ed. W Becker, G Contopoulos, p.~351.  Dordrecht: Reidel

\nhi Freeman KC. 1975.  In {\it IAU Symposium 69, Dynamics of Stellar
     Systems}, ed. A Hayli, p.~367.  Dordrecht: Reidel

\nhi Freeman KC. 2000. In {\it Toward a New Millennium in Galaxy Morphology},
     ed. DL Block, I Puerari, A Stockton, D Ferreira, p.~119.
     Dordrecht: Kluwer

\nhi Frei Z, Guhathakurta P, Gunn JE, Tyson JA. 1996. {\it Astron. J.}
     111:174

\nhi Frenk CS, White SDM, Davis M, Efstathiou G. 1988.
     {\it Ap.~J.} 327:507

\nhi Friedli D. 1996. In {\it Barred Galaxies}, ed. R Buta, DA Crocker,
     BG Elmegreen, p. 378. San Francisco: ASP

\nhi Friedli D. 1999. In {\it The Evolution of Galaxies on Cosmological 
     Timescales}, ed. JE Beckman, TJ Mahoney, p. 88. San Francisco: 
     Astron. Soc. Pacific

\nhi Friedli D, Benz W. 1993. {\it Astron. Astrophys.} 268:65

\nhi Friedli D, Benz W. 1995. {\it Astron. Astrophys.} 301:649

\nhi Friedli D, Martinet L. 1993. {\it Astron. Astrophys.} 277:27


\nhi Friedli D, Pfenniger D. 1991. In {\it IAU Symposium 146, Dynamics of 
     Galaxies and Their Molecular Cloud Distributions}, ed. F Combes, 
     F Casoli, p. 362.  Dordrecht: Kluwer


\nhi Friedli D, Wozniak H, Rieke M, Martinet L, Bratschi P. 1996. 
     {\it Astron. Astrophys. Suppl.} 118:461


\nhi Gadotti DA, dos Anjos S. 2001. {\it Astron. J.} 122:1298


\nhi Gallagher JS, Goad JW, Mould J. 1982. {\it Ap.~J.} 263:101


\nhi Gao Y, Solomon PM. 2004. {\it Ap. J. Suppl.} in press (astro-ph/0310341)

\nhi Garc\'\i a-Barreto JA, Dettmar R-J, Combes F, Gerin M, Koribalski B. 1991.
     {\it Rev. Mexicana Astron. Astrof.} 22:197

\nhi Garc\'\i a-Burillo S, Combes F, Hunt LK, Boone F, Baker AJ, et al. 2003.
     {\it Astron. Astrophys.} 407:485

\nhi Garc\'\i a-Burillo S, Combes F, Neri R. 1999. {\it Astron. Astrophys.} 
     343:740

\nhi Garc\'\i a-Burillo S, Sempere MJ, Combes F, Hunt LK, Neri R. 2000.
  {\it Astron. Astrophys.} 363:869

\nhi Garc\'\i a-Burillo S, Sempere MJ, Combes F, Neri R. 1998. 
     {\it Astron. Astrophys.} 333:864

\nhi Geha M, Guhathakurta P, van der Marel RP. 2002. {\it Astron. J.} 124:3073







\nhi Gerin M, Combes F, Athanassoula E. 1990. {\it Astron. Astrophys.} 230:37

\nhi Gerin M, Nakai N, Combes F. 1988. {\it Astron. Astrophys.} 203:44

\nhi Gerssen J. 2002. In {\it Disks of Galaxies: Kinematics, Dynamics and
     Perturbations}, ed. E Athanassoula, A Bosma, R Mujica, p. 197.
     San Francisco: ASP

\nhi Gerssen J, Kuijken K, Merrifield MR. 1999. {\it MNRAS} 306:926

\nhi Gerssen J, Kuijken K, Merrifield MR. 2003. {\it MNRAS} 345:261

\nhi Gnedin OY, Goodman J, Frei Z. 1995. {\it Astron. J.} 110:1105

\nhi Goad JW, Roberts MS. 1981. {\it Ap.~J.} 250:79

\nhi Gordon KD, Hanson MM, Clayton GC, Rieke GH, Misselt KA. 1999. {\it Ap. J.}
     519:165

\nhi Graham AW. 2001. {\it Ap. J.} 121:820


\nhi Graham A, Lauer TR, Colless M, Postman M. 1996. {\it Ap. J.} 465:534

\nhi Greusard D, Friedli D, Wozniak H, Martinet L, Martin P. 2000. 
     {\it Astron. Astrophys. Suppl.} 145:425

\nhi Griv E, Chiueh T. 1998. {\it Ap. J.} 503:186


\nhi Hackwell JA, Schweizer F. 1983. {\it Ap. J.} 265:643


\nhi Hasan H, Norman C. 1990. {\it Ap.~J.} 361:69

\nhi Hasan H, Pfenniger D, Norman C. 1993. {\it Ap.~J.} 409:91





\nhi Helfer TT, Thornley MD, Regan MW, Wong T, Sheth K, et al. 2003.
     {\it Ap.~J.~Suppl.} 145:259


\nhi Heller CH, Shlosman I. 1996. {\it Ap. J.} 471:143


\nhi Ho LC, Filippenko AV, Sargent WLW. 1997. {\it Ap.~J.} 487:568 





\nhi Hughes MA, Alonso-Herrero A, Axon D, Scarlata C, Atkinson J, et al. 2003.
     {\it Astron. J.} 126:742

\nhi Hummel E, van der Hulst JM, Keel WC. 1987 
     {\it Astron. Astrophys.} 172:32

\nhi Illingworth G. 1977. {\it Ap.~J.} 218:L43


\nhi Illingworth G, Schechter PL. 1982. {\it Ap.~J.} 256:481

\nhi Inoue AK, Hirashita H, Kamaya H. 2000. {\it A.~J.} 120:2415


\nhi Jaffe W, Ford HC, O'Connell RW, van den Bosch FC, Ferrarese L. 1994. 
     {\it Astron. J.} 108:1567

\nhi Jarrett TH, Chester T, Cutri R, Schneider SE, Huchra JP.
     2003. {\it Astron. J.} 125:525

\nhi Jarvis B. 1987. {\it Astron.~J.} 94:30

\nhi Jarvis B. 1990. In {\it Dynamics and Interactions of Galaxies}, 
      ed. R Wielen, p.~416.  New York: Springer
 
\nhi Jarvis BJ. 1986. {\it Astron.~J.} 91:65

\nhi Jarvis BJ, Dubath P. 1988. {\it Astron. Astrophys.} 201:L33 

\nhi Jarvis BJ, Dubath P, Martinet L, Bacon R. 1988. 
      {Astr.~Ap.~Suppl.} 74:513


\nhi Jarvis BJ, Freeman KC. 1985. {\it Ap. J.} 295:324 

\nhi Jogee S. 1998. In {\it Molecular Gas and Star Formation in the Inner
     kpc of Starbursts and Non-Starbursts}, Ph.D. thesis, Yale University

\nhi Jogee S., Scoville NZ, Kenney JDP. 2004. {\it Ap.~J.} in press
     (astro-ph/0402341)

\nhi Jogee S, Shlosman I, Laine S, Englmaier P, Knapen J, et al. 2002.
     {\it Ap.~J.} 575:156


\nhi Jones DH, Mould JR, Watson AM, Grillmair C, Gallager JS, et al. 1996.
     {\it Ap. J.} 466:742

\nhi Jungwiert B, Combes F., Axon DJ. 1997. {\it Astron. Astrophys. Suppl.}
     125:479

\nhi Kalnajs A. 1973. {\it Proc.~Astron.~Soc. Aust.} 2:174

\nhi Kalnajs A. 1996. In {\it Proceedings of the Nobel Symposium 98,
     Barred Galaxies and Circumnuclear Activity\/}, ed. Aa Sandqvist, 
      PO Lindblad, p. 165 Berlin: Springer

\nhi Kannappan SJ, Jansen RA, Barton EJ. 2004. {\it Astron.~J.} 127:1371

\nhi Kenney JDP, Wilson CD, Scoville NZ, Devereux NA, Young JS. 1992. 
  {\it Ap. J. Letters} 395:L79

\nhi Kennicutt RC. 1983. {\it Ap. J.} 272:54


\nhi Kennicutt RC. 1994.  In {\it Mass-Transfer Induced Activity in
     Galaxies}, ed. I Shlosman, p. 131.  Cambridge: Cambridge Univ. Press

\nhi Kennicutt RC. 1998a. {\it Annu. Rev. Astr. Astrophys.} 36:189

\nhi Kennicutt RC. 1998b. {\it Ap. J.} 498:541


\nhi Kennicutt RC, Chu Y-H. 1988. {\it Astron. J.} 95:720

\nhi Kennicutt RC, Edgar BK, Hodge PW. 1989a. {\it Ap. J.} 337:761

\nhi Kennicutt RC, Keel WC, Blaha CA. 1989b. {\it Astron. J.} 97:1022

\nhi Kennicutt RC, Keel WC, van der Hulst JM, Hummel E, Roettiger KA.
     1987.  {\it Astron. J.} 93:1011

\nhi Kennicutt RC, Schweizer F, Barnes JE. 1998 {\it Galaxies:
     Interactions and Induced Star Formation} Berlin: Springer

\nhi Kent SM. 1985. {\it Ap.~J.~Suppl.} 59:115 

\nhi Kent SM. 1987a. {\it Astron. J.} 93:816   

\nhi Kent SM. 1987b. {\it Astron. J.} 93:1062  


\nhi Kent SM. 1988. {\it Astron.~J.} 96:514    




\nhi Kent SM, Dame TM, Fazio G. 1991. {\it Ap. J.} 378:131


\nhi Kewley LJ, Geller MJ, Jansen RA, Dopita MA. 2002. {\it Astron. J.}
     124:3135



\nhi Knapen JH, Beckman JE, Heller CH, Shlosman I, 
     de Jong RS. 1995a. {\it Ap. J.}, 454:623

\nhi Knapen JH, Beckman JE, Shlosman I, Peletier RF, 
     Heller CH, de Jong RS. 1995b. {\it Ap. J.}, 443:L73


\nhi Knapen JH, P\'erez-Ram\'\i rez D, Laine S. 2002. {\it MNRAS}
     337:808

\nhi Knapen JH, Shlosman I, Peletier RF. 2000. {\it Ap. J.} 529:93

\nhi Kohno K, Ishizuki S, Matsushita S, Vila-Vilar\'o B, Kawabe R. 2003.
     {\it PAS Japan} 55:L1


\nhi Kormendy J. 1979a. In {\it Photometry, Kinematics and Dynamics of
      Galaxies}, ed. DS Evans, p.~341. Austin: Dept.~of Astronomy, 
      Univ.~of Texas at Austin

\nhi Kormendy J. 1979b. {\it Ap.~J.} 227:714

\nhi Kormendy J. 1981. In {\it The Structure and Evolution of Normal 
      Galaxies}, ed. SM Fall, D Lynden-Bell, p.~85. Cambridge:
      Cambridge Univ. Press

\nhi Kormendy J. 1982a. In {\it Morphology and Dynamics of Galaxies, Twelfth
     Advanced Course of the Swiss Society of Astronomy and Astrophysics}, 
     ed. L Martinet, M Mayor, p. 113. Sauverny: Geneva Obs.

\nhi Kormendy J. 1982b. {\it Ap.~J.} 257:75


\nhi Kormendy J. 1984. {\it Ap. J.} 286:116






\nhi Kormendy J. 1988. {\it Ap.~J.} 335:40


\nhi Kormendy J. 1993. In {\it IAU Symposium 153, Galactic Bulges},
      ed. H Habing, H Dejonghe, p. 209. Dordrecht: Kluwer


\nhi Kormendy J, Bender R. 1996. {\it Ap. J.} 464:L119

\nhi Kormendy J, Bender R, Ajhar EA, Dressler A, Faber SM, et al. 1996a.
     {\it Ap.~J.} 473:L91   

\nhi Kormendy J, Bender R, Richstone D, Ajhar EA, Dressler A, et al. 1996b. 
     {\it Ap.~J.} 459:L57   

\nhi Kormendy J., Cornell ME. 2004. In {\it Penetrating Bars Through Masks
     of Cosmic Dust:
     The Hubble Tuning Fork Strikes a New Note}, ed. DL Block, KC Freeman,
     I Puerari, G Groess, in press.  Dordrecht: Kluwer

\nhi Kormendy J, Djorgovski S. 1989. {\it Annu. Rev. Astr. Astrophys.}
     27:235

\nhi Kormendy J, Dressler A, Byun Y-I, Faber SM, Grillmair C, et al. 1994. 
     In {\it ESO/OHP Workshop on Dwarf Galaxies}, ed. G Meylan, P Prugniel, 
     p. 147. Garching: ESO

\nhi Kormendy J, Gebhardt K. 2001. In {\it 20$^{th}$ Texas
      Symposium on Relativistic Astrophysics}, ed. JC Wheeler, H Martel,
      p. 363. Melville: AIP

\nhi Kormendy J, Gebhardt K, Macchetto FD, Sparks WBS. 2004, {\it Astron. J.}
     submitted

\nhi Kormendy J, Illingworth G. 1982. {\it Ap. J.} 256:460

\nhi Kormendy J, Illingworth G. 1983. {\it Ap.~J.} 265:632

\nhi Kormendy J, McClure RD. 1993. {\it Astron. J.} 105:1793

\nhi Kormendy J, Norman CA. 1979. {\it Ap. J.} 233:539

\nhi Kormendy J, Richstone D. 1992 {\it Ap. J.} 393:559



\nhi Krajnovi\'c D, Cappellari M, Emsellem E, McDermid R, de Zeeuw PT. 2004.
     In {\it IAU Symposium 220, Dark Matter in Galaxies}, ed. S Ryder, 
     DJ Pisano,
     M Walker, KC Freeman, in press.  San Francisco: ASP (astro-ph/0310645)


\nhi Kuijken K, Merrifield MR. 1995. {\it Ap. J.} 443:L13

\nhi Kuijken K, Rich RM. 2002. {\it Astron. J.} 124:2054

\nhi Kuntschner H. 2000. {\it MNRAS} 315:184

\nhi Kuntschner H, Davies RL. 1998. {\it MNRAS} 295:L29

\nhi Kuntschner H, Smith RJ, Colless M, Davies RL, Kaldare R, Vazdekis A. 2002.
     {\it MNRAS} 337:172


\nhi Laine S, Shlosman I, Knapen JH, Peletier RF. 2002. {\it Ap. J.} 567:97

\nhi Larson RB, Tinsley BM. 1978. {\it Ap. J.} 219:46

\nhi Lauberts A, Valentijn EA. 1989. {\it The Surface Photometry Catalogue of 
     the ESO-Uppsala Galaxies}. Garching: ESO

\nhi Lauer TR. 1985. {\it Ap. J. Suppl.} 57:473
 

\nhi Lauer TR. 1987. In {\it Nearly Normal Galaxies: From the Planck Time to the
     Present}, ed. SM Faber, p 207. New York: Springer

\nhi Lauer TR, Ajhar EA, Byun Y-I, Dressler A, Faber SM, et al. 1995.
      {\it Astron. J.} 110:2622

\nhi Lauer TR, Faber SM, Ajhar EA, Grillmair CJ, Scowen PA. 1998, 
     {\it Astron. J.}, 116, 2263   

\nhi Lauer TR, Faber SM, Currie DG, Ewald SP, Groth EJ, et al. 1992. 
     {\it Ap. J.} 104:522   

\nhi Lauer TR, Faber SM, Groth EJ, Shaya EJ, Campbell B. et al. 1993. 
     {\it Astron. J.} 106:1436     


\nhi Laurikainen E, Salo H. 2002. {\it MNRAS} 337:1118

\nhi Leitherer C, Schaerer D, Goldader JD, Gonz\'alez Delgado RM, 
     Robert C, et al. 1999. {\it Ap. J. Suppl.} 123:3


\nhi Lindblad B. 1958. {\it Stockholms Obs.~Ann.} 20:No.~6

\nhi Lindblad B. 1959. {\it Handbuch der Physik} 53:21

\nhi Lindblad PAB, Lindblad PO, Athanassoula E. 1996, {\it Astron.
     Astrophys.} 313:65

\nhi Lindblad PO. 1999. {\it Astron. Astrophys. Rev.} 9:221

\nhi Long KS, Charles PA, Dubus G. 2002. {\it Ap. J.} 569:204

\nhi Lourenso S, Vazdekis A, Peletier RF, Beckman JE. 2001. 
     {\it Astrophys. Space Sci.} 276:651

\hfuzz = 55pt

\nhi L\"utticke R, Dettmar R-J, Pohlen M. 2000a. {\it Astron. Astrophys. Suppl.}
     145:405

\nhi L\"utticke R, Dettmar R-J, Pohlen M. 2000b. {\it Astron. Astrophys.}
     362:435

\hfuzz = 5pt

\nhi Lynden-Bell D. 1979. {\it MNRAS} 187:101

\nhi Lynden-Bell D. 1996. In {\it Proceedings of the Nobel Symposium 98,
     Barred Galaxies and Circumnuclear Activity\/}, ed. Aa Sandqvist,
     PO Lindblad, p.~8.  New York: Springer

\nhi Lynden-Bell D, Kalnajs AJ. 1972. {\it MNRAS} 157:1

\nhi Lynden-Bell D, Wood, R. 1968, {\it MNRAS} 138:495

\nhi MacArthur LA, Courteau S, Holtzman JA. 2003. {\it Ap. J.} 582:689

\nhi Maciejewski W. 2003. (astro-ph/0302250)

\nhi Maciejewski W, Sparke, LS. 2000. {\it MNRAS} 313:745

\nhi Maciejewski W, Teuben PJ, Sparke LS, Stone JM. 2002. {\it MNRAS} 329:502

\nhi Maoz D. 2002. {\it Ap. J.} 455:L131  

\nhi Maoz D, Barth AJ, Ho LC, Sternberg A, Filippenko AV. 2001.
     {\it Astron. J.} 121:3048  

\nhi Maoz D, Barth AJ, Sternberg A, Filippenko AV, Ho LC, et al. 1996.
     {\it Astron. J.} 111:2248  

\nhi M\'arquez I, Durret F, Gonz\'alez Delgado RM, Marrero I, Masegosa J,
     et al. 1999. {\it Astron. Astrophys. Suppl.} 140:1

\nhi Martinet L. 1995. {\it Fund. Cosmic Physics} 15:341

\nhi Martini P, Pogge RW. 1999. {\it Astron. J.} 118:2646

\nhi Martini P, Regan MW, Mulchaey JS, Pogge RW. 2003.
     {\it Ap. J. Suppl.} 146:353


\nhi Matthews LD, Gallagher JS. 2002. {\it Ap. J. Suppl.} 141:429

\nhi Matthews LD, Gallagher JS, Krist JE, Watson AM, Burrows CJ., et al. 1999a.
     {\it Astron. J.} 118:208

\nhi Matthews LD, Gallagher JS, van Driel W. 1999b. {\it Astron.~J.}
     118:2751

\nhi McWilliam A, Rich RM. 1994. {\it Ap. J. Suppl.} 91:749

\nhi Mehlert D, Thomas D, Saglia RP, Bender R, Wegner G. 2003. 
     {\it Astron. Astrophys.} 407:423

\nhi Merrifield MR. 1996. In {\it IAU Colloquium 157, Barred Galaxies},
     ed. R Buta, DA Crocker, BG Elmegreen, p. 179.  San Francisco: ASP

\nhi Merrifield MR, Kuijken K. 1995, {\it MNRAS} 274:933

\nhi Merrifield MR, Kuijken K. 1999, {\it Astron. Astrophys.} 345:L47

\nhi Merritt D, Sellwood JA. 1994. {\it Ap. J.} 425:551


\nhi Mihalas D, Routly PM. 1968. {\it Galactic Astronomy}. San Francisco:
     Freeman

\nhi Miller RH, Smith, BF. 1979. {\it Ap. J.} 227:785

\nhi Minkowski R. 1962. In {\it IAU Symposium 15, Problems of Extra-Galactic
     Research}, ed. GC McVittie, p. 112.  New York: Macmillan

\nhi Minniti D, Olszewski EW, Rieke M. 1993. {\it Ap. J.} 410:L79

\nhi M\"ollenhoff C, Heidt J. 2001. {\it Astron. Astrophys.} 368:16

\nhi M\"ollenhoff C, Matthias M, Gerhard OE. 1995. {\it Astron. Astrophys.}
     301:359

\nhi Moore B. 1994. {\it Nature} 370:629

\nhi Morgan WW. 1951. {\it Publ. Obs. Univ. of Michigan} 10:33

\nhi Morgan WW. 1958. {\it Publ. Astron. Soc. Pacific} 70:364

\nhi Moriondo G, Giovanardi C, Hunt LK. 1998. {\it Astron. Astrophys. Suppl.} 
     130:81

\nhi Mulchaey JS, Regan MW. 1997. {\it Ap. J.} 482:L135

\nhi Mulchaey JS, Regan MW, Kundu A. 1997. {\it Ap. J. Suppl.} 110:299

\nhi Navarro JF, Eke VR, Frenk CS. 1996. {\it MNRAS} 283:L72

\nhi Navarro JF, Frenk CS, White SDM. 1996. {\it Ap. J.} 462:563   

\nhi Navarro JF, Frenk CS, White SDM. 1997. {\it Ap. J.} 490:493   


\nhi Nilson P. 1973. {\it Uppsala General Catalogue of Galaxies}.  Uppsala:
     Uppsala Astron. Obs.

\nhi Noguchi N. 1988. {\it Astron. Astrophys.} 203:259

\nhi Norman CA. 1984. In {\it Formation and Evolution of Galaxies and Large
     Structures in the Universe}, ed. J Audouze, JTT Van, p. 327. 
     Dordrecht: Reidel

\nhi Norman C, Hasan H. 1990. In {\it Dynamics and Interactions
      of Galaxies}, ed. R Wielen, p.~479. New York: Springer

\nhi Norman CA, Sellwood JA, Hasan H. 1996, {\it Ap. J.} 462:114

\nhi O'Connell RW. 1983. {\it Ap.~J.} 267:80

\nhi Ortolani S, Renzini A, Gilmozzi R, Marconi G, Barbuy B, et al. 1995.
     {\it Nature} 377:701


\nhi Paglione TAD, Wall WF, Young JS, Heyer MH, Richard M, et al. 2001. 
     {\it Ap. J. Suppl.} 135:183

\nhi Patsis PA, Athanassoula E, Grosb\o l P, Skokos Ch. 2002. {\it MNRAS}
     335:1049  


\nhi Patton DR, Pritchet CJ, Carlberg RG, Marzke RO, Yee HKC, et al. 2002.
     {\it Ap. J.} 565:208



\nhi Peletier RF, Balcells M. 1996. {\it Astron. J.} 111:2238

\nhi Peletier RF, Balcells M, Davies RL, Andredakis Y, 
     Vazdekis A, et al. 1999. {\it MNRAS} 310:703

\nhi Pence WD, Blackman CP. 1984. {\it MNRAS} 207:9

\nhi P\'erez-Ram\'\i rez D, Knapen JH, Laine S. 2001. {\it Astrophys.
     Space Sci.} 276:625

\nhi P\'erez-Ram\'\i rez D, Knapen JH, Peletier RF, Laine S, 
     Doyon R, Nadeau D. 2000. {\it MNRAS} 317:234

\nhi Pfenniger D. 1984. {\it Astron. Astrophys.} 134:373

\nhi Pfenniger D. 1985. {\it Astron. Astrophys.} 150:112

\nhi Pfenniger D. 1993. In {\it IAU Symposium 153, Galactic Bulges},
      ed. H Habing, H Dejonghe, p.~387. Dordrecht: Kluwer

\nhi Pfenniger D. 1996a. {\it In Barred Galaxies}, ed. JA Sellwood, p.~273.
     San Francisco: Astron. Soc. Pacific

\nhi Pfenniger D. 1996b. In {\it Proceedings of the Nobel Symposium 98,
     Barred Galaxies and Circumnuclear Activity\/}, ed. Aa Sandqvist,
     PO Lindblad, p.~91.  New York: Springer

\nhi Pfenniger D. 2000. In {\it Toward a New Millennium in Galaxy Morphology},
     ed. DL Block, I Puerari, A Stockton, D Ferreira p.~149. Dordrecht: Kluwer

\nhi Pfenniger D, Friedli D. 1991. {\it Astron. Astrophys.} 252:75

\nhi Pfenniger D, Norman C. 1990. {\it Ap.~J.} 363:391

\nhi Phillips AC. 1993. {\it Star Formation in Barred Spiral Galaxies},
     Ph.D. thesis, University of Washington

\nhi Phillips AC, Illingworth GD, MacKenty JW, Franx M. 1996. 
     {\it Astron.~J.} 111:1566

\nhi Piner BG, Stone JM, Teuben PJ. 1995. {\it Ap. J.} 449:508

\nhi Pogge RW. 1989. {\it Ap. J. Suppl.} 71:433

\nhi Pogge RW, Eskridge PB. 1993. {\it Astron. J.} 106:1405

\nhi Pompea SM, Rieke GH. 1990. {\it Ap. J.} 356:416

\nhi Prendergast KH. 1964. {\it Astron. J.} 69:147


\nhi Prendergast KH. 1983. In {\it IAU Symposium 100, Internal Kinematics 
     and Dynamics of Galaxies}, ed. E Athanassoula, p.215.  Dordrecht: 
     Reidel


\nhi Pritchet C. 1979. {\it Ap. J.} 231:354

\nhi Proctor RN, Sansom AE. 2002. {\it MNRAS} 333:517


\nhi Quillen AC, Frogel JA, Kuchinski LE, Terndrup DM. 1995.
     {\it Astron. J.} 110:156

\nhi Quillen AC, Kuchinski LE, Frogel JA, DePoy DL. 1997. {\it Ap. J.} 481:179

\nhi Raha N, Sellwood JA, James RA, Kahn FD. 1991. 
      {\it Nature} 352:411


\nhi Ravindranath S, Ho LC, Peng CY, Filippenko AV, Sargent WLW. 2001. 
     {\it Astron. J.} 122:653

\nhi Regan MW. 2000. {\it Ap. J.} 541:142

\nhi Regan MW, Sheth K, Vogel SN. 1999. {\it Ap. J.} 526:97

\nhi Regan MW, Teuben P. 2003. {\it Ap. J.} 582:723

\nhi Regan MW, Teuben P, Vogel SN, van der Hulst T. 1996.
     {\it Astron. J.} 112:2549

\nhi Regan MW, Vogel SN, Teuben PJ. 1997, {\it Ap. J.} 482:L143

\nhi Regan MW, Thornley MD, Helfer TT, Sheth K, Wong T, et al. 2001. 
     {\it Ap.~J.} 561:218

\nhi Renzini A. 1999. In {\it The Formation of Galactic Bulges}, ed. 
     CM Carollo, HC Ferguson, RFG Wyse, p.~9. Cambridge:
     Cambridge Univ. Press

\nhi Rest A, van den Bosch FC, Jaffe W, Tran H, Tsvetanov T, et al. 2001.
    {\it Astron. J.} 121:2431

\nhi Rich RM. 1999. In {\it The Formation of Galactic Bulges}, ed. 
     CM Carollo, HC Ferguson, RFG Wyse, p.~54. Cambridge:
     Cambridge Univ. Press





\nhi Rix H-W, Zaritsky D. 1995. {\it Ap. J.} 447:82

\nhi Rix H-WR, Kennicutt RC, Braun R, Walterbos RAM. 1995. {\it Ap. J.} 438:155

\nhi Roberts WW, Huntley JM, van Albada GD. 1979. {\it Ap.~J.}
     233:67

\nhi Roberts WW, Roberts MS, Shu FH. 1975. {\it Ap. J.} 196:381

\nhi Rocca-Volmerange B, Guideroni B. 1987. {\it Astron. Astrophys.} 175:15

\nhi Rubin VC. 1994. {\it Astron. J.} 107:173

\nhi Rubin VC, Ford WK, Peterson CJ. 1975. {\it Ap. J.} 199:39




\nhi Ryder SD, Knapen JH, Takamiya M. 2001. {\it MNRAS} 323:663

\nhi Saglia RP, Bender R, Dressler A. 1993. {\it Astron. Astrophys.} 279:75

\nhi Sakamoto K, Baker AJ, Scoville NZ. 2000. {\it Ap. J.} 533:149

\nhi Sakamoto K, Okamura SK, Ishizuki S, Scoville NZ. 1999. 
     {\it Ap. J.} 525:691

\nhi Sakamoto K, Okamura S, Minezaki T, Kobayashi Y, Wada K. 1995.
     {\it Astron. J.} 110:2075

\nhi Salo H, Rautiainen P, Buta R, Purcell GB, Cobb ML, et al. 1999. 
     {\it Astron. J.} 117:792

\nhi Sandage A. 1961. {\it The Hubble Atlas of Galaxies}. Washington: Carnegie
     Inst. of Washington

\nhi Sandage A. 1975. In {\it Stars and Stellar Systems, Vol.~9, Galaxies and
     the Universe}, ed. A Sandage, M Sandage, J Kristian, p.1.  Chicago:
     Univ. of Chicago Press

\nhi Sandage A. 1986. {\it Annu. Rev. Astr. Astrophys.} 24:421

\nhi Sandage A. 1990. {\it JRASC\/} 84:70

\nhi Sandage A, Bedke J. 1994. {\it The Carnegie Atlas of Galaxies}.
     Washington: Car- negie Inst. of Washington

\nhi Sandage A, Brucato R. 1979. {\it Astron. J.} 84:472





\nhi Sanders DB, Mirabel IF. 1996. {\it Ann. Rev. Astron. Astrophys.} 34:749

\nhi Sanders RH, Prendergast KH. 1974. {\it Ap. J.} 188:489

\nhi Sanders RH, Tubbs AD. 1980. {\it Ap. J.} 235:803

\nhi Sandqvist Aa, Lindblad PO. 1996. {\it Proceedings of the Nobel 
     Symposium 98, Barred Galaxies and Circumnuclear Activity\/}.  New York:
     Springer


\nhi Schinnerer E, B\"oker T, Meier DS. 2003. {\it Ap.~J.} 591:L115

\nhi Schinnerer E, Maciejewski W, Scoville N, Moustakas LA. 2002.
     {\it Ap. J.} 575:826

\nhi Schmidt AA, Bica E, Alloin D. 1990. {\it MNRAS} 243:620

\nhi Schmidt M. 1959. {\it Ap. J.} 129:243

\nhi Schwarz MP. 1981. {\it Ap.~J.} 247:77

\nhi Schwarz MP. 1984. {\it MNRAS} 209:93

\nhi Schweizer F. 1980. {\it Ap. J.} 237:303

\nhi Schweizer F. 1990.  In {\it Dynamics and Interactions of Galaxies\/}, 
     ed. R Wielen, p. 60. New York: Springer


\nhi Scorza C, Bender R. 1995. {\it Astron. Astrophys.} 293:20

\nhi Scoville NZ, Matthews K, Carico DP, Sanders DB. 1988. {\it Ap. J.} 327:L61

\nhi Seifert W, Scorza C. 1996. {\it Astron. Astrophys.} 310:75

\nhi Seigar M, Carollo CM, Stiavelli M, de Zeeuw PT, Dejonghe H.
     2002. {\it Astron. J.} 123:184

\nhi Seigar MS, James PA. 1998. {\it MNRAS} 299:672

\nhi Sellwood JA. 1981. {\it Astron. Astrophys.} 99:362

\nhi Sellwood JA. 1985. {\it MNRAS} 217:127

\nhi Sellwood JA. 1993a. In {\it IAU Symposium 153, Galactic Bulges},
     ed. H Habing, H Dejonghe, p.~391. Dordrecht: Kluwer

\nhi Sellwood JA. 1993b. In {\it Back to the Galaxy}, ed. SS Holt,
     F Verter, p. 133.  New York: AIP

\nhi Sellwood JA, Debattista VP. 1996. In {\it Proceedings of Nobel
     Symposium 98, Barred Galaxies and Circumnuclear Activity\/}, 
     ed. Aa Sandqvist, PO Lindblad, p.~43.  New York: Springer

\nhi Sellwood JA, Moore EM. 1999. {\it Ap. J.} 510:125

\nhi Sellwood JA, Shen J. 2004. In {\it Carnegie Observatories Astrophysics
     Series, Vol. 1: Coevolution of Black Holes and Galaxies}, ed. LC Ho,
     in press.  Cambridge: Cambridge Univ. Press (astro-ph/0303611)

\nhi Sellwood JA, Sparke LS. 1988. {\it MNRAS} 231:25P
 
\nhi Sellwood JA, Wilkinson A. 1993. {\it Rep. Prog. Phys.}, 56:173

\nhi Sempere MJ, Garc\'\i a-Burillo S, Combes F, Knapen JH. 1995. 
     {\it Astron. Astrophys.} 296:45

\nhi S\'ersic JL. 1968. {\it Atlas de Galaxias Australes}.  C\'ordoba:
     Obs. Astron. Univ. Nac. C\'ordoba

\nhi S\'ersic JL. 1973. {\it PASP} 85:103

\nhi S\'ersic JL, Pastoriza M. 1965. {\it PASP} 77:287

\nhi S\'ersic JL, Pastoriza M. 1967. {\it PASP} 79:152

\nhi Shaw M. 1993a. {\it Astron. Astrophys.} 280:33

\nhi Shaw M. 1993b. {\it MNRAS} 261:718

\nhi Shaw M, Axon D, Probst R, Gatley I. 1995. {\it MNRAS} 274:369

\nhi Shaw M, Dettmar R-J, Barteldrees A. 1990. {\it Astron. Astrophys.} 240:36

\nhi Shaw M, Wilkinson A, Carter D. 1993a. {\it Astron.~Astrophys.} 268:511

\nhi Shaw MA. 1987. {\it MNRAS} 229:691

\nhi Shaw MA, Combes F, Axon DJ, Wright GS. 1993b. {\it Astron. Astrophys.}
     273:31

\nhi Shen J, Sellwood JA. 2004. In {\it Carnegie Observatories Astrophysics
     Series, Vol. 1: Coevolution of Black Holes and Galaxies}, ed. LC Ho,
     in press.  Cambridge: Cambridge Univ. Press (astro-ph/0303130)

\nhi Sheth K, Vogel SN, Regan MW, Teuben PJ, Harris AI, Thornley MD.
     2002. {\it Astron. J.} 124:2581

\nhi Shields JC, Filippenko AV. 1992. In {\it Relationships Between Active
     Galactic Nuclei and Starburst Galaxies}, ed. AV Filippenko, p. 267.
     San Francisco: ASP



\nhi Shlosman I, Frank J, Begelman MC. 1989. {\it Nature} 338:45


\nhi Simien F, de Vaucouleurs G. 1986. {\it Ap. J.} 302:564

\nhi Simien F, Prugniel Ph. 1997. {\it Astron. Astrophys. Suppl.} 126:15

\nhi Simkin SM, Su HJ, Schwarz MP. 1980. {\it Ap. J.} 237:404


\nhi Sofue Y, Tomita A, Honma M, Tutui Y. 1999. {\it PASJ} 51:737

\nhi S\o rensen S-A, Matsuda T, Fujimoto M. 1976. {\it Astrophys. 
     Space Sci.} 43:491

\nhi Sparke LS, Gallagher JS. 2000. {\it Galaxies in the Universe: 
     An Introduction} Cambridge: Cambridge Univ. Press


\nhi Sparke LS, Sellwood JA. 1987. {\it MNRAS} 225:653

\nhi Spillar EJ, Oh SP, Johnson PE, Wenz M. 1992. {\it Astron. J.} 103:793

\nhi Spitzer L, Schwarzschild M. 1951. {\it Ap. J.} 114:385

\nhi Spitzer L, Schwarzschild M. 1953. {\it Ap. J.} 118:106


\nhi Talbot RJ, Jensen EB, Dufour RJ. 1979. {\it Ap. J.} 229:91







\nhi Terndrup DM. 1993. In {\it The Minnesota Lectures on Clusters of Galaxies
     and Large-Scale Structure}, ed. RM Humphreys, p. 9. San Francisco: ASP

\nhi Thomas D, Greggio L, Bender R. 1999. {\it MNRAS} 302:537

\nhi Thomas D, Maraston C, Bender R. 2002. {\it Astrophys. Space Sci.} 281:371

\nhi Thronson HA, Hereld M, Majewski S, Greenhouse M, Johnson P, et al. 1989.
     {\it Ap.~J.} 343:158

\nhi Toomre A. 1964. {\it Ap. J.} 139:1217

\nhi Toomre A. 1966. In {\it Geophysical Fluid Dynamics, Notes on the 1966
     Summer Study Program at the Woods Hole Oceanographic Institution},
     No. 66--46, p. 111

\nhi Toomre A. 1977a.  In {\it The Evolution of Galaxies and Stellar 
     Populations\/}, ed. BM Tinsley, RB Larson, p.~401.  New Haven:
     Yale University Observatory

\nhi Toomre A. 1977b. {\it Annu. Rev. Astr. Astrophys.} 15:437

\nhi Toomre A, Toomre J. 1972. {\it Ap.~J.} 178:623

\nhi T\'oth G, Ostriker JP. 1992. {\it Ap.~J.} 389:5

\nhi Trager SC. 2004. In {\it Carnegie Observatories Astrophysics Series,
     Vol.~4, Origin and Evolution of the Elements}, ed. A McWilliam,
     M Rauch, in press Cambridge: Cambridge Univ. Press (astro-ph/0307069)

\nhi Tremaine S. 1989. In {Dynamics of Astrophysical Disks}, 
     ed. JA Sellwood, p.~231.  Cambridge: Cambridge Univ.~Press

\nhi Tremaine S, Ostriker JP. 1982. {\it Ap. J.} 256:435

\nhi Tremaine SD, Ostriker JP, Spitzer L. 1975. {\it Ap. J.} 196:407

\nhi Trujillo I, Asensio Ramos A, Rubi\~no-Mart\'\i n JA, Graham AW,
     Aguerri JAL, et al. 2002. {\it MNRAS} 333:510

\nhi Usui T, Saito M, Tomita A. 1998. {\it A.~J.} 116:2166

\nhi Usui T, Saito M, Tomita A. 2001. {\it A.~J.} 121:2483


\nhi van den Bergh S. 1960a. {\it Ap.~J.} 131:215

\nhi van den Bergh S. 1960b. {\it Ap.~J.} 131:558

\nhi van den Bergh S. 1976a. {\it Ap.~J.} 203:764  

\nhi van den Bergh S. 1976b. {\it Ap.~J.} 206:883  

\nhi van den Bergh S. 1976c. {\it Astron. J.} 81:795 

\nhi van den Bergh S. 1986. {\it Astron. J.} 91:271

\nhi van den Bergh S. 1991. {\it PASP} 103:609

\nhi van den Bergh S. 1995. {\it Astron. J.} 110:613

\nhi van den Bergh S. 2002. {\it PASP} 114:797

\nhi van den Bergh S, Abraham RG, Whyte LF, Merrifield MR, Eskridge PB, et al.
     2002. {\it Astron. J.} 123:2913

\nhi van den Bosch FC, Jaffe W, van der Marel RP. 1998. {\it MNRAS} 293:343

\nhi van der Kruit PC. 1974. {\it Ap. J.} 188:3

\nhi van der Kruit PC. 1976. {\it Astron. Astrophys.} 52:85

\nhi van der Kruit PC, Jim\'enez-Vicente J, Kregel M, 
     Freeman KC. 2001. {\it Astron.~Astrophys.} 379: 374

\nhi van der Kruit PC, Searle L. 1981a. {\it Astron. Astrophys.} 95:105

\nhi van der Kruit PC, Searle L. 1981b. {\it Astron. Astrophys.} 95:116

\nhi van der Kruit PC, Searle L. 1982. {\it Astron. Astrophys.} 110:61

\nhi van Dokkum PG, Franx M. 1995. {\it Astron. J.} 110:2027

\nhi van Driel W, Rots AH, van Woerden H. 1988. {\it Astron.~Astrophys.}
     204:39

\nhi Vega Beltr\'an JC, et al. 1998. {\it Astron. Astrophys. Suppl.} 
     131:105

\nhi Vega Beltr\'an JC, Pizella A, Corsini EM, Funes JG, Zeilinger WW,
     et al. 2001. {\it Astron. Astrophys.} 374:394

\nhi Verolme EK, Cappellari M, Copin Y, van der Marel RP, Bacon R., et al.
     2002. {\it MNRAS} 335:517  

\nhi Wada K. 2003. In {\it Carnegie Observatories Astrophysics Series, Vol.~1,
     Coevolution of Black Holes and Galaxies}, ed. LC Ho, in press
     Cambridge: Cambridge Univ. Press (astro-ph/0308134)


\nhi Wagner SJ, Dettmar R-J, Bender R. 1989. {\it Astron. Astrophys.} 215:243

\nhi Wakamatsu K-I, Hamabe M. 1984. {\it Ap. J. Suppl.} 56:283

\nhi Waller WH, Fanelli MN, Keel WC, Bohlin R, Collins NR,
     at al. 2001. {\it Astron. J.} 121:1395

\nhi Walterbos RAM, Braun R, Kennicutt RC. 1994. {\it Astron. J.} 107:184

\nhi Warren MS, Quinn PJ, Salmon PJ, Zurek WH. 1992. {\it Ap.~J.} 399:405


\nhi Weiland JL, Arendt RG, Berriman GB, Dwek E, Freudenreich HT., et al.
     1994. {\it Ap. J.} 425:L81

\nhi Weiner BJ, Sellwood JA, Williams TB. 2001a. {\it Ap. J.} 546:931

\nhi Weiner BJ, Williams TB, van Gorkom JH, Sellwood JA. 2001b.
     {\it Ap. J.} 546:916





\nhi Whyte LF, Abraham RG, Merrifield MR, Eskridge PB, Frogel JA, Pogge RW. 
     2002. {\it MNRAS} 336:1281

\nhi Wilson CD. 1995. {\it Ap. J.} 448:L97


\nhi Windhorst RA, Taylor VA, Jansen RA, Odewahn SC, 
     Chiarenza CAT, et al. 2002. {\it Astrophys. J. Suppl.} 143:113

\nhi Wong T, Blitz L. 2000. {\it Ap. J.} 540:771

\nhi Wong T, Blitz L. 2001. {\it Astrophys. Space Sci.} 276:659

\nhi Worthey G, Faber SM, Gonzalez JJ. 1992. {\it Ap. J.} 398:69

\nhi Wozniak H, Friedli D, Martinet L, Martin P, Bratschi P. 1995. 
     {\it Astron. Astrophys. Suppl.} 111:115



\nhi Wyse RFG, Gilmore G, Franx M. 1997. 
     {\it Annu.~Rev.~Astr.~Astrophys.} 35:637

\nhi Young JS, Devereux NA. 1991. {\it Ap. J.} 373:414

\nhi Young JS, Scoville NZ. 1991. {\it Annu. Rev. Astron. Astrophys.}
     29:581

\nhi Yun MS, Ho PTP, Lo KY. 1994. {\it Nature} 372:530

\nhi Zhang Q, Fall SM. 1999. {\it Ap. J.} 527:L81


\nhi Zhang X. 1996. {\it Ap. J.} 457:125

\nhi Zhang X. 1998. {\it Ap. J.} 499:93

\nhi Zhang X. 1999. {\it Ap. J.} 518:613

\nhi Zhang X. 2003. {\it J.~Korean Astr.~Soc.} 36:223 (astro-ph/0301655)

\nhi Zoccali M, Renzini A, Ortolani S, Greggio L, Saviane I, et al. 2003.
     {\it Astron. Astrophys.} 399:931

\nhi Zwicky F. 1957. {\it Morphological Astronomy\/} Berlin: Springer

\vfill\eject

\end

\end